\documentclass[a4paper,12pt,oneside]{book}
\usepackage[dvips]{graphics,color}
\newcommand{\Sc}{{\mathcal S}^\dagger}
\newcommand{\Sa}{{\mathcal S}}
\newcommand{\Nb}{{\mathcal N}_B}
\newcommand{\nf}{{\mathcal N}_F}
\newcommand{\Hil}{{\mathcal H}}
\newcommand{\En}{{\mathcal E}}
\newcommand{\com}[2]{\left[ #1\, , #2 \right]}
\newcommand{\acom}[2]{\left\{ #1\, , #2 \right\}}
\newcommand{\Salg}{{\mathcal S}(C,\tilde{C},\chi,\tilde{\chi})}
\newcommand{\Salgf}{{\mathcal S}(C,C^*,\chi,\chi^*)}
\newcommand{\SRep}[1]{{\mathcal R_S}(C,\tilde{C},\chi,\tilde{\chi},#1)}
\newcommand{\SRepf}{{\mathcal R_S}(C,C^*,\chi,\chi^*, g\delta_i^j)}
\newcommand{\Ou}{{\mathcal O}^\uparrow}
\newcommand{\Od}{{\mathcal O}^\downarrow}
\newcommand{\hw}{\left|i \right>}
\newcommand{\hwb}{\left<f \right|}
\newcommand{\vf}{\left| 0 \right>}
\newcommand{\vfb}{\left< 0 \right|}

\newcommand{\opu}{\hat{O}^\uparrow}
\newcommand{\opd}{\hat{O}^\downarrow}
\newcommand{\op}{\hat{O}}
\newcommand{\rf}[1]{{\left(#1\right)_F}}
\newcommand{\rg}[2]{{\left(#1\right)_{#2}}}

\newcommand{\rbf}[1]{{\left(#1\right)_{BF}}}
\newcommand{\rx}[1]{{\left(#1\right)_{X,BF}}}
\newcommand{\D}[1]{{\textstyle #1}}
\newcommand{\I}{{\rm I}}
\newcommand{\uvec}[2]{\left|#1\right)\otimes\left|#2\right>}
\newcommand{\ufun}[2]{\left(#1\right|\otimes\left<#2\right|}
\newcommand{\wt}[1]{\widetilde{#1}}
\newcommand{\mc}[1]{{\mathcal #1 }}
\newcommand{\eig}{\left|\psi,\lambda\right)}
\newcommand{\tx}[1]{\tilde{X}_{#1}(\lambda)}
\newcommand{\Gf}{{\mathcal G_F}}
\newcommand{\Gbf}{{\mathcal G_{BF}}}
\newcommand{\MF}{V_F}
\newcommand{\MBF}{V_{BF}}
\newcommand{\fvec}[1]{\left|#1\right>}
\newcommand{\bvec}[1]{\left|#1\right)}
\newcommand{\rvec}[1]{\left|#1_R\right)}
\newcommand{\lvec}[1]{\left|#1_L\right)}
\newcommand{\lfunc}[1]{\left(#1^{(0)}_L\right|}

\newcommand{\nb}{{\mathcal N}_B}

\newtheorem{theorem}{Theorem}
\newtheorem{corollary}{Corollary}
\title{Analysis and Applications of the\\ Generalised Dyson Mapping}
\author{Izak Snyman\thanks{The author gratefully aknowledges the generous financial support 
received from the NRF and the Harry Crossley Foundation during the writing of this work.}\\
Institute of Theoretical Physics\\
University of Stellenbosch}
\begin{document}
\frontmatter
\begin{titlepage}
\begin{center}
\vspace*{2cm}
{\huge Analysis and Applications of the\\\vspace{1mm}Generalised Dyson Mapping}\\ 
\vspace{2mm}
{\large by\\}
\vspace{2mm}
{\sc Izak Snyman\footnote{E-mail: izak@sun.ac.za}}\\
\vspace{9cm}
\normalsize Thesis presented in partial fulfilment of the requirements for the degree of\\
MASTER OF SCIENCE at the University of Stellenbosch.\\
\vspace{5mm}
\begin{tabular}[h]{ll}
Supervisors:&Prof. H.B. Geyer\\
&Prof. F.G. Scholtz
\end{tabular}
\\ \vspace{5mm}
December 2004
\end{center}
\end{titlepage}
\newpage
\begin{center}
{\Large ABSTRACT}
\end{center}
In this thesis, generalized Dyson boson-fermion mappings are considered. These are techniques
used in the analysis of the quantum many-body problem, and are instances of so-called boson
expansion methods. A generalized Dyson boson-fermion mapping, or a Dyson mapping for short,
is a one-to-one linear but non-unitary operator that can be applied to vectors representing the 
states of a 
many-fermion system. A vector representing a fermion system maps onto a vector that is most 
naturally interpreted as representing
a state of a many-body system that contains both bosons and fermions. The motivation for
doing such a mapping is the hope that the mapping will reveal some property of the
system that simplifies its analysis and that was hidden in the original form.
The aims of this thesis are
\begin{enumerate}
\item{to review the theory of generalized Dyson boson-fermion mappings,}
\item{by considering a tutorial example, to demonstrate that it is feasible to implement the
theory and}
\item{to find a useful application for a generalized Dyson boson-fermion mapping, by considering a 
non-trivial model, namely the Richardson model for superconductivity.}
\end{enumerate}
The realization of the first two aims mainly involve the collecting together of ideas that have 
already
appeared in the literature, into one coherent text. Some subtle points that were treated only 
briefly due to space restrictions in the journal publications where the theory was first expounded,
are elaborated on in the present work. On the other hand, the ana\-lysis of the Richardson 
Hamiltonian that uses a Dyson mapping, goes beyond what has already appeared in the literature. It
is the first time that a boson expansion technique is implemented for a system where the roles of
both collective and non-collective fermion pairs are important. (The Dyson mapping associates 
bosons with Cooper pairs, while the fermions not bound in Cooper pairs result in fermions being 
present in the mapped system as well.) What is found is that the Dyson mapping uncovers 
non-trivial properties of the system.  These properties aid the construction of time-independent 
perturbation expansions 
for the stationary states of the system, as well as time-dependent expansions for transition
amplitudes between states. The time-independent expansions agree with results that other authors
obtained through methods other than boson expansions. The time-dependent expansions, that one
would be hard-pressed to develop without a Dyson mapping, might in 
future prove useful in understanding aspects of the dynamics of ultra-cold fermi gases, when
time-dependent magnetic fields are used to vary the atom-atom interaction strenght.
\newpage
\vspace*{3cm}
\begin{center}
{\Large ACKNOWLEDGEMENT}\\
\end{center}
The author gratefully acknowledges the generous financial support received from the NRF
and the Harry Crossley Foundation while writing this thesis. Views expressed herein are those
of the author and not necessarily those of either the NRF or the Harry Crossley Foundation 
(although, if they examined carefully these pages, both institutions, the author believes, 
would agree with all they read).
\tableofcontents
\newpage
\begin{minipage}[t][18.5cm][c]{14cm}
\begin{quote}
{\bf Molecule,} {\em n}. The ultimate, indivisible unit of matter. It is distinguished from the
corpuscle, also the ultimate, indivisible unit of matter, by a closer resemblance to the atom,
also the ultimate, indivisible unit of matter. Three great scientific theories of the structure
of the universe are the molecular, the corpuscular and the atomic. A fourth affirms, with 
Haeckel, the condensation or precipitation of matter from ether - whose existence is proved by the 
condensation or precipitation. The present trend of scientific thought is toward the theory of 
ions. The ion differs from the molecule, the corpuscle and the atom in that it is an ion. A fifth
theory is held by idiots, but it is doubtful if they know any more about the matter than the others.

{\sc Ambrose Bierce} The Devil's Dictionary (1911)
\end{quote}
\end{minipage}
\mainmatter
\section*{Introduction and Orientation}
\addcontentsline{toc}{chapter}{Introduction and Orientation}
This thesis is concerned with a class of linear transforms that can be applied to the state
vectors of some many-body quantum systems. Transforms belonging to this class are known
as `generalized Dyson boson-fermion mappings'.\footnote{This nomenclature is an example
of a variant of the 
Matthew effect \cite{Me1}. In the Gospel According to Matthew (Chapter 25, verse 29) there is 
written: ``For unto every one that hath shall be
given, and he shall hath abundance: but from him that hath not shall be taken away even that which
he hath.'' In scientific research, the Matthew effect refers to the tendency to attach an eminent 
name 
to every result, even when the eminent person to whom the name belongs has little to do with the
result. In this case, Freeman Dyson cannot be said to have had much to do with the development of
the theory expounded in these pages. (In a gesture of East-West reconciliation the name `Maleev' is
sometimes appended after Dyson's name. Maleev's contribution \cite{Ma1} to the theory expounded in 
these pages is on par with that of Dyson \cite{Dy1}.)} I make no attempt in the introduction to 
define or 
explain precisely what such a mapping is. This is done in the main body of the text 
(in Section \ref{SS1.1}). Here I simply try to indicate the position occupied by the content of
this text, in relation to other works on the same or similar subjects, and in relation to larger 
fields of study.

Dyson mappings have been around in various incarnations for roughly forty years. These mappings 
form part of a larger field of study called `Boson Expansion Methods'. The main motivation in
the development of boson expansion methods were the need for approximations to analyse the 
nuclear many-body problem. A very complete review of this program can be found in \cite{KM1}.
Also, a detailed introductory exposition is contained in Chapter 9 of \cite{RS1}. However, 
the space dedicated in these texts to mappings of the generalized Dyson type is relatively little.
To get a better idea of the specific foundation of generalized Dyson type mappings, as well as 
their relation to other branches of the field of boson expansion methods, one can consult 
\cite{Do1}.

The work reported on and extended in this thesis, concerning generalized Dyson boson-fermion 
mappings, were published in the decade after the last of the previously mentioned texts appeared.
As a result, the theory described in these pages has not yet undergone the same `canonization'
as the earlier results. This leaves me with the following tasks: In Chapter One the theory of
generalized Dyson boson-fermion mappings is presented. The argument is developed in a self-contained
manner that closely follows that described in \cite{DSG1}-\cite{CG1}. Because the space constraints
inherent in journal publications are not an issue here, the argument can be given in full, with
no or few open questions and unproven results.

Since I have not yet defined what a generalized Dyson boson-fermion mapping is, it is hard at
this point to explain why these mappings are studied or where they are useful. (This problem
is only partially resolved after these mappings are properly defined.) The course I therefore take
is first to develop the theory properly, as if for its own sake, rather than for its usefulness. 
Afterwards we can then look for properties of the mapping that might make it useful in some
situations. What we will find is that, unlike other boson expansion methods, a general Dyson
mapping does in itself not constitute an approximation method. Rather, it is a transformation that 
may cast a system into an equivalent form, that is more convenient for certain types of exact or
approximate analyses.

Chapters Two and Three are concerned with applications. In Chapter Two, a nuclear toy model with
$SO(5)$ symmetry is considered. Here the emphasis is mostly on demonstrating that it is feasible
to replace the abstract symbols of the general theory with concrete numbers, vectors and operators,
and that the resulting mapped system can be analysed in much the same way as the original. Readers
will however probably not take from this chapter a sense of what makes a general Dyson mapping a
useful analytical tool.

In Chapter Three, I try to remedy this by analysing a highly non-trivial `realistic' model that
describes Cooper pairing in a small superconductor. After applying a generalized Dyson mapping,
I do a perturbative analysis of the mapped system, using ordinary perturbation theory. 
(Here I mean the standard techniques of Rayleigh-Schr\"odinger perturbation theory to
find the approximate stationary states of the system, and Dirac perturbation theory to determine
its approximate time-evolution.) It turns out that perturbation expansions are easier to make
for the mapped system than for the original.

\chapter{The Generalized Dyson Mapping: Theory}
\label{Ch1}
\section{Bird's Eye View}
\markright{\bf Section \thesection: Bird's Eye View} 
\label{SS1.1}
This chapter is concerned with the process of translating the mathematical description 
of a quantum system in a Hilbert space $\Hil$ into an equivalent description in 
another Hilbert space $\tilde{\Hil}$. To loosen our limbs and set our terminology, we start by
mentioning a few things that every reader of this text is probably familiar with already.
It is convenient to distinguish three
components in the description of quantum systems, 
namely (1) state vectors belonging to a Hilbert space $\Hil$, (2)
linear operators that map $\Hil$ into itself and (3) linear functionals defined on $\Hil$.
All information about the system can be derived by combining these components to form 
so-called matrix elements of operators. 

The following concerns the construction of such matrix elements: 
In $\Hil$, each vector $\left|\psi\right>$ can be mapped onto a linear functional 
$f_\psi:\Hil_F\rightarrow{\bf C}$, where, for each $\left|\phi\right>$ in $\Hil$, the action 
of $f_\psi$ is defined as
$f_\psi\left|\phi\right>=\left<\psi\right|\left.\phi\right>$. Here 
$\left<\psi\right|\left.\phi\right>$ is the inner product between $\left|\psi\right>$ and 
$\left|\phi\right>$ in $\Hil$. As everyone knows, we often write $\left<\psi\right|$ instead 
of $f_\psi$. The mapping
$\left|\psi\right>\mapsto\left<\psi\right|$ is called the canonical map.
An arbitrary matrix element of an operator 
$\hat{O}:\Hil\rightarrow\Hil$ is a complex number $\left<\psi\right|\hat{O}\left|\phi\right>$.  
The process of mapping should preserve these numbers.
It is accomplished in a straight forward 
manner if a one-to-one and onto linear operator
$\mathcal T:\Hil\rightarrow\tilde{\Hil}$ can be found. $\mathcal T$ is called an isomorphism.
For the various components, the mapping is implemented as follows:
\begin{enumerate}
\item {\em Vectors}: Every vector $\left|\psi\right>$ in $\Hil$ is mapped onto a vector
$\mathcal T\left|\psi\right>$ in $\tilde{\Hil}$.
\item{\em Linear Operators}: Every linear operator $\hat{O}:\Hil\rightarrow\Hil$ is
mapped onto a linear operator $\mathcal T\hat{O}\mathcal T^{-1}:\tilde{\Hil}\rightarrow
\tilde{\Hil}$. 
Note that the mapping
$\hat{O}\mapsto \mathcal T\hat{O}T^{-1}$ is linear and invertible.
\item{\em Linear Functionals}: Every linear functional $f:\Hil\rightarrow{\bf C}$ maps onto
a linear functional $f\,\mathcal T^{-1}:\tilde{\Hil}\rightarrow{\bf C}$. 
\end{enumerate}
The objects $\mathcal T\left|\psi\right>$, $\mathcal T\hat{O}\mathcal T^{-1}$ and $f\,\mathcal 
T^{-1}$ are said to be the images of respectively
$\left|\psi\right>$, $\hat{O}$ and $f$ under the action of $\mathcal T$. 
Since $\left<\psi\right|\hat{O}\left|\phi\right>
=\left(\left<\psi\right|\mathcal T^{-1}\right)\left(\mathcal T\hat{O}\mathcal T^{-1}\right)
\left(\mathcal T\left|\phi\right>\right)$
it is immaterial whether we use $\left|\psi\right>$, $\left|\phi\right>$ and $\hat{O}$ or their
images to calculate matrix elements. The reason for insisting on the linearity of $\mathcal T$ is 
to ensure a duplication of the linear structure of the original system in its mapped counterpart.
For instance, if the system is described by a superposition 
$\alpha\left|\psi\right>+\beta\left|\phi\right>$
then the mapped system is described by an equivalent superposition
$\alpha \mathcal T\left|\psi\right>+\beta \mathcal T\left|\phi\right>$. Operators and functionals 
also preserve their linearity during mapping. 

We will deal with mappings $\mathcal T$ that are not unitary. On this topic a small aside: 
It has to be remembered that the hermiticity of operators is not preserved by non-unitary
mappings. Also, consider a vector $\left|\psi\right>$ in $\Hil$ that maps onto a vector 
$\left|\psi\right)=\mathcal T\left|\psi\right>$ in $\tilde{\Hil}$. 
$\left<\psi\right|$ is the functional 
obtained from $\left|\psi\right>$ through the canonical map in $\Hil$, and 
$\left(\psi\right|$ is the functional obtained from $\left|\psi\right)$ through the canonical
map in $\tilde{\Hil}$.\footnote{ This notation will be used throughout
the text: In $\Hil$, the space from which we map, vectors and functionals will be indicated with 
triangular brackets. In the space $\tilde{\Hil}$ into which we map, round brackets will be used.}
When $\mathcal T$ is not unitary, then in general 
$\left(\psi\right|\not=\left<\psi\right|\mathcal T^{-1}$.

Let us now become more specific. In this chapter we will derive one particular type of mapping 
$\mathcal T$, called a generalized Dyson boson-fermion mapping, or a Dyson mapping for 
short. A Dyson mapping $\mathcal T$ has as its domain 
the many-fermion Hilbert space $\Hil_F$ (or a subspace $V_F$ thereof), that is constructed 
from the one-particle Hilbert space 
$\Hil_{1,F}$, where $B_{1,F}=\left\{\left|\nu\right>\right\}_{\nu=1}^{\Omega}$ is an orthonormal 
basis for $\Hil_{1,F}$. With the states that comprise this basis are associated fermion 
creation operators $a^\nu;\:\nu=1,2,\ldots,\Omega$, that are conjugated to annihilation operators
$a_\nu=(a^\nu)^+$, where $+$ denotes hermitian conjugation in $\Hil_F$. The usual anti-commutation
relations between these operators hold. $\Hil_F$ is finite dimensional. The vacuum in $\Hil_F$
is denoted $\left|0\right>$. A general vector in $\Hil_F$ is denoted by $\left|\psi\right>$. The
functional related to this vector through the canonical map of $\Hil_F$ is denoted 
$\left<\psi\right|$. The subspace $V_F$ of $\Hil_F$ that serves as the domain of the mapping
$\mathcal T$ might also be the whole $\Hil_F$. 

The mapping $\mathcal T$ maps vectors in the domain $V_F$, representing states of a many-fermion 
system, onto vectors in the Hilbert space $\Hil_{BF}$, representing states of a many-body system
where some particles are fermions and others are bosons.
The Hilbert space $\Hil_{BF}$ is defined as follows: Let $\Hil_B$ be the many-boson Hilbert space 
constructed from the one-particle Hilbert space 
$\Hil_{1,B}$, where $B_{1,B}=\left\{\left|k\right>\right\}_{k=1}^{M}$ is an orthonormal 
basis for $\Hil_{1,B}$. With the states that comprise this basis are associated boson 
creation operators $b^k;\:k=1,2,\ldots,M$ that are conjugated to annihilation operators
$b_k=(b^k)^+$, where we use the same symbol as before to denote hermitian conjugation, but this 
time in $\Hil_B$. The standard commutation relations between these operators hold. 
$\Hil_B$ is infinite dimensional. The vacuum in $\Hil_B$ is denoted $\left|0\right>_B$.
The infinite dimensional many-body Hilbert space $\Hil_{BF}$ is constructed by setting
$\Hil_{BF}=\Hil_B\otimes\Hil_F$. In this space, the operators $B^k=b^k\otimes{\rm I}$ create 
bosons. 
Conjugated to these are $B_k=b_k\otimes{\rm I}=(B^k)^\dagger$ which annihilate bosons. Here 
$\dagger$ denotes hermitian conjugation in $\Hil_{BF}$.\footnote{A different symbol is used 
as a precaution to avoid confusion, because we study isomorphisms between
$V_F$ and subspaces of $\Hil_{BF}$, that are not unitary. In other words, we work
with vector space isomorphisms, rather than Hilbert space isomorphisms.} The usual
commutation relations between these operators still hold. Similarly 
$\alpha^\nu={\rm I}\otimes a^\nu$ and $\alpha_\nu={\rm I}\otimes a_\nu=(\alpha^\nu)^\dagger$
respectively create and annihilate fermions in $\Hil_{BF}$ with the standard 
anti-commutators in place. The vacuum in $\Hil_{BF}$ is 
$\left|0\right)=\left|0\right>_B\otimes\left|0\right>$. A general vector in
$\Hil_{BF}$ is denoted $\left|\psi\right)$ and the functional related to it by the 
canonical map of $\Hil_{BF}$ by $\left(\psi\right|$. In $\Hil_{BF}$ we refer to 
bosons and fermions as ideal bosons and ideal fermions. Note that the ideal boson operators
commute with the ideal fermion operators.

In short then, a Dyson mapping $\mathcal T$ is a non-unitary, one-to-one, linear operator that
maps vectors representing states of a many-fermion system onto vectors representing states of
a many-body system, where some of the particles are bosons and some are fermions. (Do not read this
as a definition. These are just properties Dyson mappings happen to have.)
Since the mapping $\mathcal T$ maps vectors in the domain  $V_F$, that is finite dimensional, 
onto vectors in the space $\Hil_{BF}$, that is infinite dimensional, the range of $\mathcal T$ 
must be a finite dimensional subspace $V_{BF}=T(V_F)$ 
of the infinite dimensional space $\Hil_{BF}$. $V_{BF}$ is called the physical subspace.

In the notation of the introductory discussion above, this means that $\tilde{\Hil}$ is 
embedded in a larger Hilbert space. For the mapping of operators 
$\hat{O}:\Hil_F\rightarrow\Hil_F$ and functionals $\left<\psi\right|:\Hil_F\rightarrow{\bf C}$
complications then arise. 
Let us first deal with the functionals. The image of $\left<\psi\right|$ is 
$\left<\psi\right|\mathcal T^{-1}$. It is only defined on the physical subspace $V_{BF}$. 
However, the most natural way to express
linear functionals acting on states in $\Hil_{BF}$ of which $V_{BF}$ is a subspace, 
is as functionals related to 
vectors in $\Hil_{BF}$ through the canonical map of $\Hil_{BF}$. Such functionals are then
defined on the whole $\Hil_{BF}$. In order to express the mapped functionals in this way
we therefore have to extend the domain of the image of each functional to the whole $\Hil_{BF}$.
Stated in symbols, for each linear functional $\left<\psi\right|$ on $\Hil_F$, we have to
find a linear functional $\left(\psi\right|$ on $\Hil_{BF}$ such that 
$\left.\left(\psi\right|\right|_{V_{BF}}=\left<\psi\right|\mathcal T^{-1}$. Here the suffixed 
$\left.\cdot\right|_{V_{BF}}$ indicates that the object it is preceded by has its domain 
restricted to the physical subspace $V_{BF}$.

A similar problem is encountered for operators. In the Hilbert space $\Hil_{BF}$ 
we want to express all operators of importance in terms
of ideal bosonic and fermionic creation and annihilation operators. Such operators that are 
functions of the ideal boson and fermion operators are
defined on the whole $\Hil_{BF}$. However the image of any operator 
$\hat{O}:\Hil_F\rightarrow\Hil_F$ 
is $\mathcal T \hat{O} \mathcal T^{-1}$, which is only defined on the physical subspace $V_{BF}$. 
Therefore, we have to 
find a linear extension $\hat{O}_{BF}:\Hil_{BF}\rightarrow\Hil_{BF}$, expressed in terms of ideal 
boson and fermion operators, for the image of every operator $\hat{O}:\Hil_F\rightarrow\Hil_F$,
so that $\left.\hat{O}_{BF}\right|_{V_{BF}}=\mathcal T\hat{O}\mathcal T^{-1}$. Note that 
$\hat{O}_{BF}$ leaves the physical subspace $V_{BF}$ invariant.

In this discussion we have already touched upon the answer to the following question:
Is it not trivial to establish an isomorphism between the many-fermion Hilbert space 
$\Hil_{F}$ (or a subspace $V_F$ thereof) and a subspace $V_{BF}$ of the space $\Hil_{BF}$?
After all, we could just do the following: 
If we have a basis 
$B_F=\left\{\left|\phi_i\right>\right\}_{i=1}^Q$ for $\Hil_F$, (implying $\Hil_F$ is 
$Q$-dimensional), we choose any $Q$-dimensional subspace $V_{BF}$ of $\Hil_{BF}$. 
For this subspace a basis
$B_{V_{BF}}=\left\{\left|\psi_i\right)\right\}_{i=1}^Q$ can be found. A mapping 
$\tilde{\mathcal T}$ is then specified 
by defining the image of the basis elements in $B_F$ under $\tilde{\mathcal T}$ as 
$\tilde{\mathcal T}\left|\phi_i\right>=\left|\psi_i\right)$. 
What makes the Dyson mapping special? A significant motivation for mapping fermion systems
onto boson-fermion systems is the following. For an operator like the Hamiltonian $H$ of a
fermion system, we want to find an operator $H_{BF}$ that is a function of ideal boson and fermion
operators and is equivalent to the original fermion Hamiltonian in the sense that 
$\left. H_{BF}\right|_{V_{BF}}=\mathcal T H \mathcal T^{-1}$. In other words, the mapping $\mathcal
T$ must be of a type that allows us to derive extensions to the whole space $\Hil_{BF}$ of the 
images of the fermion operators, {\em expressed as functions of ideal boson and fermion operators.}
A mapping as arbitrary as the above $\tilde{\mathcal T}$ suggests no clear way of doing this.
In contrast, the procedure that is used to construct the Dyson mapping allows one to find
extensions for the images of all relevant operators of the fermion system and explicitly express 
them as functions of the ideal boson and fermion operators.

It is my aim in this chapter to go through the procedure by means of which a Dyson mapping is 
constructed in detail. However, the danger of such a detailed analysis is that the reader looses
sight of the forest for the trees. Therefore I try now to give an intuitive overview of the 
argument that comprises the next few sections.

The argument hinges on the properties of a certain type of superalgebra. For reasons that become 
clear only in hindsight, this type of superalgebra and the properties common to certain classes of 
its representations are introduced in Sections \ref{SS1.2} and \ref{SS1.3}. In Section \ref{SS1.4}
we construct a representation of the superalgebra using the fermion operators defined in $\Hil_F$.
Then, using the commutation rules inherent to the superalgebra, matrix elements of operators of
physical interest can be computed. In these computations the details of the specific 
representation of the superalgebra we are dealing with is unimportant. A whole class of 
representations would have given the same answer. This inspires us to try and build a 
representation of the superalgebra in the boson-fermion space $\Hil_{BF}$ that also belongs to 
this class of representations that all give the same matrix elements. 
The way we go about constructing this boson-fermion
representation is as follows. We define a linear operator $T$ that we suspect might be a 
one-to-one mapping between $\Hil_F$ or a subspace $V_F$ thereof and a subspace $V_{BF}$ of the 
boson-fermion space $H_{BF}$. We do not know, however, if the operator $T$ is indeed invertible.
The operator $T$ has a very convenient property though. For operators $\hat{O}_F$ acting in 
$\Hil_F$, that are functions of operators in the fermion representation of the superalgebra,
the nature of the definition of $T$ allows us to find operators $\hat{O}_{BF}$ that are defined
on $\Hil_{BF}$ and are expressed as functions of the ideal boson and fermion operators, such that
$T\hat{O}_F=\hat{O}_{BF}T$. If $T$ is indeed invertible, then $\hat{O}_{BF}$ is the sought-after
extension of the image of $\hat{O}_F$ under the mapping $T$. We
find operators $\hat{O}_{BF}$, expressed as functions of the ideal boson and fermion operators,
corresponding to all operators $\hat{O}_F$ in the fermion 
representation of the superalgebra. The operators $\hat{O}_{BF}$ are then shown to form a 
new representation of the same superalgebra, and specifically one that belongs to the same class of
representations as the fermion representation, for which matrix elements of operators come out the
same no matter the specific representation. This fact is then used to construct a simple proof of
the invertibility of $T$. This all happens in Sections \ref{SS1.5} and \ref{SS1.6}. In short then,
the reasons for introducing the superalgebra are, firstly that the operator $T$ must be defined 
in terms of operators belonging to the fermion representation of the superalgebra and secondly,
the superalgebra is a substantial enough structure that, if the operator $T$ takes it as input 
on the fermion side and reproduces it as output on the ideal boson-fermion side, the invertibility
of $T$ is guaranteed. 

The reason that I denote this mapping $T$ instead of $\mathcal T$ is that, for reasons mentioned
in Section \ref{SS1.7}, the mapping $T$ has some undesirable properties that we want to get rid 
of. Two strategies are proposed, one in Section \ref{SS1.7} and the other in Section \ref{SS1.8}, 
that are applicable in different circumstances. They both involve
acting on $T$ with another linear, invertible operator, that for the moment, we will just denote
$\mathcal O$. The Dyson mapping is then given by $\mathcal T=\mathcal O T$. In Section \ref{SS1.9}
we discuss why and how the subspace $V_{BF}$ of $\Hil_{BF}$, that is the physical subspace, is 
identified. Finally in Section \ref{SS1.10} we look back on the derivation of the mapping. It
is pointed out how this derivation fits into the larger literature on the subject. Also, some
features of and problems with the mapping are mentioned.

The work that Sections \ref{SS1.2} - \ref{SS1.7} are concerned with, first appeared in \cite{DSG1}
- \cite{NGD3} and was reviewed in \cite{CG1}. The methods for finding the physical subspace, that
are discussed in Section \ref{SS1.9}, are also set out in \cite{DGH1} and \cite{GEH1}.

\section{The Abstract Superalgebra}
\markright{\bf Section \thesection: The Abstract Superalgebra}
\label{SS1.2}
A type of superalgebra is defined in this paragraph. For the purposes of this text, a superalgebra
may be viewed as a vector space, the elements of which are operators, 
that can be partitioned into two subspaces, called the even and
odd sectors respectively. The commutator of any two operators belonging to the even sector must
be equal to an operator that belongs to the even sector. Similarly, the
anti-commutator of any two operators in the odd sector must be equal to an operator in the even 
sector. The commutator of an operator from the even sector with an 
operator from the odd sector must be equal to an operator from the odd 
sector. No further knowledge of the theory of superalgebras is required for the reader to follow 
the arguments presented. In fact, the definition given here exhausts the author's knowledge
of the general properties of superalgebras.
A general
representation is considered of the superalgebra, so that we may be sure that properties derived 
from the definitions 
are not specific to only one representation of the superalgebra. This will be useful when we 
construct a new representation of the superalgebra from a given one, because properties derived
for a general representation will then hold for both representations. To indicate that we are 
working with a general representation of the superalgebra, over-bars
will be superimposed on operators.

Let $\Hil$ be a Hilbert space on which $2M$ operators $\bar{A}_i$ and $\bar{A}^j$
, $i,j=1,2,\ldots,M$ are 
defined, such that operators with upper indices commute amongst themselves, and similarly operators 
with lower indices commute amongst themselves:
\begin{eqnarray}
\com{\bar{A}^i}{\bar{A}^j}&=&0,\nonumber\\
\com{\bar{A}_i}{\bar{A}_j}&=&0.\label{BF1}
\end{eqnarray}
Set
\begin{equation}
\bar{K}_i^j=\com{\bar{A}_i}{\bar{A}^j}.\label{BF2}
\end{equation}
We assume that the operators $\bar{A}_i$, $\bar{A}^j$ and $\bar{K}_i^j$ are 
generators of a representation of a 
Lie algebra, the closure conditions being
\begin{eqnarray}
\com{\bar{K}_i^j}{\bar{A}_k}&=&C^{jl}_{ki}\bar{A}_l,\nonumber\\
\com{\bar{A}^i}{\bar{K}_j^k}&=&\tilde{C}^{ik}_{jl}\bar{A}^l,\label{BF3}
\end{eqnarray}
where repeated indices, one upper and one lower, imply summation. 
(When we say that a set of operators are
generators of a representation of a Lie algebra, we mean the following. The representation of the
Lie algebra in question is a vector space, the elements of which are operators. The set of 
generators of the representation is just a basis for the vector space of operators. Hence, every
element of the representation can be expressed as a linear combination of the generators.) 
The Jacobi identity can be used to show that $C^{jl}_{ik}$ is symmetric in its lower indices 
and $\tilde{C}^{ik}_{jl}$ is symmetric in its upper indices. Using these relations it is possible 
to compute the commutator
\begin{equation}
\com{\bar{K}_i^j}{\bar{K}_m^n}=C_{im}^{jl}\bar{K}_l^n-\tilde{C}_{il}^{jn}\bar{K}_m^l,\label{BF4}
\end{equation}
so that we see the assumed commutation relations in eq. \ref{BF3} do indeed ensure that 
$\bar{A}_i$, $\bar{A}^i$ and $\bar{K}_i^j$ close to a Lie algebra under commutation.
Let there further be defined $2\Omega$ operators $\bar{a}_\mu$ and $\bar{a}^\nu$, 
$\mu,\nu=1,2,\ldots,\Omega$ 
on $\Hil$ that obey the anti-commutator relations
\begin{eqnarray}
\acom{\bar{a}^\mu}{\bar{a}^\nu}&=&0,\nonumber\\
\acom{\bar{a}_\mu}{\bar{a}_\nu}&=&0,\nonumber\\
\acom{\bar{a}_\mu}{\bar{a}^\nu}&=&\delta_\mu^\nu.\label{BF5}
\end{eqnarray}
If the commutator relations 
\begin{equation}
\com{\bar{A}^i}{\bar{a}^\nu}=\com{\bar{A}_i}{\bar{a}_\mu}=0,\label{BF6}
\end{equation}
and
\begin{eqnarray}
\com{\bar{A}^i}{\bar{a}_\nu}&=&\chi^i_{\mu \nu}\bar{a}^\mu,\nonumber\\
\com{\bar{a}^\nu}{\bar{A}_i}&=&\tilde{\chi}_i^{\mu \nu}\bar{a}_\mu,\label{BF7}
\end{eqnarray}
hold, then it follows that
\begin{eqnarray}
\com{\bar{a}_\nu}{\bar{K}_i^j}&=&\chi^j_{\mu \nu}\tilde{\chi}_i^{\rho \mu}\bar{a}_\rho,\nonumber\\
\com{\bar{a}^\nu}{\bar{K}_i^j}&=&\chi^j_{\rho \mu}\tilde{\chi}_i^{\mu \nu}\bar{a}^\rho.\label{BF8}
\end{eqnarray}
In the derivation of the last two expressions, use has to be made of the Jacobi identity 
and the commutators in eq. \ref{BF7}.

To summarize, if we assume that the commutation relations in eq. \ref{BF1} and eq. \ref{BF3} hold 
then the commutation relation eq. \ref{BF4} holds automatically. 
If the anti-commutation relations eq. \ref{BF5} and the
commutation relations eq. \ref{BF6} and eq. \ref{BF7} are assumed, then the commutation 
relations eq. \ref{BF8}
hold automatically.
Therefore, under the above assumptions, the operators 
$\left\{\bar{A}_i,\bar{A}^j,\bar{K}_i^j\right\}$ 
and $\left\{\bar{a}_\mu,\bar{a}^\nu,I\right\}$ are 
respectively the generators (read basis elements) for the even and odd sectors of a representation 
of a superalgebra. The
properties of the superalgebra are determined by the structure coefficients $C^{jl}_{ik}$, 
$\tilde{C}^{ik}_{jl}$, $\chi^i_{\mu \nu}$ and $\tilde{\chi}_i^{\mu \nu}$. We therefore indicate 
the abstract superalgebra as $\Salg$.
\section{The Calculation of Matrix Elements}
\markright{\bf Section \thesection: The Calculation of Matrix Elements} 
\label{SS1.3}
Group the generators that were discussed in the previous section into three sets
\begin{eqnarray}
\Ou&=&\left\{\bar{A}^i,\bar{a}^\nu:i=1,2,\ldots,M;\nu=1,2,\ldots,\Omega\right\},\nonumber\\
\Od&=&\left\{\bar{A}_i,\bar{a}_\nu:i=1,2,\ldots,M;\nu=1,2,\ldots,\Omega\right\},\nonumber\\
{\mathcal K}&=&\left\{\bar{K}_i^j:i,j=1,2,\ldots,M\right\}.\label{BF9}
\end{eqnarray} 
Any two elements of $\Ou$ either commute or anti-commute. The same holds for the elements of $\Od$.
Assume there is a vector $\hw \in \Hil$, called a foundational vector
and a linear functional $\hwb$ on $\Hil$, called the foundational functional  
that have the following properties:
\begin{itemize}
\item $\opd\hw=0$ for all $\opd \in \Od$,
\item $\hwb\opu=0$ for all $\opu \in \Ou$,
\item Coefficients $\lambda_i^j$ exist so that $K_i^j\hw=\lambda_i^j\hw$ and 
$\hwb K_i^j=\hwb \lambda_i^j$ for all $K_i^j \in {\mathcal K}$.
\end{itemize}
This definition seems rather arbitrary. I therefore mention in passing that it is motivated by
the concept of the highest weight state of an irreducible representation of a semi-simple Lie
algebra.\footnote{In fact, I was tempted 
to use the term `highest weight' instead of `foundational', but resisted because of two reasons: 
Firstly `highest' suggests uniqueness, which does not have to be the case here and 
secondly, the structure here is a superalgebra, not a semi-simple Lie algebra, in which context
the term `highest weight' was coined.} 
Beyond this justification of the definition, I can only say that it will turn out to
be a useful definition in the arguments that are to follow.
The set of all representation of $\Salg$ that, for a given set of coefficients $\lambda_i^j$ 
possess such a foundational vector and a foundational functional is denoted
$\SRep{\lambda}$. 
A very useful property of the representations in $\SRep{\lambda}$ can now be stated:

For any representation in $\SRep{\lambda}$, let $\opd$ be any operator in $\Od$ and let $\opu_k$ 
be arbitrary elements of $\Ou$, then there exist $\opu_{k,j} \in \Ou$ and complex coefficients $w_j$
such that 
\begin{equation}
\opd\left(\prod_{k=1}^N \opu_k\right)\hw=\sum_j\omega_j\left(\prod_{k=1}^{N_j}\opu_{k,j}\right)
\hw.\label{BF10}
\end{equation}
The $w_j$ can be chosen such that they are valid for all representations in $\SRep{\lambda}$.
The proof can be found in Appendix A.

This result has an important corollary: Let $\op_k,\: k=1,2,\ldots,N$ be
operators from the set $\Ou \cup \Od \cup \mathcal K$. Then complex numbers $q_j$ and operators 
$\opu_{j,k} \in \Ou$ exist, such that
\begin{equation}
\left(\prod_{k=1}^N \op_k\right)\hw=\sum_j q_j\left(\prod_{k=1}^{N_j}\opu_{k,j}\right)\hw.
\label{BF19}
\end{equation}
Again, the coefficients $q_k$ can be chosen so that they are equally valid for all representations 
in $\SRep{\lambda}$.

To prove eq. \ref{BF19} one simply has to note that, in the expression 
$\left(\prod_{k=1}^N \op_k\right)\hw$ the operators belonging to $\mathcal K$ are commutators (and
therefore linear combinations of products) of elements belonging to $\Ou$ and $\Od$, and that one
can systematically eliminate all operators 
$\op_k\in\Od$, by working from right to left using eq. \ref{BF10}.

If we take the inner product of eq. \ref{BF19} with $\left|f\right>$ we find
\begin{equation}
\hwb\left(\prod_{k=1}^N \op_k\right)\hw=\sum_{j\in\{k:N_k=0\}} q_j,
\label{BF20}
\end{equation}
since the operators $\opu_{j,k}$, when present, yield zero if acted on by $\hwb$. This is the main 
result from this section and might be restated as follows: 

{\em The matrix elements with respect to the foundational states of products of 
generators in any representation in $\SRep{\lambda}$ 
can be calculated solely from the commutation and anti-commutation relations of the superalgebra.  
These matrix elements are the same in all representations in $\SRep{\lambda}$.} 
\section{The Single- plus Bifermion Representation of the 
Superalgebra}
\markright{\bf Section \thesection: The Fermionic Representation} 
\label{SS1.4}
Up to this point we have not encountered any concrete representation of $\Salg$, but only 
postulated the existence of a representation. We now construct a concrete 
representation carried by  $\Hil_F$. This representation will be used to define an operator $T$ 
that maps state vectors in the fermion space $\Hil_F$ onto state vectors in the boson-fermion
space $\Hil_{BF}$. As was mentioned in the introduction, we can, with the aid of the operator $T$,
construct a new representation of the superalgebra $\Salg$, this time in $\Hil_{BF}$, where the
generators of the representation are explicitly expressed as functions of the ideal boson and 
fermion operators. Both the fermion and boson-fermion representations lie in the
same class $\SRep{\lambda}$ of representations. This fact will be used to demonstrate the 
invertibility of the mapping $T$. 

Let $\chi_{\mu\nu}^i$, $i=1,\dots M;\; \mu,\nu=1,\ldots,\Omega$ be complex
numbers such that $\chi_{\nu\mu}^i=-\chi_{\mu\nu}^i$.\footnote{It will shortly become clear that 
this is not the only restriction we have to place on these numbers for the subsequent construction
to work.} Define $\chi_i^{\mu\nu}=\left(\chi_{\mu\nu}^i\right)^*$.
We adopt the following notation: to indicate that the generators we 
construct belong to the representation of $\Salg$ that is carried by $\Hil_F$, the generators will 
be placed inside round brackets subscripted by $F$. Define
\begin{eqnarray}
\rf{a_\nu}&=&a_\nu,\nonumber\\
\rf{a^\nu}&=&a^\nu,\nonumber\\
\rf{A_i}&=&\D{\frac{1}{2}}\chi_i^{\mu\nu}a_\nu a_\mu,\nonumber\\
\rf{A^i}&=&\D{\frac{1}{2}}\chi^i_{\mu\nu} a^\mu a^\nu,\nonumber\\
\rf{K_i^j}&=&\com{\rf{A_i}}{\rf{A^j}}.\label{BF21}
\end{eqnarray}
It is assumed that 
\begin{equation}
\D{\frac{1}{2}}\chi_i^{\mu\nu}\chi^j_{\mu\nu}=g\delta_i^j,\label{BF22}
\end{equation}
where $g$ is a positive real constant independent of $i$ and $j$. 
With the aid of the orthogonality condition
eq. \ref{BF22}, $\rf{K_i^j}$ when calculated explicitly, yields:
\begin{equation}
\rf{K_i^j}=g\delta_i^j-\chi_i^{\nu\rho}\chi^j_{\mu\rho}a^\mu a_\nu.\label{BF23}
\end{equation}
The reason for assuming the orthogonality condition eq. \ref{BF22} is to
ensure the linear independence of the operators $\rf{A_i}$, $i=1,\ldots,M$. Strictly speaking we 
could 
therefore have proceeded on the weaker assumption that the $\chi_i^{\mu\nu}$ are linearly 
independent
rather 
than orthogonal. However, if they are linearly independent then there is an invertible 
transformation
that will make them orthogonal. On the level of the algebra this merely represents a transformation
of basis for the linearly independent operators $A_i$. All we therefore assume in enforcing 
the orthogonality condition, is that we have chosen a particular kind of basis for the algebra, 
which represents no 
real loss of generality, since, for any algebra such a choice is possible. Note that since there are
$\D{\frac{1}{2}}\Omega(\Omega-1)$ linearly independent operators $a_\mu a_\nu$ in $\Hil_F$ the 
linear independence of the $\rf{A_i}$ operators places the upper bound 
$M\leq\D{\frac{1}{2}}\Omega(\Omega-1)$ on $M$. 
We now have to investigate under which conditions eq. \ref{BF21} is a representation of $\Salg$. The 
only real work that is required pertains to the closure condition 
$\com{\rf{K_i^j}}{\rf{A_k}}=C_{ik}^{jl}\rf{A_l}$. We must check under which conditions coefficients
 $C_{ik}^{jl}$ exist that satisfy this equation. Since 
\begin{equation}
\com{\rf{K_i^j}}{\rf{A_k}}=\chi_i^{\mu\rho}\chi_{\lambda\rho}^j\chi_k^{\lambda\eta}\rf{a_\eta}
\rf{a_\mu},\label{BF24}
\end{equation}
holds, enforcing the closure condition is equivalent to demanding that $C_{ik}^{jl}$ exist 
that solve
\begin{equation}
\chi_i^{\mu\rho}\chi_{\lambda\rho}^j\chi_k^{\lambda\eta}=\D{\frac{1}{2}}C_{ik}^{jl}\chi_l^{\mu\eta}.
\label{BF25}
\end{equation}
This is a matrix problem of the form: find $x$ such that $Ax=y$ where $x$ is an $M$ dimensional
column vector, $A$ is matrix with $M$ columns and $\D{\frac{1}{2}}\Omega(\Omega-1)$ rows, and 
$y$ is an $\D{\frac{1}{2}}\Omega(\Omega-1)$ dimensional column vector. This system of linear 
equations may be over-specified and does not 
always have a solution. Thus there is a further restriction on the $\chi$'s: they should be chosen
such that the system of linear equations eq. \ref{BF25} has a solution.
For our work it turns out that it is not necessary to have a more explicit formulation of this 
restriction.

When eq. \ref{BF25} does have a solution, we can multiply the equation by 
$\frac{1}{g}\chi_{\mu\eta}^l$ and sum over $\mu$ and $\eta$ to find\footnote{It must be stressed
that eq. \ref{BF26} follows from eq. \ref{BF25} but the reverse is not true. Therefore, simply 
choosing
the $C_{ik}^{jl}$ as in eq. \ref{BF26} does not automatically take care of the closure condition.}
\begin{equation}
C_{ik}^{jl}=\frac{1}{g}\chi_i^{\mu\rho}\chi_{\lambda\rho}^j\chi_k^{\lambda\eta}\chi_{\mu\eta}^l.
\label{BF26}
\end{equation}
From the fact that, in the present representation, operators with subscripts are
transformed into their superscripted partners through hermitian conjugation, it follows that
$\tilde{C}_{jl}^{ik}=\left(C_{ik}^{jl}\right)^*$ holds, and from eq. \ref{BF26} that the $C$'s are 
symmetric in both upper and lower indices:
\begin{equation}
C_{ik}^{jl}=C_{ki}^{jl}=C_{ik}^{lj}=C_{ki}^{lj}.\label{BF27}
\end{equation}

It is straight forward to check all other commutation and anti-commutation relations that 
the operators in eq. \ref{BF21} should satisfy. We conclude that the operators in eq. \ref{BF21}
are indeed the generators of a representation of $\Salg$, 
with $\tilde{\chi}_i^{\mu\nu}=\left(\chi^i_{\mu\nu}\right)^*$
and $\tilde{C}^{ik}_{jl}=\left(C_{ik}^{jl}\right)^*$. 
We denote this representation $\mathcal F$.
For ease of reference we define a set 
$\Gf=\left\{\rf{a_\nu};\rf{a^\nu};\rf{A_i};\rf{A^i};\rf{K_i^j}\right\}$.
The vacuum $\vf$ in $\Hil_F$ has all the properties that are required of the foundational vector 
$\hw$. The functional $\vfb$ has all the properties required of foundational functional $\hwb$.
In particular $\lambda_i^j=g\delta_i^j$. Therefore the representation $\mathcal F$ is a member of
$\SRepf$.
\section{What is Required of a Mapping Procedure?}
\markright{\bf Section \thesection: What is Required of the Mapping Procedure?}
\label{SS1.5}
In the representation $\mathcal F$ of the superalgebra $\Salgf$, the operators $\rf{A^i}$ create
fermion pairs. We shall call these operators collective pair creation operators, and their 
conjugates collective pair annihilation operators.\footnote{The term `collective' is used
because, when many $\chi$ coefficients are non-zero, states such as $\rf{A^i}\vf$ consist of
a linear combination of many Slater determinants. However, we do not reserve this terminology
for the case where many Slater determinants contribute.} We earmark these operators for
association with ideal boson operators.

Let us consider what is required to map a fermion system in which the operators in $\Gf$ and 
$\rf{A^i}$ and $\rf{A_j}$ in particular,  are the
building blocks of the physically important states and operators. Note that
\begin{enumerate}
\item any state in $\Hil_F$ can be written (in more than one way) in the form 
\begin{equation}
\left|\phi\right>=\sum_{j}\eta_j\prod_{k=1}^{N_j}\rf{\op_{j,k}}\left|0\right>,\label{BF27A}
\end{equation}
where the $\eta_j$ are complex numbers and are $\rf{\op_{j,k}}$ elements of $\Gf$.
\item any linear operator $\rf{\op}:\Hil_F\rightarrow\Hil_F$ can be written (in more than one way) 
in the form 
\begin{equation}
\rf{\op}=\sum_{j}\eta_j\prod_{k=1}^{N_j}\rf{\op_{j,k}},\label{BF27B}
\end{equation}
where the $\eta_j$ are complex numbers and are $\rf{\op_{j,k}}$ elements of $\Gf$.
\item any linear functional on $\Hil_F$ can be written (in more than one way) in the form 
\begin{equation}
\left<\psi\right|=\left<0\right|\sum_{j}\eta_j\prod_{k=1}^{N_j}\rf{\op_{j,k}},\label{BF27C}
\end{equation}
where the $\eta_j$ are complex numbers and are $\rf{\op_{j,k}}$ elements of $\Gf$.
\end{enumerate}
Firstly, we must find a suitable isomorphism $T:\Hil_F\rightarrow V_{BF}\subset \Hil_{BF}$, that
maps states in the fermion space onto states in the boson-fermion space.
Then we have to find the image of the foundational vector $\left|0\right>$ and an extension
of the image of the foundational functional $\left<0\right|$. If we can then find extensions of 
the images of
the operators in $\Gf$ and express them in terms of ideal boson and fermion operators, 
we see from the expressions for general vectors, operators and functionals in $\Hil_F$, 
eq. \ref{BF27A}, eq. \ref{BF27B} and eq. \ref{BF27C} that we are done.

A remark on notation: An extension of the image under $T$ of an operator 
$\rf{\op}\in\Gf$ will be denoted $\rg{\op}{BF}$. It is related to $\rf{\op}$ through
\begin{equation}
\rf{\op}=T^{-1}\left.\rg{\op}{BF}\right|_{V_{BF}}T.\label{BF27D}
\end{equation}
This is the only restriction on $\rg{\op}{BF}$, and does not fix it uniquely, as the condition in 
eq. \ref{BF27D}
only specifies the properties of $\rg{\op}{BF}$ in the domain $V_{BF}$. Outside this domain it 
is undetermined save that it should be linear.\footnote{The extensions will have to be linear
if we want to express them in terms of ideal boson and fermion operators.}
We therefore indicate the
relationship between $\rf{\op}$ and $\rg{\op}{BF}$ as
\begin{equation}
\rf{\op}\longleftarrow\rg{\op}{BF},\label{27E}
\end{equation}
to stress that, while there exists a linear mapping from the extensions $\rg{\op}{BF}$ to 
the operators $\rf{\op}$ stated in eq. \ref{BF27D}, there need not be a linear mapping in the
opposite direction.
\section{Concerning the Usui Operator}
\markright{\bf Section \thesection: Concerning the Usui Operator}
\label{SS1.6}
I now describe a method for finding an isomorphism $T$ between $\Hil_F$ and the physical subspace
$V_{BF}\subset\Hil_{BF}$.
We start by defining several operators in the space $\Hil_U=\Hil_{BF}\otimes\Hil_F$:
\begin{eqnarray}
\widetilde{B^i}&=&B^i\otimes \I,\nonumber\\
\widetilde{B_i}&=&B_i\otimes \I,\nonumber\\
\widetilde{\alpha^\nu}&=&\alpha^\nu\otimes \I,\nonumber\\
\widetilde{\alpha_\nu}&=&\alpha_\nu\otimes \I.\label{BF28}
\end{eqnarray}
Further linear operators on $\Hil_U$ are defined via their action on a specific type of basis for
$\Hil_U$: It is possible to find a basis $B_U$ for $\Hil_U$ such that each element of $B_U$ has 
the form 
$\left|\psi,n\right)\otimes\left|\phi\right>$ 
where $\left|\psi,n\right)$ is a vector
in $\Hil_{BF}$ with a well-defined number of ideal fermions $n$, i.e. 
$\alpha^\nu\alpha_\nu\left|\psi,n\right)=n\left|\psi,n\right)$, and $\left|\phi\right>$ belongs to
$\Hil_F$. Define the images of $\widetilde{\rf{a^\nu}}$ and $\widetilde{\rf{a_\nu}}$ on such 
basis vectors as
\begin{equation}
\widetilde{\rf{a^\nu}}\uvec{\psi,n}{\phi}=(-1)^n\left|\psi,n\right)\otimes
a^\nu\left|\phi\right>,\nonumber
\end{equation}
and
\begin{equation}
\widetilde{\rf{a_\nu}}\uvec{\psi,n}{\phi}=(-1)^n\left|\psi,n\right)\otimes
a_\nu\left|\phi\right>.\label{BF29}
\end{equation}
Because these operators leave the number of ideal fermions unaltered, any product of an even 
number of them looses the $(-1)^n$ that appears in eq. \ref{BF29}. Thus, for example,
$\widetilde{\rf{a^\mu}}\widetilde{\rf{a^\nu}}=\I\otimes\rf{a^\mu}\rf{a^\nu}$. Hence, if we define
\begin{eqnarray}
\widetilde{\rf{A^i}}&=&
\D{\frac{1}{2}}\chi^i_{\mu\nu}\widetilde{\rf{a^\mu}}\widetilde{\rf{a^\nu}},\nonumber\\
\widetilde{\rf{A_i}}&=&
\D{\frac{1}{2}}\chi_i^{\mu\nu}\widetilde{\rf{a_\nu}}\widetilde{\rf{a_\mu}}.\label{BF30}
\end{eqnarray}
then it is also true that 
\begin{eqnarray}
\widetilde{\rf{A^i}}&=&
\D{\frac{1}{2}}\chi^i_{\mu\nu}\I\otimes\rf{a^\mu}\rf{a^\nu},\nonumber\\
\widetilde{\rf{A_i}}&=&
\D{\frac{1}{2}}\chi_i^{\mu\nu}\I\otimes\rf{a_\nu}\rf{a_\mu}.\label{BF31}
\end{eqnarray}
We conclude that the operators $\wt{\rf{A_i}}$, $\wt{\rf{A^i}}$, 
${\com{\wt{\rf{A_i}}}{\wt{\rf{A^j}}}}$, 
$\wt{\rf{a_\nu}}$
and $\wt{\rf{a^\nu}}$ obey the same commutation and anti-commutation relations as their 
counterparts without tildes. Furthermore, from eq. \ref{BF31} it is clear that $\wt{\rf{A_i}}$ and
$\wt{\rf{A^i}}$ commute with $\wt{B_j}$, $\wt{B^j}$, $\wt{\alpha_\nu}$ and $\wt{\alpha^\nu}$. 
Since $\wt{B_i}$ and $\wt{B^i}$ do not alter the ideal fermion number, 
$\wt{\rf{a_\nu}}$ 
and $\wt{\rf{a^\nu}}$ 
commute with these operators. However, 
$\wt{\rf{a_\nu}}$ 
and $\wt{\rf{a^\nu}}$ 
were defined in precisely such a way that the fact that $\wt{\alpha_\mu}$ and $\wt{\alpha^\nu}$
change the ideal fermion number by one results in the anti-commutation of $\wt{\rf{a}}$'s 
and $\wt{\rf{\alpha}}$'s.

The Usui operator $U:\Hil_U\rightarrow\Hil_U$ is defined as
\begin{equation}
U=\exp\left(\wt{B^i}\wt{\rf{A_i}}+\wt{\alpha^\nu}\wt{\rf{a_\nu}}\right).\label{BF32}
\end{equation}
The Usui operator allows us to specify a linear operator $T:\Hil_F\rightarrow\Hil_{BF}$ in the 
following way: For each $\left|\phi\right>$ in $\Hil_F$ let 
$\left|\phi\right)=T\left|\phi\right>$ be the unique vector in $\Hil_{BF}$ such that, for all
$\left|\psi\right)$ in $\Hil_{BF}$
\begin{equation}
\left(\psi\right|\left.\phi\right)=\ufun{\psi}{0}U\uvec{0}{\phi},\label{BF33}
\end{equation}
holds. For the operator $T$ to be an isomorphism between $\Hil_F$ and the physical subspace 
$V_{BF}=T(\Hil_F)$ we need to show that $T$ is $1:1$. This matter is deferred for a while. 
Until then we must be careful not to construct arguments that assume the existence of 
$T^{-1}:V_{BF}\rightarrow\Hil_F$.

The advantage of defining $T$ along this route lies in the exponential form of $U$, which 
suggests a systematic way of deriving (extensions of) the images of operators 
under this mapping. Suppose there exists an operator $\rbf{\op}:\Hil_{BF}\rightarrow\Hil_{BF}$ such
that $\rf{\op}\leftarrow\rbf{\op}$, where $\rf{\op}\in{\mathcal G_F}$ is a generator of the 
fermion representation of the superalgebra $\Salg$. Then it holds that
\begin{equation}
T\rf{\op}=\rbf{\op}T,\label{BF34}
\end{equation}
from which follows 
\begin{eqnarray}
\left(\psi\right|\rbf{\op}\left|\phi\right)&
=&\ufun{\psi}{0}U\left|0\right)\otimes\rf{\op}\left|\phi\right>\nonumber\\
&=&\ufun{\psi}{0}U\,\left\{\I\otimes\rf{\op}\right\}\uvec{0}{\phi}\nonumber\\
&=&\ufun{\psi}{0}U\wt{\rf{\op}}\uvec{0}{\phi}.\label{BF35}
\end{eqnarray}
with $\left|\phi\right>$ and 
$\left|\psi\right)$ arbitrary vectors in $\Hil_F$ and $\Hil_{BF}$ respectively and where  
$\left|\phi\right)=T\left|\phi\right>$.
In the last line of eq. \ref{BF35}, $\I\otimes\rf{\op}$ could be replaced 
by $\wt{\rf{\op}}$ without worrying about
the $(-1)^n$ that appears in the definition eq. \ref{BF29} because the vacuum $\left|0\right)$ in 
$\Hil_{BF}$ contains no ideal fermions. The BCH formulas
\begin{eqnarray}
\exp(P)Q&=&\left(\sum_{k=0}^\infty\frac{1}{k!}(P,Q)_k\right)\exp(P),\label{BF36}\\
Q\exp(P)&=&\exp(P)\left(\sum_{k=0}^\infty\frac{(-1)^k}{k!}(P,Q)_k\right).\label{BF37}
\end{eqnarray}
where $(P,Q)_0=Q$ and $(P,Q)_{k+1}=\com{P}{(P,Q)_k}$ can now be used to find an operator 
${\mathcal O}:\Hil_U\rightarrow\Hil_U$ such that ${\mathcal O}U=U\wt{\rf{\op}}$ where 
${\mathcal O}$ is expressed in terms of the $\wt{\rf{A}}$, $\wt{\rf{a}}$, $\wt{B}$ 
and $\wt{\alpha}$ operators. From there on a few tricks are pulled out of the hat to find 
an operator $\rbf{\op}:\Hil_{BF}\rightarrow\Hil_{BF}$ expressed in terms of the ideal
boson and fermion operators, such that 
\begin{equation}
\ufun{\psi}{0}\left\{\rbf{\op}\otimes\I\right\} U\uvec{0}{\phi}=
\ufun{\psi}{0}{\mathcal O}U\uvec{0}{\phi}.\label{BF38}
\end{equation}
These tricks involve, among other things, using the fact that some terms in $\mathcal O$ annihilate
vectors like $\uvec{0}{\phi}$ and functionals like $\ufun{\psi}{0}$ because they respectively
contain the vacua in $\Hil_{BF}$ and $\Hil_F$. The operator $\rbf{\op}$ that, thanks to this
procedure, is expressed in terms of the $B$ and $\alpha$ operators, satisfies the requirements
contained in eq. \ref{BF34}. 

A slight complication in this process is that it sometimes works 
better in reverse, so that for a given generator $\rf{\op}$ in $\mathcal G$ we first have to guess 
the corresponding operator $\rbf{\op}$ and then work our way backwards from eq. \ref{BF38} to
eq. \ref{BF35}, thereby testing whether our guess is correct.

In Appendix B the images of all the generators in $\mathcal G_F$ are calculated explicitly. The 
final result is:
\begin{eqnarray}
\rf{a_\nu}&\longleftarrow&\rbf{a_\nu}=\alpha_\nu,\nonumber\\
\rf{A_j}&\longleftarrow&\rbf{A_j}=B_j,\nonumber\\
\rf{a^\nu}&\longleftarrow&\rbf{a^\nu}=\alpha^\nu+\chi_i^{\nu\mu}B^i\alpha_\mu,\nonumber\\
\rf{A^j}&\longleftarrow&\rbf{A^j}=
B^i\mc{K}_i^j-\D{\frac{1}{2}}C_{ik}^{jl}B^iB^kB_l+\mc{A}^j,\nonumber\\
\rf{K_i^j}&\longleftarrow&\rbf{K_i^j}=\mc{K_i^j}-C_{ik}^{jl}B^kB_l.\label{BF39}
\end{eqnarray}
Here $\mc{A}^i$, $\mc{A}_i$ and $\mc{K}_i^j$ refer to the ideal fermion counterparts of 
$\rf{A^i}$, $\rf{A_i}$ and $\rf{K_i^j}$:
\begin{eqnarray}
\mc{A}_i&=&\D{\frac{1}{2}}\chi_i^{\mu\nu}\alpha_\nu\alpha_\mu,\nonumber\\
\mc{A}^i&=&\D{\frac{1}{2}}\chi^i_{\mu\nu}\alpha^\mu\alpha^\nu,\nonumber\\
\mc{K}_i^j&=&\com{\mc{A}_i}{\mc{A}^j}.\label{BF40}
\end{eqnarray}
Furthermore, although the mapping is not unitary, the identities
\begin{eqnarray}
T\left|0\right>&=&\left|0\right),\nonumber\\
\left<0\right|&\longleftarrow&\left(0\right|.\label{BF40A}
\end{eqnarray}
do hold.
In Appendix C, I show that these images comprise the generators of a representation, 
denoted $\mc{B}$, of $\Salg$
{\em in the whole} $\Hil_{BF}$, which justifies the notation $\rbf{\cdot}$, by showing that the 
desired
commutation and anti-commutation relations are satisfied. This result is non-trivial, since 
the definition of the mapping eq. \ref{BF34} of the generators says nothing about their properties
outside the physical subspace $V_{BF}$, which we know to be a finite dimensional subspace of the 
infinite dimensional $\Hil_{BF}$.

Furthermore, even in the subspace $V_{BF}$, we have not yet proved that $T$ is invertible. 
This has the following implication: Say that in the representation $\mc{F}$ the following 
commutator holds:
\begin{equation}
\com{\rf{\op_1}}{\rf{\op_2}}=\rf{\op_3},\label{BF64}
\end{equation}
where $\rf{\op_1}$, $\rf{\op_2}$ and $\rf{\op_3}$ are generators of $\mc{F}$. Then because
$T$ is a linear operator and $T{\rf{\op}}=\rbf{\op}T$, we know that 
\begin{equation}
\com{\rbf{\op_1}}{\rbf{\op_2}}T=\rbf{\op_3}T.\label{BF65}
\end{equation}
However, we do not know if $T$ is invertible in $V_{BF}$ so that we can multiply with the 
inverse from the right, to establish that the commutator holds in $V_{BF}$. 

The fact that we do have a representation of $\Salg$ carried by $V_{BF}$ therefore strongly
suggests the invertibility of $T$. In fact, as a consequence of this, the invertibility of $T$
is proved without much effort:
Firstly, 
it is easy to establish that the foundational states $\left|0\right)$ 
and $\left(0\right|$ in $\Hil_{BF}$ have exactly the same properties as those possessed by 
the foundational states of the representation $\mc{F}$. This implies that 
both $\mc{F}$ and $\mc{B}$ belong to $\SRep{g\delta_i^j}$. This in turn implies that for
$\rf{\op_i}$, $i=1,2,\ldots,N$ generators of $\mc{F}$, and $\rbf{\op_i}$ their extended 
counterparts in $\mc{B}$ the equality of matrix elements 
\begin{equation}
\left<0\right|\prod_{i=1}^N\rf{\op_i}\left|0\right>
=\left(0\right|\prod_{i=1}^N\rbf{\op_i}\left|0\right),\label{BF66}
\end{equation}
holds, as we have seen in Section \ref{SS1.3}.
Let $\left|\phi\right>$ be any vector in $\Hil_F$. Then $\left|\phi\right>$ can be expressed as
\begin{equation}
\left|\phi\right>=\sum_{j}\eta_j\prod_{k=1}^{N_j}\rf{\op_{j,k}}\left|0\right>,\label{BF67}
\end{equation}
where $\eta_j$ are complex constants and $\rf{\op_{j,k}}$ are generators of $\mc{F}$. If 
$\left|\phi\right>\not=0$ then 
\begin{eqnarray}
\left<\phi\right|\left.\phi\right>&=&
\left<0\right|\left(\sum_{j}\eta_j^*\prod_{k=N_j}^{1}\rf{\op_{j,k}}^+\right)
\left(\sum_{j}\eta_j\prod_{k=1}^{N_j}\rf{\op_{j,k}}\right)\left|0\right>\nonumber\\
&\not=&0,\label{BF68}
\end{eqnarray}
where the order of the factors in the first product is now reversed. The representation 
$\mc{F}$ is such that if $\rf{\op}$ is a generator, so is its conjugate $\rf{\op}^+$. 
Hence it follows
that there are operators $\rbf{\op_{j,k}}:\Hil_{BF}\rightarrow\Hil_{BF}$
and $\rbf{\op_{j,k}^+}:\Hil_{BF}\rightarrow\Hil_{BF}$ such that
$\rbf{\op_{j,k}}T=T\rf{\op_{j,k}}$, $\rbf{\op_{j,k}^+}T=T\rf{\op_{j,k}}^+$ holds, for which 
it is true that
\begin{eqnarray}
0&\not=&\left<0\right|\left(\sum_{j}\eta_j^*\prod_{k=N_j}^{1}\rf{\op_{j,k}}^+\right)
\left(\sum_{j}\eta_j\prod_{k=1}^{N_j}\rf{\op_{j,k}}\right)\left|0\right>\nonumber\\
&=&\left(0\right|\left(\sum_{j}\eta_j^*\prod_{k=N_j}^{1}\rbf{\op_{j,k}^+}\right)
\left(\sum_{j}\eta_j\prod_{k=1}^{N_j}\rbf{\op_{j,k}}\right)\left|0\right).\label{BF69}
\end{eqnarray}
In the last line of eq. \ref{BF69} we can replace
$\left(\sum_{j}\eta_j\prod_{k=1}^{N_j}\rbf{\op_{j,k}}\right)\left|0\right)$ with
$\left|\phi\right)=T\left|\phi\right>$. Furthermore 
$\left(0\right|\left(\sum_{j}\eta_j^*\prod_{k=N_j}^{1}\rbf{\op_{j,k}^+}\right)$ is some 
linear functional (in general not $\left(\phi\right|$, since $T$ is not unitary) on $\Hil_{BF}$.
Therefore eq. \ref{BF69} says that there is a linear functional on $\Hil_{BF}$ that does not map
$T\left|\phi\right>$ onto zero. Thus $T\left|\phi\right>$ cannot be the zero vector in 
$\Hil_{BF}$. Thus if $\left|\phi\right>\not=0$ then $T\left|\phi\right>\not=0$ and
$T$ is invertible.

Define the set
$\Gbf=\left\{\rbf{a_\nu};\rbf{a^\nu};\rbf{A_i};\rbf{A^i};\rbf{K_i^j}\right\}$.
Because $T$ is invertible, the operators in $\Gbf$, together with the foundational states
$\left(0\right|$ and $\left|0\right)$ comprise the essential elements of the mapping of
systems in $\Hil_F$ to $V_{BF}$. 

However, consider the operators 
$\rf{A_i}=\D{\frac{1}{2}}\chi_i^{\mu\nu}\rf{a_\nu}\rf{a_\mu}$ and $\rf{a_\nu}$ and the 
extensions of their respective images: $\rbf{A_i}=B_i$ and $\rbf{a_\nu}=\alpha_\nu$.\\
Clearly $\rbf{A_i}=B_i\not=\mc{A_i}=\D{\frac{1}{2}}\chi_i^{\mu\nu}\rbf{a_\nu}\rbf{a_\mu}$,
even though, in the fermion representation, it is true that 
$\rf{A_i}=\frac{1}{2}\chi_i^{\mu,\nu}\rf{a_\nu}\rf{a_\mu}$. Is
this not in contradiction with the requirement that the images of operators should be
related to the operators through a similarity transformation? There is no contradiction:
$\rf{A_i}$ and $\rf{a_\nu}$ need not be related to $\rbf{A_i}$ and $\rbf{a_\nu}$ 
through a similarity transformation. It is the restrictions of the latter to the physical
subspace that are related to the former through a similarity transformation. In other words,
while $B_i$ is not equal to $\mc{A_i}$ in $\Hil_{BF}$, they are equal when restricted
to the physical subspace, $\left.\mathcal A_i\right|_{V_{BF}}=\left.B_i\right|_{V_{BF}}$. 
Consider for instance their respective actions on the state
\begin{equation}
\left|\psi\right)=\rbf{A^i}\left|0\right)=gB^i\left|0\right)+\mc{A^i}\left|0\right).\label{BF69A}
\end{equation}
It is easily seen that $B_i\left|\psi\right)=\mc{A_i}\left|\psi\right)=g\left|0\right)$ as it 
should be for a state in the physical subspace.

Although there is no contradiction, the above discussion illustrates an undesirable feature of
the mapping $T$. The association between the collective fermion pair 
annihilation operator $\rf{A_i}$ on the fermion side and the ideal boson annihilation operator 
$B_i$ is not
as strong or exclusive as might be expected. The nature of the physical subspace is such that 
ideal boson
annihilation is equivalent to ideal collective fermion pair annihilation. In the next section
we attempt to address this problem.
\section{A Similarity Transformation}
\markright{\bf Section \thesection: A Similarity Transformation}
\label{SS1.7}
Consider again the action of $\rbf{A^i}$ on the vacuum:
\begin{equation}
\rbf{A^i}\left|0\right)=\left(gB^j+\mc{A}^j\right)\left|0\right).\label{BF70}
\end{equation}
So, the collective state $\rf{A^j}\left|0\right>$ is mapped by $T$ onto a state, that 
although it contains a boson, retains a collective fermion component $\mc{A}^j\left|0\right)$. 
If we cannot get rid of the collective fermions through the mapping, then there is little
profit in mapping, speaking from a computational perspective. 
Also, the intuitive notion that some fermion pairs are similar to bosons
demand that we get rid of the collective fermion component in the mapped states.
We have jumped the gun by settling on $T$ as our isomorphism.

Instead of the images in eq. \ref{BF40}, let us rather postulate a mapping that results in 
extended operator images of the form
\begin{eqnarray}
\rf{A_j}&\longleftarrow&\rbf{A_j}=B_j,\nonumber\\
\rf{A^j}&\longleftarrow&R^j=
B^i\mc{K}_i^j-\D{\frac{1}{2}}C_{ik}^{jl}B^iB^kB_l\nonumber\\
\rf{K_i^j}&\longleftarrow&\rbf{K_i^j}=\mc{K_i^j}-C_{ik}^{jl}B^kB_l.\label{BF76}
\end{eqnarray}
From the calculations in Appendix C, it is easily seen that altering the above operators
in this way does not alter their commutators with each other. (What does change is the 
commutators of these operators with $\rbf{a^\nu}$ and $\rbf{a_\nu}$.) This inspires the notion that 
there is some similarity transformation $X$ so that the operators in eq. \ref{BF76} are the
transformed operators of the original mapped operators,
\begin{eqnarray}
X^{-1}\rbf{A_j}X&=&B_j=\rbf{A_j},\nonumber\\
X^{-1}\rbf{A^j}X&=&R^j=\rbf{A^j}-\mc{A}^j,\nonumber\\
X^{-1}\rbf{K_i^j}X&=&\rbf{K_i^j},\label{BF77}
\end{eqnarray}
because similarity transformations preserve commutators (and anti-commutators). This can be viewed
as a new isomorphism between $\Hil_F$ and a subspace of $\Hil_{BF}$, which is established by 
the operator $X^{-1}\circ T$, where `$\circ$' indicates the conjunction of operators, i.e.
$X^{-1}\circ T\left|\psi\right)=X^{-1}\left\{T\left|\psi\right)\right\}$. 
Under this new mapping, the images of the single fermion operators become $X^{-1}\rbf{a^\nu}X$
and $X^{-1}\rbf{a_\nu}X$. These operators, together with those in eq. \ref{BF77} form a new 
representation of $\Salg$ in $\Hil_{BF}$. This representation will be denoted $\mc{X}$.
It is also a member of $\SRep{g\delta_i^j}$, this time with  foundational states 
$X^{-1}\left|0\right)$ and $\left(0\right|X$.
The generators of this representation will be denoted $\rx{\op}=X^{-1}\rbf{\op}X$. The new 
physical subspace is $V_{BF}=X^{-1}\circ T(\Hil_F)$. (We use the same symbol as before to
denote the physical subspace.)

I now prove the existence of an operator $X$ such that
\begin{eqnarray}
B_jX&=&XB_j,\nonumber\\
\left(R^j+\mc{A}^j\right)X&=&XR^j.\label{BF78}
\end{eqnarray}
If this $X$ is invertible, it will perform the duties required of the 
similarity transform described above.
We start by defining $C_F=\mc{A}^k\mc{A}_k$, which is an hermitian operator in $\Hil_{BF}$. 
It is easily verified that $C_F$ commutes with all the operators $\mc{K}_i^j$. Since 
$C_F$ is hermitian, we can find a complete basis for $\Hil_{BF}$ consisting of eigenvectors
of $C_F$. A generic member of this basis will be denoted $\eig$, with $C_F\eig=\lambda\eig$. 
A linear operator $X$ is defined by stating its action on this basis:
\begin{equation}
X\eig=\sum_{n=0}^\infty\left\{\left(C_F-\lambda\right)^{-1}\mc{A}^iB_i\right\}^n\eig.\label{BF79}
\end{equation}
Note that we do not have to worry about the convergence of the infinite sum. Since each term
$\left\{\left(C_F-\lambda\right)^{-1}\mc{A}^iB_i\right\}^n$ creates $2n$ ideal fermions, and
the fermionic sector of $\Hil_{BF}$ is saturated after the creation of $\Omega$ ideal 
fermions, terms in eq. \ref{BF79} with $2n>\Omega$ are identically zero. 

However, the definition is still problematic: since $C_F$ has an eigenvalue $\lambda$, 
$C_F-\lambda$ has a zero eigenvector, which means that it is not invertible on the whole
$\Hil_{BF}$. However, $C_F-\lambda$ is invertible on the orthogonal complement in $\Hil_{BF}$
of the $\lambda$-eigenspace of $C_F$. This orthogonal complement is  denoted 
$\Lambda_\bot$. It is therefore necessary that the 
operator $\mc{A}^iB_i$ maps every vector it acts on in eq. \ref{BF79} onto 
$\Lambda_\bot$, in which case we take 
$\left(C_F-\lambda\right)^{-1}$ to be the inverse of $C_F-\lambda$ on the invariant subspace
$\Lambda_\bot$. I do not know of a result that shows under which general conditions $\mc{A}^iB_i$
possesses this property, so that it has to be treated on a case by case basis. In Chapter Three
we discuss an instance where it can be demonstrated that the similarity transformation is 
well-defined. 

Assuming then that eq. \ref{BF79} does represent a bona fide definition of the operator $X$, I
prove that it satisfies eq. \ref{BF78}. First notice that, if $\eig$ is an eigenvector of $C_F$
associated with eigenvalue $\lambda$, then so is $B_j\eig$. Therefore
\begin{eqnarray}
\com{X}{B_j}\eig&=&
\com{\sum_{n=0}^\infty\left\{\left(C_F-\lambda\right)^{-1}\mc{A}^iB_i\right\}^n}
{B_j}\eig,\nonumber\\
&=&0.\label{BF80}
\end{eqnarray}
Because $\eig$ is an arbitrary member of a basis for $\Hil_{BF}$, this proves the first 
part of eq. \ref{BF78}. The second part is proved by showing that
\begin{equation}
\com{X}{R^j}=\mc{A}^jX.\label{BF81}
\end{equation}
Define, for convenience, operators
\begin{equation}
\tilde{X}_n(z)=\left\{\left(C_F-z\right)^{-1}\mc{A}^iB_i\right\}^n,\label{BF82}
\end{equation}
where $z$ is a complex number and $n=0,1,2,\ldots$ Now define linear operators $X_n$ by
means of their action on the basis vectors $\eig$:
\begin{equation}
X_n\eig=\tilde{X}_n(\lambda)\eig.\label{BF83}
\end{equation}
From this definition it follows that $X=\sum_{n=0}^\infty X_n$. By induction I now show that 
\begin{equation}
\com{X_{n+1}}{R^j}=\mc{A}^jX_n\label{BF84}
\end{equation}
for $n=0,1,2,\ldots$, so that
\begin{eqnarray}
\com{X}{R^j}&=&\sum_{n=0}^\infty\com{X_{n+1}}{R^j}\nonumber\\
&=&\sum_{n=0}^\infty\mc{A}^jX_n\nonumber\\
&=&\mc{A}^jX.\label{BF85}
\end{eqnarray}
We shall often refer back to the following two intermediate results: Firstly
\begin{equation}
\com{C_F}{R^j}=0,\label{BF86}
\end{equation}
so that if $\eig$ is an eigenvector of $C_F$, with eigenvalue $\lambda$, then so too is 
$R^j\eig$. Secondly
\begin{eqnarray}
\com{\tilde{X}_1(z)}{R^j}&=&\left(C_F-z\right)^{-1}\com{\mc{A}^iB_i}{R^j},\nonumber\\
&=&\left(C_F-z\right)^{-1}\com{C_F}{\mc{A}^j}.\label{BF87}
\end{eqnarray}
Now we can start the proof proper, by setting $n=0$, then
\begin{equation}
\com{X_1}{R^j}\eig=\com{\tilde{X}_1(\lambda)}{R^j}\eig,\label{BF88}
\end{equation}
where the expression on the right follows because both $\eig$ and $R^j\eig$ are eigenvectors
of $C_F$ with eigenvalue $\lambda$. Continue by substituting eq. \ref{BF87} into 
eq. \ref{BF88} to find
\begin{eqnarray}
\com{\tilde{X}_1(\lambda)}{R^j}\eig&=&\left(C_F-\lambda\right)^{-1}\com{C_F}{\mc{A}^j}\eig
\nonumber\\
&=&\left(C_F-\lambda\right)^{-1}\left(C_F-\lambda\right)\mc{A}^j\eig\nonumber\\
&=&\mc{A}^j\eig.\label{BF89}
\end{eqnarray}
We conclude that 
$\com{X_1}{R^j}=\mc{A}^j$ holds, which is what we wanted to prove, since $X_0={\rm I}$.
Take the next step, by assuming eq. \ref{BF84} is true for $n=k-1$, i.e.
\begin{equation}
\com{X_k}{R^j}=\mc{A}^jX_{k-1}.\label{BF90}
\end{equation}
Then
\begin{eqnarray}
&&\com{X_{k+1}}{R^j}\eig\nonumber\\
&=&\com{\tx{k+1}}{R^j}\eig\nonumber\\
&=&\com{\tx{1}\tx{k}}{R^j}\eig\nonumber\\
&=&\left(\com{\tx{1}}{R^j}\tx{k}+\tx{1}\com{\tx{k}}{R^j}\right)\eig.\label{BF91}
\end{eqnarray}
Now we take the induction step by substituting from eq. \ref{BF91}.
\begin{equation}
\com{X_{k+1}}{R^j}\eig
=\underbrace{\left(\com{\tx{1}}{R^j}\tx{1}+\tx{1}\mc{A}^j\right)}\tx{k-1}\eig.\label{BF92}
\end{equation}
The indicated factor can be simplified. Firstly, substitute into it the explicit expression for the 
commutator, that was derived in eq. \ref{BF87}, and also the explicit expression for $\tx{1}$.
This yields
\begin{equation}
\com{\tx{1}}{R^j}\tx{1}+\tx{1}\mc{A}^j
=\left(C_F-\lambda\right)^{-1}\left(\com{C_F}{\mc{A}^j}\left(C_F-\lambda\right)^{-1}
+\mc{A}^j\right)\mc{A}^iB_i.\label{BF93}
\end{equation}
Here we need a little sleight of hand: we write the commutator appearing on the right as follows:
\begin{equation}
\com{C_F}{\mc{A}^j}=(C_F-\lambda)\mc{A}^j-\mc{A}^j(C_F-\lambda),\label{BF94}
\end{equation}
from which it follows that
\begin{equation}
(C_F-\lambda)^{-1}\com{C_F}{\mc{A}^j}(C_F-\lambda)^{-1}
=\mc{A}^j(C_F-\lambda)^{-1}-(C_F-\lambda)^{-1}\mc{A}^j.\label{BF95}
\end{equation}
If we substitute this back into eq. \ref{BF93} we find
\begin{equation}
\com{\tx{1}}{R^j}\tx{1}+\tx{1}\mc{A}^j=\mc{A}^j\tx{1}.\label{BF96}
\end{equation}
This, substituted back into eq. \ref{BF92} shows that 
if eq. \ref{BF84} is valid for $n=k-1$, it is also valid for $n=k$, which completes the proof by
induction.

Next we must show that $X$ is invertible. In order to do this, define recursively
\begin{eqnarray}
Y_0&=&\I,\nonumber\\
Y_{n+1}&=&-\sum_{l=1}^{n+1}X_lY_{n+1-l}.\label{BF96A}
\end{eqnarray}
The operator $X_n$ creates $2n$ ideal fermions when $2n\leq\Omega$. If $2n>\Omega$ then $X_n$
is the zero operator. From this follows by induction that $Y_n$ creates $2n$ ideal fermions 
(and is therefore zero when $2n>\Omega$): Say it is true for all $Y_k$, $k\leq n$, then from
the definition eq. \ref{BF96A} it follows that $Y_{n+1}$ creates $2(n+1-l+l)=2(n+1)$ ideal 
fermions, which completes the proof by induction. 

This means that 
\begin{equation}
Y=\sum_{n=0}^\infty Y_n\label{BF96B}
\end{equation}
contains only a finite number of non-zero terms, because all terms with $2n>\Omega$ are identically 
zero. $Y$ is therefore well-defined (provided
of course that the operators $X_n$ are well-defined, a matter that was touched upon previously).
A similar argument, based on the fact that $X_n$ annihilates $n$ ideal bosons, shows that $Y_n$ 
also annihilates $n$ ideal bosons.

Now we calculate
\begin{eqnarray}
\left(\sum_{m=0}^\infty X_m\right)\left(\sum_{n=0}^\infty Y_n\right)&=&
\sum_{n=0}^\infty\sum_{m=0}^n X_mY_{n-m}\nonumber\\
&=&\I+\sum_{k=0}^\infty\underbrace{\left(Y_{k+1}
+\sum_{m=1}^{k+1}X_lY_{k+1-m}\right)}_{=0},\label{BF96C}
\end{eqnarray}
so that $Y=X^{-1}$, i.e. $X$ is invertible. In the last line of eq. \ref{BF96C}, we took the
$m,n=0$ term $X_0Y_0=I$ out of the summation, set $k+1=n$ and used the fact that $X_0=I$
so that $X_0Y_{k+1}=Y_{k+1}$.

Utilizing the recursive definition of $Y_n$, one finds the various terms of $X^{-1}$ 
to be
\begin{eqnarray}
Y_0&=&I,\nonumber\\
Y_1&=&-X_1,\nonumber\\
Y_2&=&-X_2+X_1X_1,\nonumber\\
Y_3&=&-X_3+X_1X_2+X_2X_1-X_1X_1X_1,\nonumber\\
Y_n&=&\sum_{m=1}^n(-)^m\sum_{{\bf p}\in{\bf P_{mn}}}X_{p_1}X_{p_2}\ldots X_{p_m},\label{BF96D}
\end{eqnarray}
where ${\bf P_{mn}}=\{{\bf p}=(p_1,p_2,\ldots,p_m): 1\leq p_j\leq n+1-m;
\;p_1+p_2+\ldots+p_m=n\}$.

The foundational states in this mapping are $X^{-1}\left|0\right)=\left|0\right)$ and 
$\left(0\right|X=\left(0\right|$.
To summarize then, if $X$ as defined in eq. \ref{BF79} is well-defined, 
we have an isomorphism $X^{-1}\circ T$ between $\Hil_F$ and a subspace $V_{BF}\subset\Hil_{BF}$
of the ideal boson-fermion space.
The essential elements of this mapping read
\begin{eqnarray}
\left|0\right>&\longrightarrow&\left|0\right),\nonumber\\
\left<0\right|&\longleftarrow&\left(0\right|,\nonumber\\
\rf{A_j}&\longleftarrow&\rx{A_j}=B_j,\nonumber\\
\rf{A^j}&\longleftarrow&\rx{A^j}=B^i\mc{K}_i^j-\D{\frac{1}{2}}C_{ik}^{jl}B^iB^kB_l\nonumber\\
\rf{K_i^j}&\longleftarrow&\rx{K_i^j}=\mc{K_i^j}-C_{ik}^{jl}B^kB_l,\nonumber\\
\rf{a_\nu}&\longleftarrow&\rx{a_\nu}=X^{-1}\alpha_\nu X,\nonumber\\
\rf{a^\nu}&\longleftarrow&\rx{a^\nu}
=X^{-1}\left(\alpha^\nu+\chi_i^{\nu\mu}B^i\alpha_\mu\right)X,\label{BF97}
\end{eqnarray}
It is hard to find more explicit formulas for the extended images $\rx{a_\nu}$ and $\rx{a^\mu}$
in this general setting,
firstly, because $X$ is defined partially by stating its action on a complete basis, 
rather than by completely expressing it in terms of other known operators, and secondly,
because of the cumbersome formula for $X^{-1}$. In some particular cases however, more 
concise expressions may be derived.

When calculating the image $X^{-1}\left|\psi,N\right)$ where $\left|\psi,N\right)$ is a 
vector with a well-defined number of ideal bosons $N$, a simplification does occur, thanks to
the structure of $X^{-1}$. Since the operator $Y_k$ annihilates $k$ ideal boson, the only
terms in $X^{-1}$ that contribute are those with $Y_k$ such that $k\leq N$.
In particular, since $T\rf{a^\nu}\left|0\right>=\rbf{a^\nu}\left|0\right)
=\alpha^\nu\left|0\right)$ contains no ideal bosons, we have
\begin{equation}
\rf{a^\nu}\left|0\right>\stackrel{X^{-1}\circ T}{\longrightarrow}
\alpha^\nu\left|0\right).\label{BF98}
\end{equation}
The image of $\rf{a^\mu}\rf{a^\nu}\left|0\right>$ contains one ideal boson. It is therefore 
not to hard too calculate what $X^{-1}$ does to this image. Only the terms $Y_0=\I$ and 
$Y_1=-X_1$ contribute. Let us therefore calculate it in detail.
Firstly, note that $T\rf{a^\mu}\rf{a^\nu}\left|0\right>
=(\alpha^\mu\alpha^\nu+\chi^{\mu\nu}_jB^j)\left|0\right)$. Therefore 
$X^{-1}\circ T\rf{a^\mu}\rf{a^\nu}\left|0\right>
=(\alpha^\mu\alpha^\nu+\chi^{\mu\nu}_jB^j)\left|0\right)-\chi^{\mu\nu}_jX_1B^j\left|0\right)$.
Since $C_FB^j\left|0\right)=0$ we have from the definition of $X_1$ that 
$X_1B^j\left|0\right)=C_F^{-1}\mc{A}^iB_iB^j\left|0\right)=C_F^{-1}\mc{A}^j\left|0\right)$.
We know that $C_F=\mc{A}^k\mc{A}_k$ so that 
$C_F\mc{A}^j\left|0\right)=\mc{A}^k\com{\mc{A}_k}{\mc{A}^j}\left|0\right)=g\mc{A^j}\left|0\right)$.
Hence $C_F^{-1}\mc{A}^j\left|0\right)=\D{\frac{1}{g}}\mc{A}^j\left|0\right)$.
The final result then is
\begin{equation}
\rf{a^\mu}\rf{a^\nu}\left|0\right>\stackrel{X^{-1}\circ T}{\longrightarrow}
(\alpha^\mu\alpha^\nu+\chi^{\mu\nu}_jB^j-\frac{1}{g}\chi^{\mu\nu}_j\mc{A}^j)\left|0\right).
\label{BF99}
\end{equation}
Note that when we multiply by $\D{\frac{1}{2}}\chi_{\mu\nu}^j$ and sum over $\mu$ and $\nu$
we find 
$\D{\frac{1}{2}}\chi^j_{\mu\nu}\rf{a^\mu}\rf{a^\nu}\left|0\right>\mapsto gB^j\left|0\right)$ as
we should.

It is also not hard to compute $\left(0\right|\rbf{a_\nu}X$:
for an arbitrary vector $\left|\psi\right)$, consider 
$\left(0\right|\rbf{a_\nu}X\left|\psi\right)=\sum_{n=0}^\infty\left(0\right|\alpha_\nu X_n
\left|\psi\right)$. Every term $\alpha_\nu X_n$, with $n\geq1$ creates $2n-1$ ideal fermions
so that $\alpha_\nu X_n\left|\psi\right)$, $n\geq1$ has zero overlap with the vacuum. The
only surviving term has $n=0$. Hence $\left(0\right|\rbf{a_\nu}X\left|\psi\right)
=\left(0\right|\alpha_\nu\left|\psi\right)$. It follows that an extension for the
image of the functional $\left<0\right|\rf{a_\nu}$ is
\begin{equation}
\left<0\right|\rf{a_\nu}\stackrel{X^{-1}\circ T}{\longleftarrow}\left(0\right|\alpha_\nu.
\label{BF100}
\end{equation}
\section{Alternatives to the General Similarity Transformation}
\markright{\bf Section \thesection: Alternatives to the General Similarity Transformation}  
\label{SS1.8}
The similarity transformation of the previous section has the draw-back of being complicated. As 
a result it is not clear how to express the mapped single fermion operators as functions of the
ideal boson and fermion operators. Also, and perhaps more alarmingly, we could not prove the 
general validity of the definition of the similarity transformation, as has already been discussed.
Because of this state of affairs, I now show an alternative route that leads to the mapping of 
eq. \ref{BF97} in certain restricted circumstances.

These circumstances occur when, instead of mapping the whole space $\Hil_{ F}$ we only want
to map a subspace $\Hil_{F,\rm col}$, which is spanned by vectors of the form
\begin{equation}
\left(\prod_{m=1}^N \rf{A^{j_m}}\right)\rf{a^\nu}\left|0\right>,\label{BF101}
\end{equation}
and
\begin{equation}
\left(\prod_{m=1}^N \rf{A^{j_m}}\right)\left|0\right>.\label{BF102}
\end{equation}
We refer to $\Hil_{F, \rm col}$ as the collective fermion subspace. 
States of the form eq. \ref{BF101} are said to belong to the odd subspace of 
$\Hil_{F,\rm col}$ because they contain an odd number of fermions and similarly 
states such as eq. \ref{BF102} are said to belong to the even subspace. Often, a system can
effectively be described by a Hamiltonian that leaves the collective fermion subspace invariant.
Furthermore it is often then the case that the states associated with the 
lowest part of the Hamiltonian's spectrum lie in the collective fermion subspace. The argument
in this section is developed with such a situation in mind.

Under the mapping that we denoted by $T$ and that results from the Usui operator, the states in
eq. \ref{BF101} and eq. \ref{BF102} are mapped onto
\begin{equation}
\left(\prod_{m=1}^N R^{j_m}+{\mathcal A}^{j_m}\right)\alpha^\nu\left|0\right),\label{BF103}
\end{equation}
and
\begin{equation}
\left(\prod_{m=1}^N R^{j_m}+{\mathcal A}^{j_m}\right)\left|0\right),\label{BF104}
\end{equation}
where we define $R^j=B^i\mc{K}_i^j-\D{\frac{1}{2}}C_{ik}^{jl}B^iB^kB_l$ as previously.
Note that $R^j$ does not change the number of ideal fermions. What we want to achieve, and 
for which we had to introduce the similarity transformation $X$ in the previous section, is 
removing the collective ideal fermion pair creation operators ${\mathcal A}^j$ from the 
above expressions. However, in the present restricted case, we see that this can
be achieved very simply. We just take the projections onto the subspace with less 
than two ideal fermions, of the states in eq. \ref{BF103} and eq. \ref{BF104}. We denote this 
projection
operator $P$.

Perhaps the reader is troubled by the use of a projection operator. After all, the mapping from
$\Hil_{ F, \rm col}$ into $\Hil_{BF}$ must be invertible and we know that
projection operators aren't invertible. However, although $P$ is not invertible on the whole
$\Hil_{BF}$, it is invertible on the range $T(\Hil_{F, \rm col})$.
I demonstrate here the equivalent statement that $P\circ T$ is invertible on the whole 
collective fermion subspace $\Hil_{F, \rm col}$. The proof is similar in spirit to
the proof of invertibility of $T$. I only show the proof for odd states. For even states 
the proof follows the same route.\footnote{Why do I not simply say ``When $X^{-1}$ is restricted
to $T(\Hil_{F, \rm col})$ it is identical to $P$, the projection operator onto the space 
with fewer than two ideal fermions, restricted to the same domain.'' 
Answer: Because $X$ and $X^{-1}$ have definitions that may be problematic,
the cleanest view to take, is that when $X$ can be defined as in the previous section, 
$X^{-1}$ is an extension of the restricted $P$ operator to the whole $\Hil_{BF}$. 
The definition and properties of $P$ restricted to $T(\Hil_{F,\rm col})$ should 
not rely on those of the sometimes trustworthy $X$.}

We construct an arbitrary non-zero state $\left|\phi\right>$ from the odd subspace of 
$\Hil_{F,\rm col}$ by choosing appropriate indices
$l_{j,k}\in\{1,2,\ldots,M\}$, $\nu_j\in\{1,2,\ldots,\Omega\}$ and complex coefficients $\eta_j$
and setting 
$\left|\phi\right>=\sum_j\eta_j\left(\prod_{k=1}^{N_j}\rf{A^{l_{j,k}}}\right)\rf{a^{\nu_j}}
\left|0\right>$. Then $0\not=\left<\phi\right|\left.\phi\right>$ holds. The 
inner product $\left<\phi\right|\left.\phi\right>$ is explicitly given by
\begin{equation}
\left<\phi\right|\left.\phi\right>=\sum_{j,j'}\eta_j^*\eta_{j'}
\left<0\right|\rf{a_{\nu_j}}\left(\prod_{k=1}^{N_j}\rf{A_{l_{j,k}}}\right)
\left(\prod_{k'=1}^{N_j'}\rf{A^{l_{j',k'}}}\right)\rf{a^{\nu_j'}}\left|0\right>.\label{BF105}
\end{equation}
The properties of the mapping $T$ allow us to replace the right hand side of the previous
expression with
\begin{equation}
\left\{\sum_j\eta_j^*\left(0\right|\alpha_{\nu_j}\left(\prod_{k=1}^{N_j} B_{l_{j,k}}\right)\right\}
\left\{\sum_{j'}\eta_{j'}\left(\prod_{k'=1}^{N_j'} R^{l_{j',k'}}+{\mathcal A}^{l_{j',k'}}\right)
\alpha^{\nu_j'}\left|0\right)\right\},\label{BF106}
\end{equation}
where we have also separated the $\Hil_{BF}$ inner product 
into a bra and a ket with curly brackets. Since the bra state contains only one ideal fermion,
the collective fermion pairs ${\mathcal A}^j$ may be omitted from the ket state. The ket, 
with these pairs omitted is equal to $P\circ T \left|\phi\right>$, while the bra is just some
linear functional on $\Hil_{BF}$, which we'll call $f$ for brevity's sake. 
We have then deduced that $0\not=\left<\phi\right|\left.\phi\right>
=f\left(P\circ T\left|\phi\right>\right)$. This means that if $\left|\phi\right>$ is not the
zero vector in $\Hil_{F,\rm col}$ then $P\circ T\left|\phi\right>$ is not the
zero vector in $\Hil_{BF}$ and hence $P\circ T$ is invertible on the odd subspace of
$\Hil_{F,\rm col}$. The same result can be obtained for the even subspace, so that
we can conclude that $P\circ T$ is invertible on the whole $\Hil_{F,\rm col}$.

Now that the isomorphism between states in $\Hil_{F,\rm col}$ and 
states in a subspace of $\Hil_{BF}$ has been established, we can find the extended
images of operators under the mapping. 
Obviously only those operators that leave $\Hil_{F,\rm col}$ 
invariant may be mapped. The operators that leave $\Hil_{F,\rm col}$ invariant
are $\rf{A^j}, \rf{A_k}$ and $\rf{K^j_k}$, i.e. the even sector of the super-algebra
so that only these operators can be said to have images under $P\circ T$. A calculation that 
is too trivial to reproduce here give these images as
\begin{eqnarray}
\rf{A_j}&\longleftarrow&B_j,\nonumber\\
\rf{A^j}&\longleftarrow&B^i\mc{K}_i^j-\D{\frac{1}{2}}C_{ik}^{jl}B^iB^kB_l,\nonumber\\
\rf{K_i^j}&\longleftarrow&\mc{K_i^j}-C_{ik}^{jl}B^kB_l,\label{BF107}
\end{eqnarray}
the same as that in the previous section. 

In passing it may be noted that, had we worked with only the even subspace of 
$\Hil_{F,\rm col}$, we could have used the projection operator onto the 
subspace of $\Hil_{BF}$ that contains no ideal fermions, instead of $P$. 
The mapping would then have simplified to
\begin{eqnarray}
\rf{A_j}&\longleftarrow&B_j,\nonumber\\
\rf{A^j}&\longleftarrow&g B^j - \D{\frac{1}{2}}C_{ik}^{jl}B^iB^kB_l,\nonumber\\
\rf{K_i^j}&\longleftarrow&g \delta_i^j-C_{ik}^{jl}B^kB_l.\label{BF108}
\end{eqnarray}
This same mapping could also have been derived by a much simpler route, that starts with an
Usui operator
$U=\exp\left(\wt{B^i}\wt{\rf{A_i}}\right)$, without ideal fermions.

\section{Finding the Physical Subspace}
\markright{\bf Section \thesection: Finding the Physical Subspace}
\label{SS1.9}
In the preceding sections we discussed two situations, firstly a mapping
$\Hil_{F,\rm col}\stackrel{P\circ T}{\longrightarrow}\Hil_{BF}$ from the collective fermion
subspace into the ideal boson-fermion space and secondly a mapping
$\Hil_{F}\stackrel{X^{-1}\circ T}{\longrightarrow}\Hil_{BF}\label{BF110}$
with a larger domain. 
It was noted that $X^{-1}\circ T$, when it exists, is an extension of $P\circ T$ from the domain
$T(\Hil_{F, \rm col})$ to the a larger domain $T(V_{F})$, where $V_F$ is a subspace of $\Hil_F$ and
might in fact be the whole $\Hil_F$. In what follows we will
employ a new notation that covers both these situations. We denote the subspace of 
$\Hil_{F}$ that is being mapped (and which may be the whole $\Hil_{F}$) by 
$\MF$. The mapping itself (i.e. the operator $X^{-1}\circ T$ or $P \circ T$) is denoted 
$\mathcal T$. The range of $\mathcal T$, which is always a subspace of $\Hil_{BF}$, is 
called the physical subspace and is denoted $\MBF=\mathcal T(\MF)$.

As was previously explained, for operators $\hat{O}_F:\MF\rightarrow\MF$ that are expressed
in terms of products of the generators of the fermion realization of a certain superalgebra,
we were able to find corresponding operators $\hat{O}_{BF}:\Hil_{BF}\rightarrow\Hil_{BF}$,
expressed in terms of ideal boson and fermion operators, and are such that
\begin{equation}
\left.\hat{O}_{BF}\right|_{\MBF}=\mathcal T\hat{O}_F \mathcal T^{-1},\label{BF111}
\end{equation}
where $|_{V_{BF}}$ indicates a restriction of the operator's domain to the physical subspace.
An operator that satisfies eq. \ref{BF111} should properly be called a physical boson-fermion 
operator. For brevity's sake we will use the less descriptive term of physical operator. 
Physical operators should not be confused with the extended images of observables, which, it
is true, form a proper subset of the set of all physical operators. (In the original fermion 
system, an observable is any hermitian operator acting in the fermion space.)
Note that, as an immediate
consequence of eq. \ref{BF111}, any physical operator leaves invariant the physical subspace $\MBF$.
When we analyse a fermion system by investigating its mapped counterpart, then the fermion 
operators are replaced by physical operators. Although these operators are defined on the whole
$\Hil_{BF}$, only their action on the physical subspace is of relevance. An 
investigation of the mapped system must therefore always be accompanied by an identification
of the physical subspace $\MBF$.

In this section we discuss techniques for identifying the physical subspace. The most obvious of
these is the following. If we have a basis for $\MF$, say 
$\left\{\left|\phi_i\right>\right\}_{i=1}^{N}$, then
by applying the mapping $\mathcal T$ to each $\left|\phi_i\right>$ we find a basis
$\left\{\left|\phi_i\right)=\mathcal T\left|\phi_i\right>\right\}_{i=1}^{N}$ for the physical 
subspace 
$\MBF$. However, it turns out that this procedure leaves us with a basis for the physical subspace
that is complicated to work with. As I hope to make clear by considering applications of 
the Dyson mapping in the next two chapters, it is impractical to work with the above basis. 
It entails dealing with structures
that so closely resemble the immediately evident structures of the original system, that it is 
hardly worth the trouble to do a mapping.
If one works in the above basis, it is fair to say that the mapping changes the analysis of the
system under consideration only cosmetically. Essentially the same objects are manipulated, they
are only represented by different symbols.
The rest of this section 
will therefore be concerned with more sophisticated procedures of characterising the physical 
subspace.

A basis such as the one above is called a physical boson-fermion basis. 
It is to be contrasted with the ideal boson-fermion basis that consists of states the form
\begin{equation}
\left(\prod_j\frac{\left(B^j\right)^{N_j}}{\left(N_j !\right)^{\frac{1}{2}}}\right)\prod_l
\alpha^{\nu_l} \left|0\right).\label{BF112}
\end{equation}
When analising operators that are defined in $\Hil_{BF}$, it is natural to use 
this ideal boson-fermion basis, and only afterwards identify the physical subspace. An objection
might be that whereas $\MF$ and therefore also $\MBF$ are finite dimensional, $\Hil_{BF}$
is infinite dimensional. Since working in infinite dimensional spaces is hard, it seems a good idea
to identify a finite dimensional subspace that still contains all the physical information before
analysing the system. This is true. However, it is possible to find such  a finite dimensional 
subspace that contains the physical subspace, but has the advantage that ideal basis states
may still be used rather than physical basis states. This is possible because the fermion systems
we consider may only contain a finite number of particles. A property of the 
mapping we consider is that if a fermion state $\left|\psi\right>$ contains $N$ fermions 
then its mapping $S\left|\psi\right>$ is an eigenstate of $2N_b+N_f$ with eigenvalue $N$, where
$N_b$ counts the number of ideal bosons and $N_f$ counts the number of ideal fermions. Therefore,
if the fermion subspace $\MF$ only contains states with $N_i,\: i=1,2,\ldots,n$ fermions, 
instead of working with the whole ideal boson-fermion basis, we can select only those basis 
elements for which the twice number of ideal bosons plus the number of ideal fermions equals
one of the $N_i$. 

When $\MF$ is the collective fermion subspace
$\Hil_{F,\rm col}$ we can use an even stronger criterium to select the physically 
relevant ideal boson-fermion basis states. Because mapping involves projection onto the component
of $\Hil_{BF}$ with fewer than two ideal fermions, the ideal subspace is further reduced to 
contain states with at most a single ideal fermion. In this 
instance it is often found that the ideal boson-fermion basis states thus selected in fact
spans just the physical subspace $\MBF$ and nothing more.
The space spanned by ideal boson-fermion states that are selected with the aid of the above 
particle number considerations is called the ideal subspace. It contains the physical subspace 
as a subspace.

Typically, analysing a physical system means diagonalizing one or more physical operators. The procedure we expound here is the following:
\begin{enumerate}
\item{Diagonalize the physical operators using the ideal boson-fermion basis states that
fulfill the correct particle number requirements, i.e. that lie in the ideal subspace.}
\item{From the eigenstates thus obtained, select those that span the physical subspace.}
\end{enumerate}
How the second step in this procedure is implemented, depends on the application we have in 
mind and on the nature of the fermion space $\MF$. I discuss two procedures.

The first procedure utilizes the fact that physical operators leave the physical subspace
invariant. The argument I present here implements an idea that can be traced back to a paper 
\cite{GEH1} by
Geyer and co-workers. What I present here does not contain any new fundamental insight that goes
beyond the content of \cite{GEH1}. However, I present in greater detail the actual algorithm that
one would have to implement to utilize the idea of Geyer and co-workers. 
The procedure is based on an algorithm for solving the following problem:

{\bf Problem:} Given are two linear operators $A:V\rightarrow V$ and $B:V\rightarrow V$ where
$V$ is a finite dimensional vector space with dimension $M$. $A$ is assumed fully diagonalizable
with a non-degenerate spectrum. Find a non-trivial subspaces (if one exists) of $V$ that 
$A$ and $B$ both leave invariant. (Non-trivial here means a subspace of $V$
that is neither $V$ itself, nor does it only contain the zero vector.)

{\bf Solution:} Let $\left|\phi_i\right)$, $i=1,\ldots,M$ be the eigenvectors of $A$, with
corresponding eigenvalues $\lambda_i$. We have assumed no degeneracy so if $i\not=j$ then
$\lambda_i\not=\lambda_j$. Let $\left(\psi_i\right|$, $i=1,\ldots,M$ be the left eigenvectors
of $A$ such that $\left(\psi_i\right|A=\lambda_i\left(\psi_i\right|$. It is easy to prove that
$\left(\psi_i\right.\left|\phi_j\right)$ equals zero if and only if $i\not=j$. By choosing the
left eigenvectors to be properly normalized we may therefore, without loss of generality take
\begin{equation}
\left(\psi_i\right.\left|\phi_j\right)=\delta_{ij}.\label{BF113}
\end{equation}
This gives rise to the unity resolution in $V$
\begin{equation}
{\rm I}=\sum_{i=1}^{M}\left|\phi_i\right)\left(\psi_i\right|.\label{BF114}
\end{equation}
Now suppose that $W$ is an $N$-dimensional subspace of $V$ that $A$ and $B$ both leave invariant. 
Because $A$ leaves $W$ invariant, there exists a basis for $W$ that consists of
$N$ of the eigenvectors of $A$. Let $C_W$ be this basis.
Then we have the following theorem, the proof of which is trivial:
\begin{theorem}
\label{BF115}
If $\left|\psi\right)$ is a vector in $W$, which has the unique expansion 
\begin{equation}
\left|\psi\right)=\sum_{k=1}^{M}c_k\left|\phi_k\right),
\end{equation}
in terms of the eigenstates of $A$, then for each $c_k\not=0$, $|\phi_k)$ is a vector in $C_W$.
\end{theorem}
Define a matrix $B_{ji}=\left(\psi_i\right|B\left|\phi_j\right)$. Thanks to the unity resolution
eq. \ref{BF114} it holds that
\begin{equation}
B\left|\phi_j\right)=\sum_{i=1}^{M}B_{ji}\left|\phi_i\right).\label{BF116}
\end{equation}
A corollary of Theorem \ref{BF115} can now be stated.
\begin{corollary}
\label{BF117}
If $\left|\phi_j\right)$ belongs to a subspace that $A$ and $B$ both leave invariant and
$B_{ji}\not=0$ then $\left|\phi_i\right)$ also belongs to that subspace.
\end{corollary}
Again the proof is trivial.

Using the tools we have gathered we can formulate an algorithm that enables us to find some
non-trivial subspaces that $A$ and $B$ both leave invariant, if any exists.
If (and only if) none exist, the algorithm will find only the trivial subspace $V$:

Assume there exists a non-trivial subspace $W$  that both $A$ and $B$ leave invariant. $W$ 
has to contain an eigenvector of $A$, because $A$ leaves $W$ invariant. Therefore, guess that
$W$ contains $\left|\phi_j\right)$. If this guess is wrong, the algorithm I am about to describe
will return $W=V$.
If this happens, we simply try another eigenvector of $A$ as belonging to $W$. If we
have exhausted all options and the algorithm still returns $W=V$, then we'll  know that no
non-trivial subspace of $V$ exists, that $A$ and $B$ both leave invariant.
Define sets of indices recursively as follows: $I_j^{(0)}=\{j\}$. For $k=1,2,\ldots$, let
$I_j^{(k+1)}$ be the set of all $l\in1,\ldots,M$ such that there exists an $i\in I_j^{(k)}$ for
which $B_{il}\not=0$. To see what I am up to, note that, since $\left|\phi_j\right)$ belongs
to $W$, if $i\in I_j^{(1)}$, it holds that $B_{ji}\not=0$ and therefore it follows from Corollary 
\ref{BF117} that $\left|\phi_i\right)$ is also an element of $W$. Then, if $i\in I_j^{(1)}$ 
and $l\in I_j^{(2)}$ it holds that $B_{il}\not=0$. Therefore, because $\left|\phi_i\right)$ 
is in $W$, from Corollary \ref{BF117} follows that so too is $\left|\phi_l\right)$.
Thus far we conclude that if $i\in I_j^{(0)}\cup I_j^{(1)}\cup I_j^{(2)}$ it holds
that $\left|\phi_i\right)\in W$.
For any $k$ define $U_j^{(k)}=\cup_{l=0}^k I_j^{(l)}$. 
Inductively we arrive at the result: For any $k$, if $i\in U_j^{(k)}$ then $\left|\phi_i\right)$
belongs to $W$. 
Furthermore $U_j^{(k)}$ is a subspace of $U_j^{(k+1)}$, or in other words, if $i$ belongs to
$U_j^{(k)}$ then it also belongs to $U_j^{(k+1)}$, but 
the number of elements in any  $U_j^{(k)}$ is at most $M$. 
This means that at some $k$, the set $U_j^{(k)}$ must stop growing.
More formally put, there always exists a $\tilde{k_j}$ such that 
$U_j^{(k)}=U_j^{(\tilde{k_j})}$ for all $k>\tilde{k_j}$. If we define 
$C_j=\left\{\left|\phi_i\right):\;i\in U_j^{(\tilde{k_j})}\right\}$ then there exists a basis
for the invariant subspace $W$ that contains all the elements of $C_j$. The result we have arrived 
at may also be
stated: {\em If $W$ is a subspace that $A$ and $B$ both leave invariant, and $W$ contains
the eigenvector $\left|\phi_j\right)$ of $A$, then $W$ contains all 
those eigenvectors of $A$ that form the set $C_j$.}

Now we show that $A$ and $B$ both leave $span(C_j)$ is itself invariant. 
Since $span(C_j)$ has a basis $C_j$ that is comprised of 
eigenvectors of $A$, we know that $A$ leaves $span(C_j)$ invariant. To prove that $B$
leaves $span(C_j)$ invariant, assume that it is not so. Then there exists an index 
$i\in U_j^{(\tilde{k_j})}$ such that $B\left|\phi_i\right)$ does not lie in $span(C_j)$.
This is only possible if there is an index $l$ such that $B_{il}\not=0$ but 
$l\not\in U_j^{(\tilde{k_j})}$. Since $i$ is a member of $U_j^{(\tilde{k_j})}$,
it must be a member of some $I_j^{(m)}$ with $m\leq\tilde{k_j}$. Suppose 
$m<\tilde{k_j}$, then because of the recursive definition of the $I_j^{(k)}$ and because
$B_{il}\not=0$, it holds that $l\in I_j^{(m+1)}$. Since $m+1\leq\tilde{k_j}$, it holds that 
$I_j^{(m+1)}$ is a subset of $U_j^{(\tilde{k_j})}$ and therefore $l$ is an element of 
$U_j^{(\tilde{k_j})}$, which is a contradiction. Suppose on the other hand $m=\tilde{k_j}$.
But then, since $B_{il}\not=0$, the index $l$ is an element of $I_j^{(\tilde{k_j}+1)}$, 
which is a subset of $U_j^{(\tilde{k_j}+1)}$. But $\tilde{k_j}$ was defined in such a way
that $U_j^{(\tilde{k_j}+1)}=U_j^{(\tilde{k_j})}$. Thus again we are led to the contradiction
that $l\in U_j^{(\tilde{k_j})}$.

The above results imply that both $A$ and $B$ leave $span(C_j)$ itself invariant 
and that $span(C_j)$ is minimal in the sense that every subspace of $V$ that contains
$\left|\phi_j\right)$ as an element, and that $A$ and $B$ leave invariant,
contains $span(C_j)$ as a subspace. If we run this algorithm, starting with $j=1$ and 
continuing until $span(C_j)\not=V$ or $j=M$, we will find a nontrivial subspace of $V$, that
$A$ and $B$ leave invariant, if such a subspace exists. If none exists, and only then, will we 
reach $j=M$ and 
$span(C_i)=V$ for $i=1,2,\ldots,M$.

I shall now explain how we can use the above algorithm to find the physical subspace. The key
insight is that all physical operators leave the physical subspace invariant. To implement the
algorithm, we set $V$ equal to the ideal subspace. For the operator $A$ we choose a physical
operator that has a nondegenerate spectrum in $V$. If degeneracy is a problem, we may simply
use a complete set of commuting physical operators that we simultaneously diagonalize. Instead
of using only one operator $B$ we choose a set of physical operators $B_\alpha$, 
$\alpha=1,2,\ldots,N$ that is large enough that they leave no subspace other than the physical 
subspace invariant. Using a set of operators $\left\{B_\alpha\right\}_{\alpha=1}^N$ 
simply requires that in the 
algorithm we change the recursive definition of the index sets $I_j^{(k)}$ to:
{\em $I_j^{(k+1)}$ is the set of all indices $l$ such that, for some $i\in I_j^{(k)}$ 
there exists an $\alpha$ such that $(B_\alpha)_{il}\not=0$.} Since we assume that the only
subspace that both $A$ and all the $B_\alpha$ leave invariant, is the physical subspace,
this is the only possible subspace that our algorithm can yield.

I fear that the above description of the algorithm looks very formal and that the simple
idea behind it is obscured. Therefore I roughly summarize. To find the physical subspace we
first diagonalize a physical operator $A$ that has a non-degenerate spectrum. A sub-set of its 
right eigenvectors spans the physical subspace. Therefore we make a guess and take one particular
eigenvector of $A$ and assume that it is in the physical subspace. The algorithm will let us know if
our guess is wrong. To find the physical subspace we now take a physical operator $B$ that we assume
is such that the only space that both $A$ and $B$ leave invariant, is the physical subspace. If
a single operator $B$ does not do the job, we can, as was explained, extend the procedure to use
several operators $B_\alpha$, such that, taken together, the only subspace that both $A$ and
all the $B$ simultaneously leave invariant, is the physical subspace. 
For simplicity, in the rest of this summary
we assume a single operator $B$ is sufficient. We now apply the operator $B$ to the 
eigenvector of $A$ that we assumed to lie in the physical subspace, and expand the result in
terms of the eigenvectors of $A$. Every eigenvector of $A$ for which there is a non-zero expansion
coefficient in this expansion must also lie in the physical subspace. We now repeat the process of 
applying $B$ and
expanding in terms of the eigenvectors of $A$ to each of these newly found elements of the basis
for the physical subspace. If non-zero expansion coefficients for eigenvectors of $A$, different 
from those already taken to be inside the physical subspace, are generated, these eigenvectors 
are
also taken to be inside the physical subspace, and the process is repeated for them too. We
continue in this way, until non-zero expansion coefficients are generated only for states 
that are already taken to be inside the physical subspace. If, at this point, what we take to
be the physical subspace does not include all the eigenstates of $A$, our original guess has
been vindicated, and we have found a basis for the physical subspace. If, on the other hand, we are 
left
with a result that says all of the eigenstates of $A$ lie in the physical subspace, this either 
means
that the state we initially guessed to be in the physical subspace is actually not in the
physical subspace, or there simply are no ghost states. To eliminate the first possibility we
repeat the process starting with a different initial guess. If, after all possible choices for the
initial guess, we still end up with the algorithm saying that all of the eigenvalues of $A$ 
belong to the physical subspace, then we conclude that this is indeed the case. 

The astute reader may have noticed that several things can go wrong in this procedure. The
root of the problem is that, assuming the physical operators are obtained through
mapping, we only have control over their behaviour inside the physical subspace. What they
do in the rest of the ideal subspace is determined by the specifics of the mapping. So, 
while it is true that we may find physical operators that satisfy all our requirements inside 
the physical subspace, i.e. no degeneracy for $A$ and irreducibility for the $B_\alpha$
together with $A$,
whether they still do this in the whole ideal subspace is not within our control. There 
might be ways around these problems, but I do not consider them because, in the application
I will discuss, we will not run into any trouble.
 
We now come to a second method for identifying the physical subspace, that goes by the 
rather uninformative name of `$\mathcal R$-projection'.\footnote{The symbol `$\mathcal R$' has 
no special significance and, strictly speaking, the procedure does not involve projection.} 
The way that this method is presented below looks superficially different from how
it was originally presented by Dobaczewski in \cite{Do1}, but the method is in fact identical to
the $\mathcal R$-projection introduced there.
We shall see that this procedure is not as widely applicable as the first procedure. It 
can only be implemented when the fermion space that gets mapped is the collective subspace
$\Hil_{F,\rm col}$.
To discuss the procedure effectively we first have to define a few operators and elaborate on 
certain of their properties.

Let $C_{BF}$ be the canonical mapping on $\Hil_{BF}$. This means that 
$C_{BF}$ maps a general vector $\left|\psi\right)$ in $\Hil_{BF}$ onto
the functional $\left(\psi\right|$, where the meaning of $\left(\psi\right|$ is fixed
through the boson-fermion inner product defined on $\Hil_{BF}$, as we have explained in
a preceding section. The operator $C_{BF}$ is anti-linear. Let $\tau$ be the 
linear operator that acts on the space of linear functionals defined on $\Hil_{BF}$, that
maps a general linear functional $\left(\psi\right|$ onto $\left(\psi\right|\mathcal T$, 
where $\mathcal T$ is
the boson-fermion mapping. The operator 
$\tau$ maps linear functionals defined on $\Hil_{BF}$ onto linear functionals
defined on the fermions space $V_{F}$ which is the domain of $\mathcal T$. Lastly, let
$C_{F}^{-1}$ be the inverse of the canonical map on $\Hil_{F}$. 
$C_{F}^{-1}$ is an anti-linear operator that maps a general functional 
$\left<\phi\right|$ onto the vector $\left|\phi\right>$, where this time, the relationship 
between $\left<\phi\right|$ and $\left|\phi\right>$ is determined by the fermion inner product
defined on $\Hil_{F}$.

We now form a linear operator $\mathcal R$ through the conjunction of the above operators: We 
define
\begin{equation}
{\mathcal R}=C_{F}^{-1}\circ \tau \circ C_{BF}.\label{BF118}
\end{equation}
Since, of the operators used in the definition of $\mathcal R$, two are anti-linear, and the other
linear, $\mathcal R$ itself is linear. It is an operator that maps the space $\Hil_{BF}$
onto $V_{F}$, the domain of the mapping $\mathcal T$, which is a subspace of fermion Fock space. 
This statement can be verified simply by keeping
track of the domains and ranges of the operators used in the construction of $\mathcal R$.

It will now be demonstrated that the operator $\mathcal R$ has the property that 
$\mathcal R\left|\psi\right)=0$ if and only if $\left|\psi\right)$ belongs to the orthogonal 
complement $V_{BF \perp}$ of the physical subspace $V_{BF}$: 
First we prove that $\mathcal R \left|\psi\right)=0$ implies that $\left|\psi\right)$ lies in 
the orthogonal complement of the physical subspace $V_{BF,\perp}$.
Let $\left|\phi\right>$ be any vector in $V_F$. 
Now we compute the inner product in $H_{BF}$ between $\mathcal R\left|\psi\right)$ and
the arbitrary vector $\left|\phi\right>$.
\begin{eqnarray}
\left<\phi\right|{\mathcal R}\left|\psi\right)&=&\left<\phi\right|C_{F}^{-1}
{\tau} C_{BF}\left|\psi\right).\label{BF120}
\end{eqnarray}
In the above equation, by the respective definitions of $\tau$ and $C_{BF}$, 
the functional $\tau C_{BF}\left|\psi\right)$ is equal to $\left(\psi\right|\mathcal T$.
Furthermore $\left<\phi\right|C_{F}^{-1}$ maps functionals defined on the 
fermion space $V_{F}$ onto the complex numbers in the following way: Let 
$\left<\chi\right|$ be any functional defined on $V_{F}$ then 
\begin{eqnarray}
\left<\phi\right|C_{F}^{-1}\left(\left<\chi\right|\right)&=&\left<\phi\right.
\left|\chi\right>,\nonumber\\
&=&\left<\chi\right|\left.\phi\right>^*.\label{121}
\end{eqnarray}
Hence we arrive at the equation
\begin{equation}
\left(\phi\right|{\mathcal R}\left|\psi\right)=\left(\psi\right|\mathcal T\left|\phi\right>^*,
\label{BF122}
\end{equation}
in which the right-hand side is an inner product of $\left|\psi\right)$ with a vector
$\mathcal T\left|\phi\right>$ which lies in the range of $\mathcal T$, i.e. the physical subspace. 
Hence, if $\left|\psi\right)$ lies in the orthogonal compliment $V_{BF \perp}$ of the physical
subspace $\left<\phi\right|\mathcal R\left|\psi\right)$ equals zero, for arbitrary 
$\left|\phi\right>$, so that $\mathcal R\left|\psi\right)$ must be zero.
Now the reverse is proved, namely if $\left|\psi\right)$ is not an element of $V_{BF\perp}$
then $\mathcal R\left|\psi\right)$ is not zero: Take a $\left|\psi\right)$ that is not an element
of $V_{BF\perp}$. Then it can be written uniquely as 
$\left|\psi\right)=\left|\psi_\parallel\right)+\left|\psi_\perp\right)$ where 
$\left|\psi_\parallel\right)$ is a nonzero element of the physical subspace $V_{BF}$ and 
$\left|\psi_\perp\right)$ is an element of $V_{BF\perp}$. Since 
$\left|\psi_\parallel\right)$ is a non-zero element of the physical subspace $V_{BF}$, 
there exists a non-zero vector $\left|\psi_\parallel\right>$ in the fermion space $V_{F}$, 
such that $\mathcal T\left|\psi_\parallel\right>=\left|\psi_\parallel\right)$. 
If we compute
the inner product between $\left|\psi_\parallel\right)$ and $\mathcal R\left|\psi\right)$, we find:
\begin{equation}
\left(\psi_\parallel\right|\mathcal R\big|\psi\big)=\left<\psi_\parallel\right|
 C_{F}^{-1}\tau C_{BF}\left\{\left|\psi_\parallel\right)+
\big|\psi_\perp\big)\right\}.\label{BF123}
\end{equation}
On the right-hand side of this expression, the component $\left|\psi_\perp\right)$ can be
omitted as it is killed off by $\mathcal R$. This leaves us with
\begin{eqnarray}
\left(\psi_\parallel\right|\mathcal R\left|\psi\right)&=&\left<\psi_\parallel\right|
C_{F}^{-1}\tau C_{BF}\left|\psi_\parallel\right)\nonumber\\
&=&\left(\psi_\parallel\right|\mathcal T\left|\psi_\parallel\right>^*\nonumber\\
&=&\left(\psi_\parallel\right|\left.\psi_\parallel\right),\label{BF124}
\end{eqnarray}
which is nonzero. This completes the proof.

We have seen that at our disposal is an operator $\mathcal R$, that has the orthogonal complement 
of the physical subspace as its null space. Before we can use this operator 
though, we must be able to evaluate its action on a generic basis element of the ideal subspace.
This evaluation, as it turns out, is only feasible when the fermion space that is mapped, is 
the collective subspace $\Hil_{\mathcal F,\rm col}$. In this case, the ideal subspace only
contains states with fewer than two ideal fermions. In what follows, I therefore evaluate
${\mathcal R}\prod_j\left(B^j\right)^{N_j}\alpha^\nu\left|0\right)$ and briefly indicate what 
goes wrong when more than one ideal fermion is present. We start from the definition of 
$\mathcal R$ in eq. \ref{BF118}.
\begin{eqnarray}
{\mathcal R}\prod_j\left(B^j\right)^{N_j}\alpha^\nu\left|0\right)&=&
C_{F}^{-1}\tau C_{BF}\prod_j\left(B^j\right)^{N_j}\alpha^\nu\left|0\right),
\nonumber\\
&=&C_{F}^{-1}\tau\left\{\left(0\right|\alpha_\nu\prod_j\left(B_j\right)^{N_j}\right\},
\nonumber\\
&=&C_{F}^{-1}\left\{\left(0\right|\alpha_\nu\prod_j\left(B_j\right)^{N_j}\mathcal T\right\}.
\label{BF125}
\end{eqnarray}
When only one ideal fermion annihilation operator is present in the last line of the previous 
expression, and only then, can the functional in curly brackets easily be evaluated. The key
is to realize that when we restrict the functional $\left(0\right|\alpha_\nu\prod_j
\left(B_j\right)^{N_j}$ to the physical subspace, as we may when we stick the operator $\mathcal T$
in front of it, we have
\begin{equation}
\left.\left(0\right|\alpha_\nu\prod_j\left(B_j\right)^{N_j}\right|_{V_{BF}}=
\left<0\right|\rf{a_\nu}\prod_j\rf{A_j}^{N_j}\mathcal T^{-1}.\label{BF126}
\end{equation}
If more than one ideal fermion were present, the similarity transformation that is part of the 
mapping would have wreaked havoc with the ideal fermions, churning out states that contain 
high numbers of non-collective fermion strings $\rf{a_{\mu_1}}\rf{a_{\mu_2}}\ldots\rf{a_{\mu_n}}$. 
Let us
therefore remain safely inside the collective subspace and substitute the most recent result
eq. \ref{BF126} into the expression we have for $\mathcal R$ acting on an ideal subspace basis
element, namely equation eq. \ref{BF125}:
\begin{eqnarray}
{\mathcal R}\prod_j\left(B^j\right)^{N_j}\alpha^\nu\left|0\right)&=&
C_{F}^{-1}\left\{\left<0\right|\rf{a_\nu}\prod_j\rf{A_j}^{N_j}\mathcal T^{-1}\mathcal T\right\}
\nonumber\\
&=&C_{F}^{-1}\left\{\left<0\right|\rf{a_\nu}\prod_j\rf{A_j}^{N_j}\right\}\nonumber\\
&=&\left\{\prod_j\rf{A^j}^{N_j}\rf{a_\nu}\left|0\right>\right\}.\label{BF127}
\end{eqnarray}
Thus, when the
fermion space that we map is the collective subspace $\Hil_{\mathcal F,\rm col}$, we
have a very simple prescription for evaluating the expression 
${\mathcal R}\prod_j\left(B^j\right)^{N_j}\alpha^\nu\left|0\right)$: Simply replace each
$B^j$ with the corresponding collective fermion operator $\rf{A^j}$ and replace the
unpaired ideal fermion with a real fermion.

To find the physical subspace using the $\mathcal R$ operator, we go about as follows. Again
we assume that at our disposal is a physical operator $A$ that is fully diagonalizable in
the ideal subspace. We assume that $A$ has a non-degenerate spectrum. (If degeneracy occurs,
the procedure I am about to describe may trivially be generalized to accommodate, instead
of a single operator $A$, a complete set of commuting physical operators, that remove
any degeneracy.) In analogy with what we did previously, we diagonalize $A$ to find
say $M$ eigenvectors $\left|\phi_i\right)$ associated with eigenvalues $\lambda_i$ such
that if $i\not=j$ then $\lambda_i\not=\lambda_j$. Similarly, a diagonalization of $A^\dagger$ 
yields $M$ eigenvectors $\left|\psi_i\right)$ associated with eigenvalues $\lambda_i^*$. The
orthonormality condition
\begin{equation}
\left(\psi_i\right.\left|\phi_j\right)=\delta_{ij},\label{BF128}
\end{equation}
holds if the correct normalization is applied.

Let us denote the dimension of the physical subspace $N$. Then because $A$ is a physical operator,
there exists an index set $\mathcal I=\left\{j_1,j_2,\ldots,j_N\right\}$ such that 
$C=\left\{\left|\phi_j\right)\right\}_{j\in \mathcal I}$ is a basis for the physical subspace. 
From the orthonormality condition in eq. \ref{BF128} it follows that 
$C_\perp=\left\{\left|\psi_j\right)\right\}_{j\not\in \mathcal I}$ is a basis for the orthogonal 
complement
of the physical subspace. Because the orthogonal complement of the physical subspace is also the
null space of the operator $\mathcal R$, we conclude that $\left|\phi_j\right)$ is a member of the
basis $C$ for the physical subspace if and only if $\mathcal R\left|\psi_j\right)\not=0$.
\section{Looking Back}
\markright{\bf Section \thesection: Looking back}
\label{SS1.10}
Now that the generalized Dyson boson-fermion mapping is derived, we are in a position to ask a
few questions and put the ideas presented into a larger perspective. For instance, 
\begin{itemize}
\item{To what end was this theory first developed?}
\item{Where can it be applied, and what do we gain by applying it?}
\item{To what extent is the presentation of the theory contained in this chapter similar to 
presentations that have appeared in the past?}
\item{How complete is the presentation? Are there aspects that are not yet fully understood?
Are there unproved conjectures on which the theory relies?}
\end{itemize}
As was mentioned in the introduction, generalized Dyson 
boson-fermion mappings are instances of what is known as boson expansion methods. Workers
in this field will probably explain the utility of boson expansion methods as follows:
A generic boson expansion will map fermion states onto boson states. (Before the development
of the generalized Dyson boson-fermion mapping, attempts have been made to include particles
that are not bosons in the mapped system. However, these `particles' were neither bosons nor
fermions, but had more complicated commutation relations amongst themselves or with the bosons.
See for instance \cite[p. 428]{KM1}.)
A boson expansion is usually such that that a bi-fermion operator $\rf{A^k}$ in the original
system maps onto a function of a few types of boson creation and annihilation operators. This
function is typically a Taylor series with infinitely many non-zero terms. The order of a term
in the series is equal to the number of normal ordered boson-operators it contains. 
The disadvantage of this (which 
partially prompted the study of Dyson-type mappings) is that for matrix elements of mapped operators
between some state vectors, the series converge very slowly. Among the advantages are that, whereas
in the original system there were say $\Omega$ different one-particle states that a fermion
could occupy, the mapped system consists of $M$ coupled bosonic oscillators, with typically
$M\ll\Omega$. Also, for some states or regions of the spectrum of the system, high-order
terms in the expansions of operators have a negligible effect and may be forgotten. Thus
one is sometimes able to approximate a complicated fermion system with many one-particle states,
with a system consisting of a few coupled bosonic oscillators. In the mapped system, the truncation
of high order terms often result in mapped Hamiltonians with no more than, say, four-body 
interactions. One might be a little skeptical about the claim that the approximate
mapped bosonic system is simpler to work with than the original fermion system. To my mind, 
it is not clear that an interacting boson system, with $M$ one-particle states and on the order
of $N$ bosons to occupy these states, is necessarily simpler than a fermion system with on the
order of $2N$ fermions that are arranged among $\Omega$ one-particle states, even when $\Omega$ is
much larger than $M$. Presumably, the simplification arises as follows: When one throws away
high order terms in an expansion, one is effectively taking a limit and reducing the description
of the system to include only what is pertinent to that limit. In the end, I think one has to
take the word of authors who found it possible to extract useful information about systems using
this technique, that the approximate mapped systems are simpler to work with than the original 
ones. 

A generalized Dyson boson-fermion mapping, on the other hand,  results in images of the fermion 
operators
that are functions of boson operators, with only a few non-zero terms. Our instinct to throw
away high order terms in power series expansions is of no use.\footnote{Sometimes this is a good 
thing.} In other words, a Dyson-type mapping does not wear on its sleeve a sign that suggests
how to proceed after the mapping is done. Perhaps the reader is puzzled by the negative tone of
the last statement: After all, Dyson type mappings give us boson-fermion systems with relatively
simple Hamiltonians. As is the case for other boson expansion methods, fermion systems with
$2N$ fermions arranged among $\Omega$ one-particle states are replaced by boson-fermion systems
where the states we are typically interested in, are mostly bosonic. There are of the order
$N$ bosons, arranged among $M\ll\Omega$ states. It seems we have all the advantages of the other
mappings, but without having to throw away terms. In my experience though, a Dyson-mapped system
is as non-trivial to analyse as the original. It is just that some techniques of analysis are
better suited to the mapped system than the original. Other analytical tools might be easier to
use if one sticks to the original system. I would venture to say that, whereas other mapping 
techniques separate out a certain limit of the problem and introduce simplification by disregarding
details unimportant to that limit, Dyson-type mappings simply constitute a way of looking from 
another angle on the original problem, with all its complicated detail still in tact.
Why then have people developed the mapping?
I think there are four reasons:

Firstly, Dyson-type mappings can be used as a stepping stone to deriving many of the other boson 
mappings that comprise the field of boson expansion methods \cite{Do1}. Secondly, there was
a need to accommodate systems with an odd number of particles. To many workers in the field, the
idea seems natural that a bosonic description should be useful for fermion pairs. However, if
there are an odd number of fermions in a system where bound fermion pairs are energetically 
favoured, one would guess the low energy states to consist of configurations with all but one of 
the fermions bound in collective pairs. Ideally, in the mapped system, one would like the
one unpaired particle to be represented as a fermion independent of the bosons. A Dyson-type mapping
has proved to be the only mapping `simple' enough to accommodate such fermions in the mapped system.
At reading the previous sentence, the reader might have felt a little uneasy. I previously
mentioned that, 
in my experience, the mapped system is as non-trivial to analyse as the original system it derives 
from. So, if it does not lead to an obvious simplification, what
does it help to have a mapping, even if it can accommodate fermions on the mapped side? The
answer to this question is the third reason: To build nuclear models one can follow one of
two approaches. The first is a microscopic approach. Here we start with a number of protons
and neutrons, whose properties and interactions we think we know something about. We then
write down a Hamiltonian that is based on this knowledge about the properties of the nucleons, and
check if indeed nucleons with these properties result in nuclei with the properties we
measure in the laboratory. The other approach is phenomenological. Here we make very few 
assumptions about the properties of nucleons themselves but focus on the known properties
of the nucleus as a whole. Our goal is to write down (almost) any Hamiltonian that reproduces
(if interpreted sympathetically), the properties of the nucleus. If we are successful in this
task, we try to deduce properties of the nucleons from the form of the Hamiltonian. 
It so happens that phenomenological models where the quanta are bosons, or for systems
with an odd number of particles, bosons and one fermion, reproduce observed nuclear spectra
very well. To understand what these models say about the fermionic nucleons, and to check
that the reproduction of spectra is not simply a coincidence, we would like to construct
boson-fermion mappings so that we can make a translation between the microscopic models
and the phenomenological models. For this translation it would be nice if the same
mapping works for systems with an even number of particles and systems with an odd number of
particles. After all, on the (microscopic) fermion side, we believe that the stationary states of 
even and odd nuclei are eigenstates with different numbers of particles, of the same Hamiltonian. 
On the phenomenological boson side we give expression to this notion by also writing down one 
Hamiltonian that treats both the even and odd case. A single mapping should link the two sides. 
Also, since the phenomenological models contain only few-body interactions, 
it should be investigated whether a mapping is possible where a correspondence can be
established without infinite series that have to be truncated. This wish list is very
compatible with Dyson-type mappings. The one problem is the non-unitarity of Dyson
mappings. Sometimes this can be overcome by a similarity transformation that preserves the
desirable properties of the generalized Dyson mapping, but still transforms the mapped Hamiltonian
into an hermitian operator \cite{KV1}. The last reason that generated interest in Dyson-type 
mappings is this:
In the calculation of matrix elements of operators, it is more pleasant to use the commutation
rules of bosons than those of the more complicated fermion superalgebra. A mapping such as
the generalized Dyson mapping is useful in this context, because it allows us to work with
bosons and their commutators, without making operators so complicated under the mapping that
the advantage of working with bosons is lost. For mappings in which the images of the original
operators are infinite series of boson operators, we might have to calculate vacuum expectation
values of products of arbitrarily large numbers of boson operators.

If we remove the search for a link between microscopic and phenomenological models from this list 
of reasons, the motivation for studying Dyson-type mappings becomes a little thread-bare. One
aim of the this thesis then, is to find new situations in which Dyson-type mappings can be useful.
Also, in the type of application mentioned above, systems are considered that have Hamiltonians 
that can be written in terms of only the operators $\rf{A^k}$ and $\rf{A_k}$ of the even sector
of the superalgebra. In such systems, the mapping is then usually only carried out for
the collective subspace, so that the mapped systems contains at most one ideal fermion. Another
aim of this thesis is then to push the generalized Dyson boson-fermion mapping beyond these
restrictions.

I conclude by making a few remarks about how the development of the theory, as presented in
this chapter, compares with the original presentation published in \cite{DSG1}-\cite{CG1}. 
As mentioned in the introduction, the present text contains more detail than the original 
publications. Therefore, for instance, when I derive the properties of the superalgebra, I use a
general representation, to be doubly sure that the original fermion representation and the
`induced' boson-fermion representation share these properties. This is in contrast to 
what is done in \cite{CG1}. Also, I take care to show a way in which meaning can be given to
the demand that, in the definition of the Usui operator, the ideal and real fermion 
operators (or more precisely 
$\widetilde{\alpha}$ and $\widetilde{a}$) anti-commute. Another example is the demonstration
of the invertibility of the mapping. However, these are small matters. A person,
like myself, who is new to the field, might struggle with them for a while, but the solutions are
simple enough that after some thought anyone can see the light. A further general remark is
that I shied away from constructions that rely heavily on group-representation theory. Rather,
I stuck to more or less elementary linear algebra of finite dimensional vector spaces.
Were I more proficient in representation theory, I might have been able to derive generalized
Dyson boson-fermion mappings rigorously from the (super)-coherent state representation of the
states of fermion systems, in a way that is analogous to the construction in \cite{Do1}. 
The disadvantage of not taking the coherent state route is that
the definition of the Usui operator seems mysterious, as if plucked from thin air. On the other
hand, I find the directness of the pedestrian approach appealing.

The one result in this chapter that I do consider a new contribution of some importance is the
fact that the similarity transform $X$ of Section \ref{SS1.7} is not necessarily well-defined,
but that trouble is avoided if we work inside the collective subspace. One may ask how serious
a problem this definition of $X$ is. Could it be that $X$ is well-defined, but the proof of this
fact eludes us? In this case, one may gain confidence in the similarity transform by constructing
concrete mappings that use $X$, checking that in these specific situations, no problems occur.
On the other hand, if $X$ is sometimes not well-defined, we might hope to demonstrate this by
an example as well. In Chapter Three we will do just that. We will look at a very special case
where the superalgebra is such that the action of $X$ and $X^{-1}$ can be probed carefully.
We will see that, for the wrong number of particles in the system, $X$ is indeed not well-defined.
Therefore, the problem with $X$ is a serious one. If we are particularly pessimistic, we might
be troubled by the fact that the example we consider is so special and ask if the situation is not
worse for more generic examples. Does our example where the similarity transform is well-defined 
represent the exception or the rule? This is an open question.
Because of this state of affairs, I think it is very important to re-emphasize that, if we stay in
the collective subspace, the mapping is always valid and does not depend on whether $X$ is
well-defined.

\section*{Appendix A}
\markright{\bf Appendix A}
\addcontentsline{toc}{section}{Appendix A}
In this appendix I derive the result stated in Section \ref{SS1.3} (Symbols used here have the
meaning assigned to them in that section.): 
For any representation in $\SRep{\lambda}$, let $\opd$ be any operator in $\Od$ and let $\opu_k$ 
be arbitrary elements of $\Ou$, then there exist $\opu_{k,j} \in \Ou$ and complex coefficients $w_j$
such that 
\begin{equation}
\opd\left(\prod_{k=1}^N \opu_k\right)\hw=\sum_j\omega_j\left(\prod_{k=1}^{N_j}\opu_{k,j}\right)
\hw.\nonumber
\end{equation}
The $w_j$ can be chosen such that they are valid for all representations in $\SRep{\lambda}$.

Consider a vector 
\begin{equation}
\left(\prod_{k=1}^N \opu_k\right)\hw,\:\: \opu_k \in \Ou.\label{BF11}
\end{equation}
Since all the elements of $\Ou$ commute or anti-commute, we can, without loss of generality, take 
this vector to be of the form
\begin{equation}
\left(\prod_{j=1}^{N_1} \bar{A}^{k_j}\right)\left(\prod_{l=1}^{N_2}\bar{a}^{\rho_l}\right)\hw.
\label{BF12}
\end{equation}
Two cases must now be distinguished:
\begin{enumerate}
\item $\opd=\bar{a}_\nu$ and
\item $\opd=\bar{A}_i$.
\end{enumerate}
We start with the first case. It is clear that a vector of the form 
\begin{equation}
\bar{a}_\nu\left(\prod_{l=1}^{N_2}\bar{a}^{\rho_l}\right)\hw,\nonumber
\end{equation}
can be written as 
\begin{equation}
\sum_j \tilde{\omega}_j \left(\prod_{l=1}^{N_j}\bar{a}^{\rho_{j,l}}\right)\hw,\nonumber
\end{equation}
by anti-commuting $\bar{a}_\nu$ to the right and using the fact that $\bar{a}_\nu\hw=0$. 
Now we are ready to compute
\begin{eqnarray}
\bar{a}_\nu\left(\prod_{j=1}^{N_1} \bar{A}^{k_j}\right)\left(\prod_{l=1}^{N_2}\bar{a}^{\rho_l}
\right)\hw
&=&\com{\bar{a}_\nu}{\left(\prod_{j=1}^{N_1} \bar{A}^{k_j}\right)\left(\prod_{l=1}^{N_2}
\bar{a}^{\rho_l}\right)}\hw,\nonumber\\
=\com{\bar{a}_\nu}{\left(\prod_{j=1}^{N_1} \bar{A}^{k_j}\right)}\left(\prod_{l=1}^{N_2}
\bar{a}^{\rho_l}\right)\hw
&+&\left(\prod_{j=1}^{N_1} \bar{A}^{k_j}\right)\com{\bar{a}_\nu}{\left(\prod_{l=1}^{N_2}
\bar{a}^{\rho_l}\right)}\hw,\nonumber\\
=\com{\bar{a}_\nu}{\left(\prod_{j=1}^{N_1} \bar{A}^{k_j}\right)}\left(\prod_{l=1}^{N_2}
\bar{a}^{\rho_l}\right)\hw
&+&\left(\prod_{j=1}^{N_1} \bar{A}^{k_j}\right)\bar{a}_\nu\left(\prod_{l=1}^{N_2}\bar{a}^{\rho_l}
\right)\hw.
\label{BF13}
\end{eqnarray}
By anti-commuting the $\bar{a}_\nu$ to the right in the second term, we can cast it in the desired 
form, 
as mentioned above, so that we only have to worry about the first term. Therefore, consider the
commutator that appears in the first term:
\begin{eqnarray}
\com{\bar{a}_\nu}{\left(\prod_{j=1}^{N_1} \bar{A}^{k_j}\right)}
&=&\sum_{j=1}^{N_1}\left(\prod_{l=1}^{j-1}\bar{A}^{k_l}\right)\com{\bar{a}_\nu}{\bar{A}^{k_j}}
\left(\prod_{l=j+1}^{N_1}\bar{A}^{k_l}\right),\nonumber\\
&=&\sum_{j=1}^{N_1}\chi^{k_j}_{\mu\nu}\left(\prod_{l=1\not=j}^{N_1} \bar{A}^{k_l}\right)
\bar{a}^\nu.\label{BF14}
\end{eqnarray}
This result, substituted into the first term of eq. \ref{BF13}, results in a sum over terms of the 
desired form and so completes the proof for the case where $\opd=\bar{a}_\nu$. 
Notice that the complex 
coefficients $\omega_j$ were calculated without any knowledge of the representation, other than 
that
it belongs to $\SRep{\lambda}$, and are therefore equally valid for all representations in 
$\SRep{\lambda}$.

Now, for the case where $\opd=\bar{A}_i$:
\begin{eqnarray}
\bar{A}_i\left(\prod_{j=1}^{N_1} \bar{A}^{k_j}\right)\left(\prod_{l=1}^{N_2}\bar{a}^{\rho_l}
\right)\hw
&=&\com{\bar{A}_i}{\left(\prod_{j=1}^{N_1} \bar{A}^{k_j}\right)\left(\prod_{l=1}^{N_2}
\bar{a}^{\rho_l}\right)}\hw,\nonumber\\
=\com{\bar{A}_i}{\left(\prod_{j=1}^{N_1} \bar{A}^{k_j}\right)}\left(\prod_{l=1}^{N_2}
\bar{a}^{\rho_l}\right)\hw
&+&\left(\prod_{j=1}^{N_1} \bar{A}^{k_j}\right)\com{\bar{A}_i}{\left(\prod_{l=1}^{N_2}
\bar{a}^{\rho_l}\right)}\hw.\label{BF15}
\end{eqnarray}
It is not hard to compute the commutator appearing in the first term. The result is
\begin{eqnarray}
&&\com{\bar{A}_i}{\left(\prod_{j=1}^{N_1} \bar{A}^{k_j}\right)}\nonumber\\
&=&\sum_{j=1}^{N_1}\left\{\left(\prod_{l=1\not=j}^{N_1} \bar{A}^{k_l}\right)\bar{K}_i^{k_j} -
\sum_{l=j+1}^{N_1}\tilde{C}^{k_j k_l}_{i n}\bar{A}^n\left(\prod_{m=1\not=j,l}^{N_1} 
\bar{A}^{k_m}\right)\right\}.
\label{BF16}
\end{eqnarray}
If we substitute eq. \ref{BF16} into eq. \ref{BF15} we see that we still need to deal with the the 
$\bar{K}_i^{k_j}$ operators. The following result takes care of that:
\begin{eqnarray}
&&\bar{K}_i^{k_j}\left(\prod_{l=1}^{N_2}\bar{a}^{\rho_l}\right)\hw\nonumber\\
&=&\com{\bar{K}_i^{k_j}}{\left(\prod_{l=1}^{N_2}\bar{a}^{\rho_l}\right)}\hw+\lambda_i^{k_j}
\left(\prod_{l=1}^{N_2}\bar{a}^{\rho_l}\right)\hw\nonumber\\
&=&-\sum_{l=1}^{N_2}\chi_{\mu\nu}^{k_j}\tilde{\chi}_i^{\mu\rho_l}\bar{a}^\nu
\left(\prod_{m=1\not=l}^{N_2}\bar{a}^{\rho_m}\right)\hw+\lambda_i^{k_j}
\left(\prod_{l=1}^{N_2}\bar{a}^{\rho_l}\right)\hw.\label{BF17}
\end{eqnarray}
If we substitute this into eq. \ref{BF16} we are done with the first term in eq. \ref{BF15}. 
Let us turn our attention to the second term of eq. \ref{BF15}. We have to compute
\begin{equation}
\com{\bar{A}_i}{\left(\prod_{l=1}^{N_2}\bar{a}^{\rho_l}\right)}=-\sum_{l=1}^{N_2}
\tilde{\chi}^{\mu\rho_l}_i
\left(\prod_{m=1}^{l-1}\bar{a}^{\rho_l}\right)\bar{a}_\mu\left(\prod_{m=l+1}^{N_2}
\bar{a}^{\rho_m}\right)\hw.
\label{BF18}
\end{equation}
We can now anti-commute the $\bar{a}_\mu$ to the right, thereby manipulating the second term of 
eq. \ref{BF15}
into the form we want. This completes the proof for the second case where $\opd=\bar{A}_i$. 
Again, note 
that the numerical coefficients appearing in the derivation are calculated without any knowledge of 
the particular representation and can be expressed in terms of the quantities $C$, $\tilde{C}$, 
$\chi$, $\tilde{\chi}$ and $\lambda$ which are common to all representations in $\SRep{\lambda}$.
\section*{Appendix B}
\markright{\bf Appendix B}
\addcontentsline{toc}{section}{Appendix B}
Here I derive the mapping of eq. \ref{BF39}, reproduced here for convenience:
\begin{eqnarray}
\rf{a_\nu}&\longleftarrow&\rbf{a_\nu}=\alpha_\nu,\nonumber\\
\rf{A_j}&\longleftarrow&\rbf{A_j}=B_j,\nonumber\\
\rf{a^\nu}&\longleftarrow&\rbf{a^\nu}=\alpha^\nu+\chi_i^{\mu\nu}B^i\alpha_\mu,\nonumber\\
\rf{A^j}&\longleftarrow&\rbf{A^j}=
B^i\mc{K}_i^j-\D{\frac{1}{2}}C_{ik}^{jl}B^iB^kB_l+\mc{A}^j,\nonumber\\
\rf{K_i^j}&\longleftarrow&\rbf{K_i^j}=\mc{K_i^j}-C_{ik}^{jl}B^kB_l.\label{BF41}
\end{eqnarray}
As extensive use will be made of the BCH formulas, they too are reproduced here for easy reference:
\begin{eqnarray}
\exp(P)Q&=&\left(\sum_{k=0}^\infty\frac{1}{k!}(P,Q)_k\right)\exp(P),\label{BF42}\\
Q\exp(P)&=&\exp(P)\left(\sum_{k=0}^\infty\frac{(-1)^k}{k!}(P,Q)_k\right).\label{BF43}
\end{eqnarray}
where $(P,Q)_0=Q$ and $(P,Q)_{k+1}=\com{P}{(P,Q)_k}$.
As has become customary in the text, we take $\left|\psi\right)$ to be an arbitrary vector in 
$\Hil_{BF}$ and $\left|\phi\right>$ to be an arbitrary vector in $\Hil_F$. We set 
$\left|\phi\right)=T\left|\phi\right>$.

We start with the image for $\rf{a_\nu}$. This is one of the instances mentioned in the text 
where the first step involves guessing the image, and then testing it afterwards. Taking 
$\rbf{a_\nu}=\alpha_\nu$ as our guess, we may write
\begin{eqnarray}
\left(\psi\right|\alpha_\nu\left|\phi\right)
&=&\left(\psi\right|\alpha_\nu T\left|\phi\right>,\nonumber\\
&=&\left\{\left(\psi\right|\alpha_\nu\right\}\otimes\left<0\right|U\uvec{0}{\phi},\nonumber\\
&=&\ufun{\psi}{0}\wt{\alpha_\nu}U\uvec{0}{\phi}.\label{BF44}
\end{eqnarray}
In the last line of eq. \ref{BF44} 
$\wt{\alpha_\nu}U$ can now be calculated, using the BCH formula eq. \ref{BF43}. The result is
\begin{equation}
\wt{\alpha_\nu}U=U\left\{\wt{\alpha_\nu}-\wt{\rf{a_\nu}}\right\}.\label{BF45}
\end{equation}
If we substitute this into $\ref{BF44}$, the $\wt{\alpha_\nu}$ term falls away because 
$\alpha_\nu$ annihilates the vacuum $\left|0\right)$, so that we have
\begin{eqnarray}
\left(\psi\right|\alpha_\nu\left|\phi\right)
&=&\ufun{\psi}{0}U\wt{a_\nu}\uvec{0}{\phi},\nonumber\\
&=&\ufun{\psi}{0}U\left|0\right)\otimes\left\{\rf{a_\nu}\left|\phi\right>\right\}\nonumber\\
&=&\left(\psi\right|T\rf{a_\nu}\left|\phi\right>,\label{BF46}
\end{eqnarray}
which proves that $\rf{a_\nu}\longleftarrow\alpha_\nu$.
The steps for deriving the image of $\rf{A_i}$ are exactly the same, this time 
starting with a guess $B_i$ as the image.

Next we derive the image for $\rf{a^\nu}$. The process in this case runs forward, but the
calculations are more involved. We start by writing
\begin{equation}
\left(\psi\right|\rf{a^\nu}\left|\phi\right)
=\ufun{\psi}{0}U\wt{\rf{a_\nu}}\uvec{0}{\phi}.\label{BF47}
\end{equation}
The BCH formula eq. \ref{BF42} is now employed to calculate
\begin{equation}
U\wt{\rf{a^\nu}}
=\left\{\underbrace{\wt{\rf{a^\nu}}}_1
+\underbrace{\chi_i^{\mu\nu}\wt{B^i}\wt{\rf{a_\mu}}}_2
+\underbrace{\wt{\alpha^\nu}}_3\right\}U.\label{BF48}
\end{equation}
When this is substituted back into eq. \ref{BF47} term (1) drops away because $\rf{a^\nu}$ 
annihilates the functional $\left<0\right|$. We turn our attention to term (2). Since
$\wt{\rf{a_\mu}}$ commutes with $U$ we have
\begin{eqnarray}
\ufun{\psi}{0}\chi_i^{\mu\nu}\wt{B^i}\wt{\rf{a_\mu}}U\uvec{0}{\phi}&=&
\ufun{\psi}{0}\chi_i^{\mu\nu}\wt{B^i}U\wt{\rf{a_\mu}}\uvec{0}{\phi},\nonumber\\
&=&\left(\psi\right|\chi_i^{\mu\nu}B^iT\rf{a_\mu}\left|\phi\right>,\nonumber\\
&=&\left(\psi\right|\chi_i^{\mu\nu}B^i\alpha_\mu T\left|\phi\right>,\label{BF49}
\end{eqnarray}
where we used the image $T\rf{a_\mu}=\alpha_\mu T$ which was proved above. We conclude that
$\rf{a^\nu}\longleftarrow\alpha^\nu+\chi_i^{\mu\nu}B^i\alpha_\mu$.

Now for deriving the image of $\rf{A^j}$, which proves to be the most taxing. 
Again we start by writing 
\begin{equation}
\left(\psi\right|\rf{A^j}\left|\phi\right)
=\ufun{\psi}{0}U\wt{\rf{A^j}}\uvec{0}{\phi}.\label{BF50}
\end{equation}
Now we employ BCH formula eq. \ref{BF42} to find
\begin{eqnarray}
U\wt{\rf{A^j}}&=&\left\{\underbrace{\wt{A^j}}_1
+\underbrace{\wt{B^i}\com{\wt{A_i}}{\wt{A^j}}}_2
-\underbrace{\chi_{\rho\mu}^j\wt{\alpha}^\mu\wt{a}^\rho}_3\right.\nonumber\\
& &\left.-\underbrace{\D{\frac{1}{2}}C_{ik}^{jl}\wt{B^i}\wt{B^k}\wt{A_l}}_4
-\underbrace{\chi_i^{\mu\nu}\chi_{\nu\lambda}^j\wt{B^i}\wt{\alpha^\lambda}\wt{a_\nu}}_5
+\underbrace{\chi_{\mu\nu}^j\wt{\alpha^\mu}\wt{\alpha^\nu}}_6\right\}U.\label{BF51}
\end{eqnarray}
Upon back substitution into eq. \ref{BF50}, terms (1) and (3) disappear because both $\rf{A^j}$
and $\rf{a^\nu}$ annihilate the functional $\left<0\right|$. 
Since $\left<0\right|\com{\rf{A_i}}{\rf{A^j}}=g\delta_i^j\left<0\right|$ we can replace 
term (2) with $g\delta_i^j\wt{B^i}$. In term (4) we can replace $\wt{\rf{A_l}}$ with $\wt{B_l}$ 
by the same argument that allowed us to replace $\rf{a_\mu}$ with $\alpha_\mu$ in eq. \ref{BF49}.
Similarly in term (5) we replace $\wt{\rf{a_\nu}}$ with $\wt{\alpha_\nu}$.
Terms (2) and (5) in their altered form may now be combined and the $\wt{B^i}$ factored out 
to find
\begin{equation}
\wt{B^i}\left\{g\delta_i^j-\chi^j_{\rho\mu}\wt{\alpha^\mu}\wt{\alpha^\rho}\right\}
=\wt{B^i}\wt{{\mathcal K}_i^j}.\label{BF52}
\end{equation}
Term (6) is simply equal to $\wt{{\mathcal A}^j}$.
Having said all this we conclude 
$\rf{A^j}\longleftarrow
B^i\mc{K}_i^j-\D{\frac{1}{2}}C_{ik}^{jl}B^iB^kB_l+\mc{A}^j.$
It is now straight forward to calculate the image of $\rf{K_i^j}$ from the commutator of the 
images of $\rf{A_i}$ and $\rf{A^j}$.
\section*{Appendix C}
\markright{\bf Appendix C}
\addcontentsline{toc}{section}{Appendix C}
In this appendix, I check that the images in eq. \ref{BF39} obey the commutator and 
anti-com\-mu\-tator relations that they should, if they are the generators of a representation of
$\Salg$ in the whole $\Hil_{BF}$. To evaluate most of the commutators and all of the 
anti-commutators involves only the mechanical use of standard manipulations. I therefore 
concentrate on the following three commutators, the calculation of which I deem to be 
less trivial: $\com{\rbf{A^i}}{\rbf{K_j^k}}=C^{ik}_{jl}\rbf{A^l}$, 
$\com{\rbf{A^i}}{\rbf{A^j}}=0$ and  $\com{\rbf{A^i}}{\rbf{a_\nu}}=\chi^i_{\nu\rho}\rbf{a^\rho}$.
For the first two of these, the whole calculation is done explicitly, for the last, the essential
hint necessary to do the calculation is given.
We have to use some properties of the structure constants $C$ and $\chi$ 
that were derived earlier, and are reproduced below for convenience:
\begin{eqnarray}
\chi_{\nu\mu}^i&=&-\chi_{\mu\nu}^i,\label{BF53}\\
C_{ik}^{jl}&=&C_{ki}^{jl}=C_{ik}^{lj}=C_{ki}^{lj},\label{BF54}\\
\chi_i^{\mu\rho}\chi_{\lambda\rho}^j\chi_k^{\lambda\eta}
&=&\D{\frac{1}{2}}C_{ik}^{jl}\chi_l^{\mu\eta},\label{BF55}\\
C_{ik}^{jl}
&=&\frac{1}{g}\chi_i^{\mu\rho}\chi_{\lambda\rho}^j\chi_k^{\lambda\eta}\chi_{\mu\eta}^l.\label{BF56}
\end{eqnarray}
Before we start evaluating commutators, it is necessary to derive an identity that the $C^{jl}_{ik}$
obey, namely that
\begin{equation}
C_{ik}^{jl}C_{ln}^{mp}=C_{nk}^{pl}C_{li}^{mj}.\label{BF57}
\end{equation}
The proof is as follows:
Consider the left-hand side $C_{ik}^{jl}C_{ln}^{mp}$ of the equation and
shuffle around the indices of the respective $C$'s, using eq. \ref{BF54} to find 
$C_{ik}^{jl}C_{ln}^{mp}=C_{ik}^{lj}C_{ln}^{pm}$. Now substitute from eq. \ref{BF56} for
$C_{ln}^{pm}$ to find 
$C_{ik}^{lj}C_{ln}^{pm}=\frac{1}{g}C_{ik}^{lj}
\chi_l^{\alpha\beta}\chi_{\gamma\beta}^p\chi_n^{\gamma\delta}\chi_{\alpha\delta}$. In this 
expression, isolate $C_{ik}^{lj}\chi_l^{\alpha\beta}$ and substitute from eq. \ref{BF55} to find
$C_{ik}^{jl}C_{ln}^{mp}=\frac{2}{g}
\chi_i^{\alpha\rho}\chi_{\lambda\rho}^j\chi_k^{\lambda\beta}
\chi_{\gamma\beta}^p\chi_n^{\gamma\delta}\chi_{\alpha\delta}^m$. Reshuffle the $\chi$'s to find 
\begin{equation}
C_{ik}^{jl}C_{ln}^{mp}=
\left(\underbrace{2\chi_k^{\lambda\beta}\chi_{\gamma\beta}^p\chi_n^{\gamma\delta}}_1\right)
\left(\underbrace{\frac{1}{g}\chi_{\alpha\delta}^m
\chi_i^{\alpha\rho}\chi_{\lambda\rho}^j}_2\right).\label{BF57a}
\end{equation}
From eq. \ref{BF55} it follows that the first factor is equal to 
$C_{kn}^{pl}\chi_l^{\lambda,\delta}$,
which when combined with the second factor, and with the aid of eq. \ref{BF56} gives
$C_{ik}^{jl}C_{ln}^{mp}=C_{nk}^{pl}C_{li}^{mj}$ as promised.

Let us now calculate
\begin{eqnarray}
&&\com{\rbf{A^i}}{\rbf{K_j^k}}\nonumber\\
&=&\com{\mc{A}^i-\D{\frac{1}{2}}C^{ip}_{mn}B^mB^nB_p+B^p\mc{K}_p^i}
{\mc{K}_j^k-C_{jr}^{ks}B^rB_s}\nonumber\\
&=&C_{jl}^{ik}\left(\mc{A}^l+B^p\mc{K}_p^l\right)-\D{\frac{1}{2}}
\left(C_{jv}^{kl}C_{lu}^{iw}+C_{kl}^{ju}C_{lv}^{iw}-C_{jl}^{kw}C_{uv}^{li}\right)B^uB^VB_w.
\label{BF58}
\end{eqnarray}
To get to the last line only the familiar commutator identities between boson operators, 
and the commutator identities associated with $\Salg$ are required. We now apply identity 
eq. \ref{BF57} to the last line of eq. \ref{BF58}. The term $C_{jl}^{kw}C_{uv}^{li}$ is equal to
$C_{ju}^{kl}C_{lv}^{iw}$ so that cancellation between the second and third terms takes place.
Only $C_{jv}^{kl}C_{lu}^{iw}$ remains and this is equal to $C_{jl}^{ik}C_{uv}^{iw}$. Thus we
find
\begin{eqnarray}
\com{\rbf{A^i}}{\rbf{K_j^k}}
&=&C_{jl}^{ik}\left(\mc{A}^l+B^p\mc{K}_p^l-\D{\frac{1}{2}}C_{uv}^{iw}B^uB^vB_w\right),\nonumber\\
&=&C_{jl}^{ik}\rbf{A^l}.\label{BF59}
\end{eqnarray}

Now for the commutator $\com{\rbf{A^i}}{\rbf{A^j}}$: if we substitute the explicit forms of 
$\rbf{A^i}$ and $\rbf{A^j}$ and distribute the commutator over the various terms thus generated
we find that we have to compute six nontrivial commutators:
\begin{eqnarray}
\com{\rbf{A^i}}{\rbf{A^j}}&=&\underbrace{B^p\com{\mc{A}^j}{\mc{K}_p^q}}_1
+\underbrace{\D{\frac{1}{4}}C_{ik}^{jl}C_{pr}^{qs}\com{B^iB^kB_l}{B^pB^rB_s}}_2\nonumber\\
&&-\underbrace{\D{\frac{1}{2}}C_{ik}^{jl}B^iB^k\com{B_l}{B^p}\mc{K}_p^q}_3
+\underbrace{B^i\com{\mc{K}_i^j}{\mc{A}^q}}_4\nonumber\\
&&-\underbrace{\D{\frac{1}{2}}C_{pr}^{qs}B^pB^r\com{B_i}{B^s}\mc{K}_i^j}_5
+\underbrace{B^iB^p\com{\mc{K_i^j}}{\mc{K_p^q}}}_6.\label{BF60}
\end{eqnarray}
It is not hard to check that terms (1) and (4) cancel, and that term (2) is zero all by itself.
Furthermore
\begin{equation}
{\rm term}\:(3)+{\rm term}\:(5)=\frac{1}{2}
\left(C_{ik}^{jl}\mc{K}_l^q-C_{ik}^{qs}\mc{K}_s^j\right)B^iB^k.\label{BF61}
\end{equation}
For calculating term (6), recall that we know
\begin{equation}
\com{\mc{K}_i^j}{\mc{K}_m^n}=C_{im}^{jl}\mc{K}_l^n-C_{il}^{jn}\mc{K}_m^l,\label{BF62}
\end{equation}
(see eq. \ref{BF4}). But we know that $\com{\mc{K}_i^j}{\mc{K}_m^n}=-\com{\mc{K}_m^n}{\mc{K}_i^j}$.
This implies another expression for $\com{\mc{K}_i^j}{\mc{K}_m^n}$, namely
\begin{equation}
\com{\mc{K}_i^j}{\mc{K}_m^n}=C_{mu}^{nj}\mc{K}_i^u-C_{mi}^{nu}\mc{K}_u^j.\label{BF63}
\end{equation}
This means that there are two equivalent ways of writing term (6):
\begin{equation}
{\rm term}\:(6)=\left(C_{ik}^{jl}\mc{K}_l^q-C_{il}^{jq}\mc{K}_k^l\right)B^iB^k,
\end{equation}
and also
\begin{equation}
{\rm term}\:(6)=\left(C_{il}^{jq}\mc{K}_k^l-C_{ik}^{lq}\mc{K}_l^j\right)B^iB^k.
\end{equation}
If we take half of the first expression plus half the second then
\begin{equation}
{\rm term}\:(6)=\frac{1}{2}\left(C_{ik}^{jl}\mc{K}_l^q-C_{ik}^{ql}\mc{K}_l^j\right)B^iB^k.
\end{equation}
Thus term (6) cancels terms (3) and (5), so that $\rbf{A^i}$ and $\rbf{A^j}$ do indeed 
commute.

I do not consider it necessary to derive in such detail the commutator
\begin{equation}
\com{\rbf{A^i}}{\rbf{a_\nu}}=\chi^i_{\nu\rho}\rbf{a^\rho}.
\end{equation}
The result follows easily when one 
uses eq. \ref{BF55} to convert a factor of the form $C\chi$ into one of the form $\chi\chi\chi$.

\chapter[A Toy Application: The $SO(5)$ Model]{A Toy Application:\\ The $SO(5)$ Model}
\label{Ch2}
\section{Introduction}
\markright{\bf Section \thesection: Introduction}
In this chapter we apply the theory of Chapter One to a specific example. By this we mean
that the abstract symbols of the previous chapter are replaced with actual numbers, vectors
and operators. In doing this, our aim is to show that the mapping procedure we advocate is
fairly straight forward and that it is feasible to analyse the mapped system along the same
lines as we are used to doing for any quantum system. We will also be afforded the opportunity to
compare the original system to its mapped counterpart. This will firstly indicate the 
correctness of the theory, as for instance the physical eigenvalues of operators will be seen to
agree in the two systems.
Secondly, we will develop some understanding of what a generalized Dyson mapping does to
a system, by examining how agreement between the two systems emerges.

The model to be studied will be properly introduced shortly. Its most important feature for
our purposes is that it may be expressed in terms of products of the generators of the 
algebra $SO(5)$. Although we consider it a toy model, and exploit it only for its non-trivial but
simple mathematical structure, it has frequently featured in the literature, sometimes with
a view to understand very `realistic' physics. I mention a few instances here to underline the
fact that, despite our opting for the tutorial advantage of simplicity, we are within range of the
coal-face where actual research is currently being done: For the early
history of the model, which involved several different applications in nuclear physics, a
brief discussion and many references may be found in \cite[p. 411]{KM1}. 
More recently, an $SO(5)$ model was analysed to understand the role of proton-neutron pairing and
four-particle correlations in nuclear wave-functions \cite{Dbs1} - \cite{DP2}. 

Outside the
field of nuclear physics, $SO(5)$ models are reported to have some relevance in non-perturbative
Quantum Chromodynamics \cite{CR1}, \cite{CHHR1}. In the field of High Temperature 
Superconductivity, there is also significant interest in $SO(5)$ models \cite{Dbs2}. 
For this last application, a substantial review has recently appeared \cite{DHZ1}. It is 
interesting to note that in all these applications, boson expansion methods also feature. 
Especially in more recent applications, the expansion method of choice is the generalized
Dyson mapping. However, to be fair, I must add that boson expansion methods, generalized Dyson 
mappings and the results obtained with their aid do not occupy a central position in any 
of the fields mentioned.  

This chapter is structured as follows: First we introduce the algebraic building blocks of
the fermion system to be studied. We then do a mapping to a boson-fermion system for the 
collective subspace of the original fermion system. We explicitly include the possibility of
having a single unpaired fermion in the mapped system. This extends the type of analysis that
is done in for instance \cite{Dbs1}, where only the even part of
the collective subspace is mapped, so that there are only bosons in the mapped system. We do
not, however, go beyond the inclusion of a single fermion for the mapped system. This would have 
entailed
working with the similarity transform of Section \ref{SS1.7}. We rather defer doing this until 
Chapter Three, where we will work with an even simpler algebraic structure and the similarity
transform can be implemented more transparently. With the mapping established, a specific 
Hamiltonian is introduced.
This Hamiltonian is interpreted as a nuclear Hamiltonian in for instance 
\cite[p. 411]{KM1} and references therein, which explains the terminology we adopt. After the mapped
counterpart of the Hamiltonian is written down, both the original and the mapped Hamiltonians are
diagonalized. It is shown that following what seems to be the simplest procedure for 
diagonalizatoin in each system, leads to virtually identical calculations in both systems. This
is perhaps not yet fully appreciated in the literature, as is briefly discussed. After 
diagonalization, ghost states are identified, and aspects of their behaviour, as a function
of the parameters of the model, are analysed. 

\section{The Algebra and the Mapping}
\markright{\bf Section \thesection: The Algebra and the Mapping} 
We consider a one-particle Hilbert space for which an orthonormal basis can uniquely be specified
using two quantum numbers $m\in\left\{-j,-j+1,\ldots,j\right\}$ and $\sigma\in\left\{-\frac{1}{2},
\frac{1}{2}\right\}$. Here $j$ is an odd multiple of a half. We define an integer 
$\Omega=\frac{1}{2}+j$. The one-particle Hilbert space therefore has dimension $4\Omega$. 
With these quantum numbers we associate anti-commuting fermion creation and annihilation 
operators $a_{m\sigma}^+$ and $a_{m\sigma}$ that operate in a fermion Fock space with
vacuum $\left|0\right>$. We will say that the operators with $\sigma=\frac{1}{2}$ create and 
annihilate protons, while the $\sigma=-\frac{1}{2}$ operators create and annihilate neutrons.
With the single-fermion operators as building blocks, we construct the following operators:
\begin{eqnarray}
E_{1\pm1}&=&\pm\sum_{m=-j}^j(-)^{j-m}a^+_{m\pm\frac{1}{2}}a^+_{-m\pm\frac{1}{2}},
\label{so1}\\
E_{10}&=&\sum_{m=-j}^j(-)^{j-m}a^+_{m\frac{1}{2}}a^+_{-m-\frac{1}{2}},
\label{so2}\\
E_{01}&=&\frac{1}{\sqrt{2}}\sum_{m=-j}^ja_{m\frac{1}{2}}^+ a_{m-\frac{1}{2}},
\label{so3}
\end{eqnarray}
and their hermitian conjugates $E_{-1\mp1}=E_{1\pm1}^+$, $E_{-10}=E_{10}^+$ and
$E_{0-1}=E_{01}^+$. Furthermore we define two hermitian operators
\begin{eqnarray}
H_1&=&\frac{1}{2}\sum_{\sigma=-\frac{1}{2}}^\frac{1}{2}\sum_{m=-j}^j a_{m\sigma}^+ 
a_{m\sigma}-\Omega\nonumber\\
&=&\frac{1}{2}N-\Omega,\label{so4}
\end{eqnarray}
and
\begin{eqnarray}
H_2&=&\sum_{\sigma=-\frac{1}{2}}^\frac{1}{2}\sum_{m=-j}^j \sigma a_{m\sigma}^+ a_{m\sigma}
\nonumber\\
&=&\frac{1}{2}\left(N_p-N_n\right).\label{so5}
\end{eqnarray}
In these last expressions, $N_p$ counts the number of protons, while $N_n$
counts the number of neutrons and of course $N=N_p+N_n$ counts the total number
of particles. The operator $E_{11}$ creates collective proton-proton pairs; $E_{10}$ creates
collective proton-neutron pairs and $E_{1-1}$ creates collective neutron-neutron pairs. The
operator $E_{01}$ changes one neutron into a proton and $E_{0-1}$ does the reverse, namely it
changes one proton into a neutron. From this the rule emerges that an operator 
$E_{\alpha_1\alpha_2}$ changes the particle number of the system by $2\alpha_1$ and the 
charge by $\alpha_1+\alpha_2$ times the charge of a proton.

The above ten bifermion operators form a basis for a Lie algebra under commutation. In fact,
they represent the so-called Cartan-Killing basis for the algebra $SO(5)$. This sounds
rather intimidating but means nothing more than that a simple recipe exists for writing down
the commutator of any two of the operators. To explain this recipe, we take up the two
indices $\alpha_1$ and $\alpha_2$ that every $E$ operator is subscripted by, into a two
dimensional vector $\alpha=(\alpha_1,\alpha_2)$. There are eight possible vectors $\alpha$, namely
$(1,1)$, $(1,0)$, $(1,-1)$, $(0,1)$, $(0,-1)$, $(-1,1)$, $(-1,0)$ and $(-1,-1)$. These are called
the roots of the algebra. The commutators between the bifermion
operators can now be listed. If we commute an $H$ and an $E$ we get
\begin{equation}
\com{H_k}{E_\alpha}=\alpha_k E_\alpha,\label{so6}
\end{equation}
so that for instance $\com{H_2}{E_{0-1}}=-E_{0-1}$. If we commute $E_\alpha$ and $E_{-\alpha}$
the result is
\begin{equation}
\com{E_\alpha}{E_{-\alpha}}=\sum_{k=1,2}\alpha_k H_k.\label{so7}
\end{equation}
So for instance we have $\com{E_{-11}}{E_{1-1}}=-H_1+H_2$. The operators $H_1$ and $H_2$ commute
with each other. Lastly, if we commute $E_\alpha$ with $E_\beta$ then the result is zero unless
$\alpha+\beta\equiv(\alpha_1+\beta_1,\alpha_2+\beta_2)$ is a root of the algebra. If $\alpha+\beta$
is a root, the commutator is proportional to the $E_{\alpha+\beta}$ operator. This result can be
stated as
\begin{equation}
\com{E_\alpha}{E_\beta}=M_{\alpha,\beta}E_{\alpha+\beta},\label{so8}
\end{equation}
where $M_{\alpha,\beta}$ is a complex number, which is zero in cases where $\alpha+\beta$ is not 
a root of the algebra. So for instance $\com{E_{11}}{E_{0-1}}=-E_{10}$ since $(1+0,1-1)=(1,0)$ is
a root of the algebra, whereas $\com{E_{1-1}}{E_{-1-1}}=0$ because $(1-1,-1-1)=(0,-2)$ is not
a root of the algebra. The constants $M_{\alpha,\beta}$ have the following symmetry properties:
\begin{equation}
M_{\alpha,\beta}=-M_{\beta,\alpha}=-M_{-\alpha,-\beta}=M_{\beta,-\alpha-\beta}=M_{-\alpha-\beta
,\alpha}.\label{so9}
\end{equation}
The value of $M_{\alpha,\beta}$ can be read off from the table:
\begin{eqnarray}
\begin{array}{cc} &\alpha\\\beta&
\begin{array}{|c||r|r|r|r|r|r|r|r|}\hline M_{\alpha,\beta}&(1,1)&(1,0)&(1,-1)&(0,1)&(0,-1)&(-1,1)
&(-1,0)&(-1,-1)\\
\hline\hline
(1,1)&0&0&0&0&1&0&-1&\\\hline
(1,0)&0&0&0&1&-1&1&&-1\\\hline
(1,-1)&0&0&0&-1&0&&1&0\\\hline
(0,1)&0&-1&1&0&&0&-1&1\\\hline
(0,-1)&-1&1&0&&0&-1&1&0\\\hline
(-1,1)&0&-1&&0&1&0&0&0\\\hline
(-1,0)&1&&-1&1&-1&0&0&0\\\hline
(-1,-1)&&1&0&-1&0&0&0&0\\\hline
\end{array}\end{array}&&\nonumber\\
\label{so9a}
\end{eqnarray}
There is an hermitian operator ${\mathcal C}$, called the second Casimir invariant, that commutes
with all the $E$ and $H$ operators that constitute the Cartan-Killing basis. This operator
is given by
\begin{equation}
{\mathcal C}=H_1^2+H_2^2+\acom{E_{11}}{E_{-1-1}}+\acom{E_{10}}{E_{-10}}+\acom{E_{1-1}}{E_{-11}}
+\acom{E_{01}}{E_{0-1}},\label{so10}
\end{equation}
where curly brackets indicate anti-commutators. Using the commutator identities for the generators
of the $SO(5)$ algebra, we will later on manipulate this expression for $\mathcal C$ into other, 
more convenient forms. 

The next step is to decide on a subspace of the fermion Fock space on which our operators are 
defined, to use as the domain of our mapping. As can be seen from the results in \cite{NGD2}, 
expressions involving the similarity
transformation of Section \ref{SS1.7} and the images of the single-fermion operators 
$a^+_{m\sigma}$ and $a_{m\sigma}$ are unpleasant to work with. We therefore stick to the
collective subspace where the complications arising from the similarity transform are avoided. 
This is the space spanned by states of the form
\begin{eqnarray}
(E_{11})^{n_1}(E_{10})^{n_0}(E_{1-1})^{n_{-1}}\left|0\right>,&&\label{so11}\\
(E_{11})^{n_1}(E_{10})^{n_0}(E_{1-1})^{n_{-1}}a_{m\sigma}^+\left|0\right>.&&\label{so12}
\end{eqnarray}
All the $E_\alpha$ and $H_k$ operators leave this space invariant as was discussed in Section
\ref{SS1.8}, and the general formulas of that section are applicable. We will set up our notation
as follows. Let $A^1=\frac{1}{\sqrt{\Omega}}E_{11}$, $A^0=\frac{1}{\sqrt{\Omega}}E_{10}$ and
$A^{-1}=\frac{1}{\sqrt{\Omega}}$, and let $A_k$ be the hermitian conjugate of $A^k$ so that 
for instance $A_{-1}=\frac{1}{\sqrt{\Omega}}E_{-11}$. It is this set of operators $A^k$ and $A_k$
that we use in the definition of the Usui operator in Section \ref{SS1.7}.\footnote{The
factors of $\frac{1}{\sqrt{\Omega}}$ in the definitions of the $A^k$ will be explained after
the mapping is written down.} The structure
constants $C_{jl}^{ik}$ such that $\com{A^i}{\com{A_j}{A^k}}=C_{jl}^{ik}A^l$ can be derived from
commutator identities for elements of the Cartan-Killing basis. For the non-zero structure 
constants we find
\begin{eqnarray}
&C_{11}^{11}=C_{-1-1}^{-1-1}=\frac{2}{\Omega}&\nonumber\\
&C^{10}_{10}=C^{10}_{01}=C^{01}_{10}=C^{01}_{01}=C^{00}_{00}=C^{0-1}_{-10}=C^{-10}_{0-1}=
C^{-10}_{-10}=\frac{1}{\Omega}&\nonumber\\
&C^{1-1}_{00}=C^{00}_{1-1}=C^{00}_{-11}=C^{-11}_{00}=-\frac{1}{\Omega}.&\label{so13}
\end{eqnarray}
Every structure constant not present in these expressions, is zero. With these structure constants
the essential part of the mapping reads
\begin{eqnarray}
A^1&\longleftarrow&B^1+\frac{1}{\Omega}\left\{\frac{1}{2}B^0B^0B_{-1}-B^1\left(N_0+N_1+
\mathcal N_p\right)-B^0\mathcal E_{01}\right\}\nonumber\\
A^0&\longleftarrow&B^0+\frac{1}{\Omega}\bigg\{\frac{1}{2}B^1B^{-1}B_0-B^0\left(N_{-1}+N_1+
\frac{N_0+\mathcal N_p+\mathcal N_n}{2}\right)\nonumber\\
&&\hspace{20mm}+B^{-1}\mathcal E_{01}-B^1\mathcal E_{0-1}\bigg\}\nonumber\\
A^{-1}&\longleftarrow&B^{-1}+\frac{1}{\Omega}\left\{\frac{1}{2}B^0B^0B_1-B^{-1}\left(N_0+N_{-1}+
\mathcal N_n\right)-B^0\mathcal E_{0-1}\right\}\nonumber\\
A_1&\longleftarrow&B_1\nonumber\\
A_0&\longleftarrow&B_0\nonumber\\
A_{-1}&\longleftarrow&B_{-1}.\label{so16}
\end{eqnarray}
The operators $\mathcal N_p$ and $\mathcal N_n$ count the numbers of ideal protons and neutrons
respectively, while $\mathcal E_{01}$ and $\mathcal E_{0-1}$ are the ideal fermion counterparts of
$E_{01}$ and $E_{0-1}$. The operator $N_k$ counts bosons of species $k\in\{-1,0,1\}$. The images of 
the operators $E_{01}$, $E_{0-1}$, $H_1$ and $H_2$ can either
be obtained as the commutators of the appropriate images in eq. \ref{so16} or by using the general
equations of Section \ref{SS1.8}. For $E_{01}$ and $E_{0-1}$ we find
\begin{eqnarray}
E_{01}&\longleftarrow&B^1B_0-B^0B_{-1}+\mathcal E_{01}\nonumber\\
E_{0-1}&\longleftarrow&B^0B_1-B^{-1}B_0+\mathcal E_{0-1}.\label{so17}
\end{eqnarray}
Instead of writing down the images for $H_1$ and $H_2$, I write down the images for the
proton number operator $N_p$ and the neutron number operator $N_n$, that are linear combinations
of $H_1$ and $H_2$ (and the identity operator). The result is
\begin{eqnarray}
N_p&\longleftarrow&2N_1+N_0+\mathcal N_p\nonumber\\
N_n&\longleftarrow&2N_{-1}+N_0+\mathcal N_n.\label{so18}
\end{eqnarray}
The reason for the factor of $\frac{1}{\sqrt{\Omega}}$ in the definition of $A^k$ is this. Say
we had defined $A^k=E_{1k}$ instead, and used this definition in the Usui operator. Then the
images of for instance $E_{11}$ and $E_{-1-1}$ would have been
$\Omega B^1+\frac{1}{2}B^0B^0B_{-1}-B^1\left(N_0+N_1+\mathcal N_p\right)-B^0\mathcal E_{01}$ and
$B_1$. The state $E_{11}\left|0\right>$ would have mapped onto $\Omega B^1\left|0\right)$ and
the functional $\left<0\right|E_{-1-1}$ would have had an image $\left(0\right|B_1$. This 
asymmetry, namely that the images of pair creation operators stretch states by a factor $\Omega$
while functionals undergo no stretching of order $\Omega$ if the images of pair annihilation 
operators are appended to them, is untidy. With the introduction of the factor 
$\frac{1}{\sqrt{\Omega}}$ into the definition of the $A^k$ used to construct the Usui operator,
the images of $E_{11}\left|0\right>$ and $\left<0\right|E_{-1-1}$ respectively become
$\sqrt{\Omega}\left(0\right|B_1$ and $\sqrt{\Omega}B^1\left|0\right)$ so that the 
$\Omega$-scaling of the images of bra's and kets are now symmetric, and the mapping `only as
non-unitary as is absolutely unavoidable'.
\section{The Model Hamiltonian}
\markright{\bf Section \thesection: The Model Hamiltonian}
To get a feel for the mapping, we now consider a toy model in both its original and mapped 
incarnations. As the terminology we have used up till now indicates, the model is ostensibly a
nuclear model. The original Hamiltonian is given by
\begin{equation}
H=\frac{g_1}{\Omega}E_{10}E_{-10}+g_2E_{01}E_{0-1}.\label{so19}
\end{equation}
There is no one-body term $\sum_{m,\sigma}a_{m\sigma}^+ a_{m\sigma}$ so that, were the model 
to describe anything, it would be the residual interaction between nucleons in a single $j$-shell.
Because of the absence of a one-body term, the Hamiltonian leaves the collective subspace 
invariant. We are allowed to restrict our analysis to the collective subspace so that the mapping 
we have just written down is applicable. 

Since there are only two independent energy-scales, the Hamiltonian can be characterised 
by one dimensionless parameter, which we will call $\theta$. If we
are interested in the behaviour of the system at every point in parameter space only up to an 
arbitrary (possibly negative) scaling of the spectrum,
we may learn all we want to know by parametrising
\begin{eqnarray}
g_1&=&G_1\cos\frac{\pi\theta}{2},\nonumber\\
g_2&=&G_2\sin\frac{\pi\theta}{2},\label{so20}
\end{eqnarray}
and considering the system for all $\theta\in\left[-1,1\right)$. (Here $G_1$ and $G_2$ are 
convenient constants whose job is to normalize the two limits where the first or second term in
the Hamiltonian dominates. We chose $G_1$ equal to one divided by the maximum eigenvalue of 
$\frac{1}{\Omega}E_{10}E_{-10}$ and $G_2$ equal to one divided by the maximum eigenvalue of
$E_{01}E_{0-1}$.) The range of $\theta$ 
includes both the limits where the first or second term dominates. Also, both the cases where the 
relative sign of $g_1$ and $g_2$ is positive and negative are included. This is the real reason
why I chose the present model, namely that we can understand the essential behavior of the spectrum
by varying a single parameter $\theta$. However, the Hamiltonian in eq. \ref{so19} is not such a
special Hamiltonian, in the sense that a rather general class of Hamiltonians reduce to it. I
briefly digress to explain this claim. For any nuclear Hamiltonian it is a reasonable 
approximation to insist
that {\em (a)} the total number of protons and neutrons should be conserved separately and 
{\em (b)} the system is invariant under the simultaneous transformation of all protons into 
neutrons and all neutrons into protons. If we add to this the unrealistic assumption that
low energy collective excitations are described by a Hamiltonian that is a quadratic function of
the $SO(5)$ generators of eq. \ref{so1}-eq. \ref{so5}, all allowed Hamiltonians are of the form
\begin{equation}
H'=\frac{g_1'}{\Omega} E_{10}E_{-10}+\frac{g_2'}{\Omega} \left(E_{11} E_{-1-1}+E_{1-1}E_{-11}
\right)+g_3'\acom{E_{01}}{E_{0-1}}
+f(H_1,H_2,\mathcal C),\label{so21}
\end{equation}
where $f$ is an arbitrary function. We can get rid of the $E_{0-1}E_{01}$ term using the commutator
$\com{E_{01}}{E_{0-1}}=H_2$. Also, we can manipulate the expression eq. \ref{so10} 
for the Casimir operator $\mathcal C$ to find
\begin{equation}
E_{11}E_{-1-1}+E_{1-1}E_{-11}=\frac{1}{2}\left(\mathcal C-H_1(H_1-3)-H_2(H_2-1)\right)
-E_{10}E_{-10}-E_{01}E_{0-1}.\label{so22}
\end{equation}
If we substitute this into the expression eq. \ref{so21} for $H'$ we find
\begin{equation}
H'=\frac{g_1''}{\Omega} E_{10}E_{-10}+g_2''E_{01}E_{0-1}
+f'(H_1,H_2,\mathcal C),\label{so23}
\end{equation}
where $g_1''$ and $g_2''$ are numbers (different from the original $g_1'$ and $g_2'$) and 
$f'$ is a function (different from $f$). The assumption that the Hamiltonian describes 
low-energy collective excitations implies that diagonalization takes place in the collective
subspace. Furthermore, since the numbers of protons and neutrons are fixed, diagonalization takes
place in a subspace of the collective subspace with a specific number of protons and 
a specific number of neutrons. This means that in the subspace where diagonalization
takes place, the operators
$H_1$ and $H_2$, that are linear combinations of the proton and neutron number operators, are
proportional to identity. Furthermore, because the proton and neutron numbers are fixed, we
either work in the subspace of the collective subspace that contains states with an odd number of
particles or the subspace that contains an even number of particles. In each of these subspaces,
the Casimir operator is also proportional to the identity operator. Thus the function 
$f'(H_1,H_2,\mathcal C)$
is proportional to the identity operator in the subspace in which we diagonalize $H'$, and we may 
therefore 
neglect it at the cost of shifting all energies by a constant. Along this route we then also
arrive at the Hamiltonian of eq. \ref{so19}.

The image of the Hamiltonian $H$ of eq. \ref{so19} under the mapping we employ is
\begin{eqnarray}
H_{BF}&=&g_1\left\{N_0-\frac{1}{2\Omega}\left(2N_{-1}+2N_1+N_0-1+\mathcal N_p+
\mathcal N_n\right)N_0\right.\nonumber\\
&&\hspace{10mm}\left.+\frac{1}{\Omega}B^1B^{-1}B_0B_0+\frac{1}{\Omega}B^{-1}B_0\mathcal E_{01}
-\frac{1}{\Omega}B^1B_0\mathcal E_{0-1}\right\}\nonumber\\
&&+g_2\left\{\mathcal E_{01} \mathcal E_{0-1}+N_1\left(1+N_0\right)+N_0\left(1+N_{-1}\right)
-B^1B^{-1}B_0B_0-B^0B^0B_1B_{-1}\right.\nonumber\\
&&\hspace{10mm}\left.+B^1B_0\mathcal E_{0-1}-B^{-1}B_0\mathcal E_{01}-B^0B_{-1}\mathcal E_{0-1} 
+B^0B_1\mathcal E_{01}\right\}.\label{so24}
\end{eqnarray}
\section{Diagonalization}
\markright{\bf Section \thesection: Diagonalization}
The obvious thing to do with a Hamiltonian is to diagonalize it. We will consider diagonalization
for both the original and mapped incarnations of the system. We will deal with the case of an odd
number of nucleons, as this is the more interesting case. An even number of nucleons can be
treated with a boson mapping that does not introduce ideal fermions into the picture. This has
already been done in the literature \cite{Dbs1} - \cite{DP2}, \cite{Dbs2} - \cite{GH1}.

Our first task is to characterize the space in which we diagonalize to such an extent that we
can label and calculate the necessary matrix elements of the Hamiltonian. On the fermion side the 
space in which we diagonalize is spanned by states of the form
\begin{equation}
\left(E_{11}\right)^{n_1}\left(E_{10}\right)^{n_0}\left(E_{1-1}\right)^{n_{-1}}
a_{m\sigma}^+\left|0\right>.\label{so25}
\end{equation}
The Hamiltonian preserves the quantum number $m$ so that we can decide on a fixed value for it,
and not explicitly indicate it in our notation. If we consider a system with $n_p$ protons and
$n_n$ neutrons, the following constraints hold:\newpage
\begin{eqnarray}
n_p&=&2n_1+n_0+\frac{1}{2}+\sigma,\nonumber\\
n_n&=&2n_{-1}+n_0+\frac{1}{2}-\sigma,\nonumber
\end{eqnarray}
\begin{equation}
n_p+n_n\;\mbox{is odd}.\label{so26}
\end{equation}
It is not hard to see that these constraints are enough so that, if $n_p$ and $n_n$ are given,
then for each permitted value of $n_0$, the number of proton-neutron pairs, the corresponding 
values of $n_{-1}$, $n_1$ and $\sigma$ are uniquely given by
\begin{eqnarray}
\sigma&=&\left(n_p+n_0\right)\mbox{mod}\,2-\frac{1}{2},\nonumber\\
n_1&=&\frac{1}{2}\left(n_p-n_0-\frac{1}{2}-\sigma\right),\nonumber\\
n_{-1}&=&\frac{1}{2}\left(n_n-n_0-\frac{1}{2}+\sigma\right).\label{so27}
\end{eqnarray}
The constraint that determines $\sigma$ simply says that if $n_p$ and $n_0$ have the same
parity, then the unpaired particle must be a neutron, whereas if they have different parity,
the unpaired particle must be a proton. The permitted values for $n_0$ are 
\begin{equation}
n_0\in\left\{0,1,2,\ldots,\min(n_p,n_n)\right\}.\label{so28}
\end{equation}
We therefore adopt the following notation. The space, in which diagonalization takes place
on the fermion side,
is characterized by specifying $n_p$, $n_n$ and $m$. For these specified values and for 
every permitted $n_0$ we understand $n_1$, $n_{-1}$ and $\sigma$ to take on the values 
given by eq. \ref{so27}, and let $\left|n_0\right>$ represent the state
\begin{equation}
\left|n_0\right>=\frac{1}{\sqrt{\left(n_{-1}\right)!\left(n_0\right)!\left(n_1\right)!}}
\left(A^{-1}\right)^{n_{-1}}\left(A^0\right)^{n_0}\left(A^1\right)^{n_1}a^+_{m\sigma}
\left|0\right>.\label{so29}
\end{equation}
The set $v=\left\{\left|n_0\right>\right\}_{n_0=0}^{\min(n_p,n_n)}$ spans the space in 
which diagonalization takes place on the fermion side. The reason for the factor 
$\frac{1}{\sqrt{\left(n_{-1}\right)!\left(n_0\right)!\left(n_1\right)!}}$ and choosing to
work with the $A$ operators instead of the $E$ operators in the definition of $\left|n_0\right>$
is that, with this choice, the set $v$ tends to an orthonormal basis in the limit 
$\Omega\gg n_p+n_n$. However, as $\Omega$ becomes of the order $\frac{n_p+n_n}{2}$, the set $v$
becomes a linearly dependent set.

For a fermion space spanned by the set $v$, the physical subspace under a boson-fermion mapping
is a subspace of the boson-fermion space spanned by states of the form
\begin{equation}
\left(B^{-1}\right)^{n_{-1}}\left(B^0\right)^{n_0}\left(B^1\right)^{n_1}\alpha_{m\sigma}^\dagger
\left|0\right).\label{so30}
\end{equation}
The same particle number constraints (eq. \ref{so26}) as for the original fermion system apply.
(This follows from the form of the images of the total proton and total neutron 
number operators as given in eq. \ref{so18}.)
Therefore we choose a basis $\tilde{v}=\left\{\left|\widetilde{n_0}\right)
\right\}_{n_0=0}^{\min(n_p,n_n)}$ where
\begin{equation}
\left|\widetilde{n_0}\right)=\frac{1}{\sqrt{\left(n_{-1}\right)!\left(n_0\right)!\left(n_1\right)!}}
\left(B^{-1}\right)^{n_{-1}}\left(B^0\right)^{n_0}\left(B^1\right)^{n_1}\alpha^\dagger_{m\sigma}
\left|0\right),\label{so31}
\end{equation}
and $n_1$, $n_{-1}$ and $\sigma$ are given by eq. \ref{so27}, for the boson-fermion space in
which we diagonalize the mapped system. The tilde in the notation $\left|\widetilde{n_0}\right)$
is meant to emphasize that the state $\left|\widetilde{n_0}\right)$ is not the image under mapping
of the fermion state $\left|n_0\right>$. The set $\tilde{v}$ always forms an orthonormal basis
for the space in which we diagonalize the mapped system.

We can now conclude that ghost states come into play for the mapped system as soon as the
set $v$ that spans the fermion space becomes linearly dependent. 
When the elements of $v$ are linearly independent, $v$ is a basis for the collective subspace. In
this case the dimension of the physical subspace is equal to the number of elements in $v$. The
dimension of the boson-fermion space spanned by $\tilde{v}$, which is the space in which the
diagonalization of the mapped system takes place, is equal to the number of elements in 
$\tilde{v}$. The sets $v$ and $\tilde{v}$ have the same number of elements. Hence the dimension
of the physical subspace is the same as the dimension of the space $span(\tilde{v})$ of which it
is a subspace. This of course implies that the physical subspace is the whole $span(\tilde{v})$. 
As soon as the elements of $v$ become linearly dependent though, the
dimension of the fermion space becomes less than the dimension of the space $span(\tilde{v})$
in which we 
diagonalize the mapped system. The space $span(\tilde{v})$ in which we diagonalize the mapped 
system must then necessarily contain more than the physical subspace.

The spaces in which the operators are to be diagonalized are now characterized and diagonalization
can begin. We first diagonalize on the fermion side. Some authors consider this a more difficult
task than diagonalizing the mapped counterpart in the boson-fermion space. They see the fact that
the set $v$, that spans the fermion space, is not a set of mutually orthogonal states, as 
a stumbling block. It seems the impression exists that an orthonormal basis should first be
found and that because this is a complicated task, the alternative i.e. doing a boson-fermion 
mapping, 
is to be recommended. However, the simple construction below shows that the complications
that arise when one works with a non-orthonormal, possibly linearly dependent set of states 
for the original system, are 
equivalent to the complications associated with the non-hermiticity and the presence of ghost 
states in the mapped system.

To explain the diagonalization procedure on the fermion side, we first look at the following
general situation: Let $\mathcal V$ be a vector space with dimension $M$. Let
$v=\left\{\left|\alpha\right>\right\}_{\alpha=1}^N$ be a set of states that span $\mathcal V$.
In general we allow the possibility that $N>M$ so that the vectors in $v$ are linearly dependent.
None the less, we want to express arbitrary vectors in terms of the elements in this set.
We are asked to diagonalize a fully diagonalizable operator 
$H:\mathcal V\longrightarrow\mathcal V$. To keep things simple, we will take it that $H$ has a 
non-degenerate spectrum. Let $w=\left\{\left|k\right>\right\}_{k=1}^M$ be the eigenvectors of
$H$, such that $H\left|k\right>=\lambda_k\left|k\right>$. We do not know the eigenvalues $\lambda$.
Nor do we know how to express the eigenvectors in $w$ in terms of the elements of $v$, or even
what the dimension $N$ of $\mathcal V$ is. Our aim is to find an expansion for each eigenvector
$\left|k\right>$ in terms of the elements of $v$. 

The set $w$ is then linearly independent
and a basis for $\mathcal V$. We assume that at our disposal is an $N\times N$ matrix $h$ with
entries $h_{\alpha,\beta}$ such that for every element $\left|\alpha\right>$ of $v$ it holds
that $H\left|\alpha\right>=\sum_{\beta=1}^N h_{\alpha,\beta}\left|\beta\right>$. 
Note that when $H$ is hermitian (which we don't have to assume), the matrix $h$ will in general
not be a hermitian matrix, the exception being the case where the elements of $v$ form an 
orthogonal basis for the space $\mathcal V$. To keep the
discussion as simple as possible we assume that the matrix $h$ is fully diagonalizable and has
a non-degenerate spectrum, which, it turns out, is consistent with the assumption that $H$
is fully diagonalisable and 
has a non-degenerate spectrum.\footnote{The simplifying assumptions we make here happen to be
valid for our $SO(5)$ example. They may however be omitted to fit more general applications,
at the cost of making the argument and results more intricate.}

Now, label the left-eigenvectors of $h$ by 
$\gamma'=1,2,\ldots,N$ and let $c_{\gamma',\alpha}$ be the component $\alpha$ of the 
left-eigenvector of $h$ that is labeled by $\gamma'$, i.e.
\begin{equation}
\sum_{\alpha=1}^Nc_{\gamma',\alpha}h_{\alpha,\beta}=\lambda_{\gamma'}'c_{\gamma',\beta},\label{so32}
\end{equation}
where $\lambda'_{\gamma'}$ is the eigenvalue associated with the left-eigenvector labeled by 
$\gamma'$. Now set $\left|\gamma'\right>=\sum_{\alpha=1}^Nc_{\gamma',\alpha}\left|\alpha\right>$.
Then 
\begin{eqnarray}
H\left|\gamma'\right>&=&\sum_{\alpha=1}^Nc_{\gamma',\alpha}H\left|\alpha\right>\nonumber\\
&=&\sum_{\beta=1}^N\sum_{\alpha=1}^Nc_{\gamma',\alpha}h_{\alpha,\beta}\left|\beta\right>
\nonumber\\
&=&\lambda_{\gamma'}'\sum_{\beta=1}^Nc_{\gamma',\beta}\left|\beta\right>\nonumber\\
&=&\lambda_{\gamma'}'\left|\gamma'\right>.\label{so33}
\end{eqnarray}
This means that if $\left|\gamma'\right>$ is not the zero vector then $\lambda'_{\gamma'}$ is
an eigenvalue of $H$ (and hence equal to one of the $\lambda_k$) and $\left|\gamma'\right>$ is
an eigenvector of $H$.

Two questions still need answers. Firstly, does the spectrum of $h$ contain all the eigenvalues
of $H$? Secondly, if $\lambda'_{\gamma'}=\lambda_k$, are we guaranteed that 
$\left|\gamma'\right>$ is the eigenvector of $H$ corresponding to the eigenvalue $\lambda_k$, and
not the zero-vector?

To answer these questions (both affirmative), there is the following argument. Let 
$d_{\alpha,k}$, $\alpha=1,2,\ldots,N$, $k=1,2,\ldots,M$ be the expansion coefficients of the
states $\left|\alpha\right>$ in terms of the eigenbasis states $\left|k\right>$: 
\begin{equation}
\left|\alpha\right>=\sum_{k=1}^Md_{\alpha,k}\left|k\right>.\label{so34}
\end{equation}
Note that, for any fixed value of $k$, there exists at least one index $\alpha$ such that
$d_{\alpha,k}$ is unequal to zero. This follows from the assumption that the set 
$\left\{\left|\alpha\right>\right\}_{\alpha=1}^N$ spans the space $\mathcal V$ for which
$\left\{\left|k\right>\right\}_{k=1}^M$ is a (linearly independent) basis.
Now construct the vector $\sum_{k=1}^M\lambda_kd_{\alpha,k}\left|k\right>$ and compute
\begin{eqnarray}
\sum_{k=1}^M\lambda_kd_{\alpha,k}\left|k\right>&=&\sum_{k=1}^Md_{\alpha,k}H\left|k\right>
\nonumber\\
&=&H\left|\alpha\right>\nonumber\\
&=&\sum_{\beta=1}^Nh_{\alpha,\beta}\left|\beta\right>\nonumber\\
&=&\sum_{k=1}^M\sum_{\beta=1}^Nh_{\alpha,\beta}d_{\beta,k}\left|k\right>\label{so35}.
\end{eqnarray}
By the linear independence of the different vectors $\left|k\right>$, the equality between the
left- and right-hand sides of the above equation must hold term for term, so that 
$\sum_{\beta=1}^Nh_{\alpha,\beta}d_{\beta,k}=\lambda_kd_{\alpha,k}$. This means that all the
eigenvalues $\lambda_k$ of $H$ are indeed contained in the spectrum of $h$. Now, take an
index $k$, associated with eigenvalue $\lambda_k$ of $H$, and let $\gamma'$ be the index that 
refers to
the unique (thanks to the non-degeneracy of the spectra of $H$ and $h$) eigenvalue 
$\lambda_{\gamma'}'$ of $h$ such that $\lambda_{\gamma'}'=\lambda_k$. Now consider the associated
state $\left|\gamma'\right>$, which we wish to show is not the zero-vector:
\begin{eqnarray}
\left|\gamma'\right>&=&\sum_{\alpha=1}^Nc_{\gamma',\alpha}\left|\alpha\right>\nonumber\\
&=&\sum_{l=1}^M\sum_{\alpha=1}^Nc_{\gamma',\alpha}d_{\alpha,l}\left|l\right>.\label{so36}
\end{eqnarray}
We now claim that $\sum_{\alpha=1}^Nc_{\gamma',\alpha}d_{\alpha,l}$ is zero if and only if 
$l\not=k$. To prove this, let $f_\alpha$, $\alpha=1,2,\ldots,N$ be the components of any 
right-eigenvector $f$, excluding $d_k$, of $h$ with eigenvalue $\lambda$. 
$f_\alpha$, $\alpha=1,2,\ldots,N$ could be equal to $d_{\alpha,l}$, 
$\alpha=1,2,\ldots,N$, for some fixed $l\not=k$, but it could also be the components of another 
right-eigenvector of $h$ that we have not yet encountered, since there are only $M$ 
right-eigenvectors denoted with $d$'s, while
there are $N$ right-eigenvectors in total. In any event, the set of all possible choices for $f$
spans an $N-1$ dimensional subspace of ${\bf C}^N$.
The non-degeneracy of the spectrum of $h$ implies that $\lambda\not=\lambda_k$. Now consider
the expression $\sum_{\alpha,\beta=1}^Nc_{\gamma',\alpha}h_{\alpha,\beta}f_\beta$. Using the
fact that $c_{\gamma'}$ is a left-eigenvector of $h$ we conclude that the expression is
equal to $\lambda_k\sum_{\alpha=1}^Nc_{\gamma',\alpha}f_\alpha$. However, using the fact that
$f$ is a right-eigenvector, we conclude that the expression is also equal to 
$\lambda\sum_{\alpha=1}^Nc_{\gamma',\alpha}f_\alpha$. Since $\lambda\not=\lambda_k$ it must hold
$\sum_{\alpha=1}^Nc_{\gamma',\alpha}f_\alpha=0$. This means that the vector with components
$c_{\gamma',\alpha}^*$, $\alpha=1,2,\ldots,N$ is orthogonal to all right-eigenvectors of $h$ that
have eigenvalues different from $\lambda_k$. If $\sum_{\alpha=1}^Nc_{\gamma',\alpha}d_{\alpha,k}$
is zero too, then this vector has to be the zero-vector, which it is not, since $c_{\gamma'}$ is
a left-eigenvector of $h$. Therefore $\sum_{\alpha=1}^Nc_{\gamma',\alpha}d_{\alpha,k}$ is
not equal to zero and $\sum_{\alpha=1}^Nc_{\gamma',\alpha}d_{\alpha,l}$ is zero if and only if 
$l\not=k$. From eq. \ref{so35} then follows that $\left|\gamma'\right>$ is proportional to 
$\left|k\right>$ with the proportionality constant not equal to zero.

The procedure for diagonalizing $H$ if we have the matrix $h$ at our disposal is clear: We
first find the left-eigenvectors of $h$. We label a left-eigenvector by an index $\gamma'$, 
and let its components be denoted $c_{\gamma',\alpha}$, $\alpha=1,2,\ldots,N$. The
eigenvalue associated with it is denoted $\lambda'_{\gamma'}$. From the  components 
$c_{\gamma',\alpha}$ we construct
states $\left|\gamma'\right>=\sum_{\alpha=1}^Nc_{\gamma',\alpha}\left|\alpha\right>$. The
non-zero $\left|\gamma'\right>$ comprise all the eigenvectors of $H$. If $\left|\gamma'\right>$ is
non-zero, and hence an eigenvector of $H$, then it has an eigenvalue $\lambda'_{\gamma'}$.

For our $SO(5)$ fermion Hamiltonian, a valid matrix $h$ of expansion coefficients can be found
by making use of the commutation rules that define the superalgebra of which the $SO(5)$ generators
are the even sector and the single fermion operators are the odd sector, as well as the action of
generators on the vacuum ket. Quite naturally, we choose to 
work with the set $\left\{\left|n_0\right>\right\}_{n_0=0}^{\min(n_p,n_n)}$ defined in 
eq. \ref{so28}. When the Hamiltonian $H$ acts on one of these states, we find the following results:
\begin{itemize}
\item{If $n_0$ has the same parity as $n_p$:
\begin{eqnarray}
\frac{1}{\Omega}E_{10}E_{-10}\left|n_0\right>&=&n_0\left(1-\frac{n_p+n_n-n_0-1}{2\Omega}\right)
\left|n_0\right>\nonumber\\
&&-\frac{1}{2\Omega}\sqrt{\left(n_p-n_0\right)\left(n_0+1\right)}\left|n_0+1\right>\nonumber\\
&&+\frac{1}{2\Omega}\sqrt{\left(n_p-n_0\right)\left(n_n-n_0-1\right)\left(n_0+1\right)
\left(n_0+2\right)}
\left|n_0+2\right>,\nonumber\\
\label{so37}
\end{eqnarray}
\begin{eqnarray}
E_{01}E_{0-1}\left|n_0\right>&=&
-\frac{1}{2}\sqrt{\left(n_p-n_0+2\right)\left(n_n-n_0+1\right)n_0\left(n_0-1\right)}
\left|n_0-2\right>\nonumber\\
&&-\frac{1}{2}\sqrt{\left(n_n-n_0+1\right)n_0}\left|n_0-1\right>\nonumber\\
&&+\left\{\frac{n_0}{2}\left(n_p+n_n-2n_0\right)+\frac{n_p}{2}\right\}\left|n_0\right>\nonumber\\
&&+\frac{1}{2}\sqrt{\left(n_p-n_0\right)\left(n_0+1\right)}\left|n_0+1\right>\nonumber\\
&&-\frac{1}{2}\sqrt{\left(n_p-n_0\right)\left(n_n-n_0-1\right)\left(n_0+1\right)
\left(n_0+2\right)}\left|n_0+2\right>.\nonumber\\
\label{so38}
\end{eqnarray}}
\newpage
\item{If $n_0$ and $n_p$ have different parity:
\begin{eqnarray}
\frac{1}{\Omega}E_{10}E_{-10}\left|n_0\right>&=&n_0\left(1-\frac{n_p+n_n-n_0-1}{2\Omega}\right)
\left|n_0\right>\nonumber\\
&&+\frac{1}{2\Omega}\sqrt{\left(n_n-n_0\right)\left(n_0+1\right)}\left|n_0+1\right>\nonumber\\
&&+\frac{1}{2\Omega}\sqrt{\left(n_p-n_0-1\right)\left(n_n-n_0\right)\left(n_0+1\right)
\left(n_0+2\right)}
\left|n_0+2\right>,\nonumber\\
\label{so39}
\end{eqnarray}
\begin{eqnarray}
E_{01}E_{0-1}\left|n_0\right>&=&
-\frac{1}{2}\sqrt{\left(n_p-n_0+1\right)\left(n_n-n_0+2\right)n_0\left(n_0-1\right)}
\left|n_0-2\right>\nonumber\\
&&+\frac{1}{2}\sqrt{\left(n_p-n_0+1\right)n_0}\left|n_0-1\right>\nonumber\\
&&+\left\{\frac{n_0}{2}\left(n_p+n_n-2n_0\right)+\frac{n_p}{2}\right\}\left|n_0\right>\nonumber\\
&&-\frac{1}{2}\sqrt{\left(n_n-n_0\right)\left(n_0+1\right)}\left|n_0+1\right>\nonumber\\
&&-\frac{1}{2}\sqrt{\left(n_p-n_0-1\right)\left(n_n-n_0\right)\left(n_0+1\right)
\left(n_0+2\right)}\left|n_0+2\right>.\nonumber\\
\label{so40}
\end{eqnarray}}
\end{itemize}
From these expressions the matrix elements $h_{n_0,n_0'}$ such that 
$H\left|n_0\right>=\sum_{n_0'=0}^{\min(n_p,n_n)}h_{n_0,n_0'}\left|n_0'\right>$ holds, may be read 
off. We see that $h$ is a matrix whose only non-zero entries lie on the main diagonal, the two 
diagonals above the main diagonal and the two diagonals below the main diagonal. At this point
we can use a computer to diagonalize the matrix $h$ numerically for given values of $n_p$
and $n_n$. From the left-eigenvectors thus found, the eigenstates of the system can be constructed 
in terms of the states $\left|n_0\right>$, $n_0=0,1,\ldots,\min(n_p,n_n)$ and the unphysical 
eigenvalues of $h$ identified.

However, before we do this, we will discuss the diagonalization of the mapped Hamiltonian
$H_{BF}$. In the mapped system we work with an orthonormal basis of states 
$\left|\widetilde{n_0}\right)$, $n_0=0,1,\ldots,\min(n_p,n_n)$, as defined in eq. \ref{so31} and 
the conventional diagonalization procedure is used. First we calculate the matrix $\tilde{h}$
whose elements are
$\tilde{h}_{m_0,n_0}=\left(\widetilde{n_0}\right|H_{BF}\left|\widetilde{m_0}\right)$. Then, with
every left-eigenvector of $\tilde{h}$ with components $f_{n_0}$ such that 
$\lambda_f f_{n_0}=\sum_{m_0}f_{m_0}\tilde{h}_{m_0,n_0}$ holds, corresponds a right-eigenvector
$\left|f_R\right)=\sum_{n_0}f_{n_0}\left|\widetilde{n_0}\right)$ of $H_{BF}$ with eigenvalue 
$\lambda_f$. Similarly, with every right-eigenvector of $\tilde{h}$ with components $g_{m_0}$ such 
that 
$\lambda_g g_{m_0}=\sum_{n_0}\tilde{h}_{m_0,n_0}g_{n_0}$ holds, corresponds a left-eigenvector
$\left|g_L\right)=\sum_{n_0}g_{n_0}^*\left|\widetilde{n_0}\right)$ of $H_{BF}$ with eigenvalue
$\lambda_g$. (When we say that $\left|g_L\right)$ is a left-eigenvector of $H_{BF}$ we mean 
that $\left(g_L\right|H_{BF}=\lambda_g \left(g_L\right|$.)

As confusion
might arise, I point out the following: In the mapped system the numbers $n_n$ and $n_p$ do
not refer to the numbers of ideal neutrons and protons. They refer, of course, to the numbers of 
neutrons and protons of the original system. 

We now simply use the rules for commuting bosons to
calculate the matrix elements $\tilde{h}_{m_0,n_0}=\left(\widetilde{n_0}\right|H_{BF}\left|
\widetilde{m_0}\right)$ of $H_{BF}$. 
The same distinction as for states of the original system must be made between the case where
$n_0$ and $n_p$ have the same parity (meaning the ideal fermion is a neutron) and the case
where $n_0$ and $n_p$ have different parity (meaning the ideal fermion is a proton). In what
follows below, we again separate the Hamiltonian into two parts. The image of the term
$\frac{1}{\Omega}E_{10}E_{-10}$ of the original Hamiltonian is denoted 
$\left(\frac{1}{\Omega}E_{10}E_{-10}\right)_{BF}$ and similarly the image of $E_{01}E_{0-1}$ by
$\left(E_{01}E_{0-1}\right)_{BF}$. The explicit forms of these images in terms of the boson and 
ideal fermion operators are given by respectively the contents of the first and second set of
curly brackets in the expression for $H_{BF}$\linebreak (eq. \ref{so24}). We find the following:
\begin{itemize}
\item{If $n_0$ has the same parity as $n_p$:
\begin{eqnarray}
\left(\frac{1}{\Omega}E_{10}E_{-10}\right)_{BF}\left|\widetilde{n_0}\right)
&=&\frac{1}{2\Omega}\sqrt{\left(n_p-n_0+2\right)\left(n_n-n_0+1\right)n_0\left(n_0-1\right)}
\left|\widetilde{n_0-2}\right)\nonumber\\
&&+\frac{1}{2\Omega}\sqrt{\left(n_n-n_0+1\right)n_0}\left|\widetilde{n_0-1}\right>\nonumber\\
&&+n_0\left(1-\frac{n_p+n_n-n_0-1}{2\Omega}\right)\left|\widetilde{n_0}\right),\label{so42}
\end{eqnarray}
\begin{eqnarray}
\left(E_{01}E_{0-1}\right)_{BF}\left|\widetilde{n_0}\right)&=&
-\frac{1}{2}\sqrt{\left(n_p-n_0+2\right)\left(n_n-n_0+1\right)n_0\left(n_0-1\right)}
\left|\widetilde{n_0-2}\right)\nonumber\\
&&-\frac{1}{2}\sqrt{\left(n_n-n_0+1\right)n_0}\left|\widetilde{n_0-1}\right)\nonumber\\
&&+\left\{\frac{n_0}{2}\left(n_p+n_n-2n_0\right)+\frac{n_p}{2}\right\}\left|\widetilde{n_0}\right)
\nonumber\\
&&+\frac{1}{2}\sqrt{\left(n_p-n_0\right)\left(n_0+1\right)}\left|\widetilde{n_0+1}\right)\nonumber\\
&&-\frac{1}{2}\sqrt{\left(n_p-n_0\right)\left(n_n-n_0-1\right)\left(n_0+1\right)
\left(n_0+2\right)}\left|\widetilde{n_0+2}\right).\nonumber\\
\label{so43}
\end{eqnarray}}
\item{If $n_0$ and $n_p$ have different parity:
\begin{eqnarray}
\left(\frac{1}{\Omega}E_{10}E_{-10}\right)_{BF}\left|\widetilde{n_0}\right)
&=&\frac{1}{2\Omega}\sqrt{\left(n_p-n_0+1\right)\left(n_n-n_0+2\right)n_0\left(n_0-1\right)}
\left|\widetilde{n_0-2}\right)\nonumber\\
&&-\frac{1}{2\Omega}\sqrt{\left(n_p-n_0+1\right)n_0}\left|\widetilde{n_0-1}\right>\nonumber\\
&&+n_0\left(1-\frac{n_p+n_n-n_0-1}{2\Omega}\right)\left|\widetilde{n_0}\right),\label{so44}
\end{eqnarray}
\newpage
\begin{eqnarray}
\left(E_{01}E_{0-1}\right)_{BF}\left|\widetilde{n_0}\right)&=&
-\frac{1}{2}\sqrt{\left(n_p-n_0+1\right)\left(n_n-n_0+2\right)n_0\left(n_0-1\right)}
\left|\widetilde{n_0-2}\right)\nonumber\\
&&+\frac{1}{2}\sqrt{\left(n_p-n_0+1\right)n_0}\left|\widetilde{n_0-1}\right)\nonumber\\
&&+\left\{\frac{n_0}{2}\left(n_p+n_n-2n_0\right)+\frac{n_p}{2}\right\}\left|\widetilde{n_0}\right)
\nonumber\\
&&-\frac{1}{2}\sqrt{\left(n_n-n_0\right)\left(n_0+1\right)}\left|\widetilde{n_0+1}\right)\nonumber\\
&&-\frac{1}{2}\sqrt{\left(n_p-n_0-1\right)\left(n_n-n_0-\right)\left(n_0+1\right)
\left(n_0+2\right)}\left|\widetilde{n_0+2}\right).\nonumber\\
\label{so45}
\end{eqnarray}}
\end{itemize}
Comparison of eq. \ref{so37} - eq. \ref{so40} where we found the entries of the matrix $h$ 
of the original
fermion system and eq. \ref{so42} - eq. \ref{so45} for the entries of matrix $\tilde{h}$ of 
the mapped system reveal that the matrices are related by $h^\dagger=\tilde{h}$. 
(Note that the entries of the matrices are 
real so that $h^\dagger$ is simply the transpose of $h$.) It is not hard to understand why
this relationship exists between the two matrices. Instead of talking about the present 
$SO(5)$ example, we keep the discussion general. The only restriction is that we work with 
the mapping of Section \ref{SS1.8} for states in the collective subspace, and not with the
mapping for a larger subspace of the fermion Fock space that is arrived at in Section \ref{SS1.7} 
by means of the similarity transformation. We use the notation 
\begin{equation}
\left|n\right>=\prod_{k=1}^M \frac{\left(A^k\right)^{n_k}}{\sqrt{n_k!}}a^\nu\left|0\right>,
\label{so46}
\end{equation}
where $n$ refers to the combined indices $n_1,\ldots,n_M$ and $\nu$
for fermion states in the collective subspace. Here the $A^k$ are the collective fermion pair 
operators. (We confine ourselves to the odd part of the collective subspace. The same argument
works for the even part.) We assume a Hamiltonian that is a function of the $A^k$, $A_k$ and
$K_i^j=\com{A_i}{A^j}$ operators. For the mapped system we use the notation
\begin{equation}
\left|\widetilde{n}\right)=\prod_{k=1}^M \frac{\left(B^k\right)^{n_k}}{\sqrt{n_k!}}\alpha^\nu
\left|0\right).
\label{so47}
\end{equation}
In the original system the matrix $h$ has matrix elements such that $H\left|n\right>=
\sum_{n'}h_{n,n'}\left|n'\right>$. As $H$ is hermitian, this can also be written
\begin{equation}
\left<n\right|H=\sum_{n'}\left<n'\right|h_{n,n'}^*.\label{so48}
\end{equation}
In calculating the numbers $h_{n,n'}^*$ we do not have to make use of any properties particular to
the fermion representation of the superalgebra. Let us suppose that we only use the commutation 
and anticommutation 
relations, and the action of various generators on the vacuum bra, which are properties possessed
by the boson-fermion representation too, for calculating $h_{n,n'}^*$.
(This is what we did in eq. \ref{so37} - eq. \ref{so40}.) In
eq. \ref{so48} we may then replace all operators by their images under the mapping and the
equality will still hold. It is here that the key lies. Thanks to the fact that the image of 
$A_k$ is $B_k$ and the image of $\left<0\right|a_\nu$ is $\left(0\right|\alpha_\nu$, 
the bra $\left<n\right|$ is replaced by $\left(\widetilde{n}\right|$.\footnote{If we venture 
outside the collective subspace and consider the bra $\left<0\right|a_\nu a_\mu$, the argument will 
break down because this bra does not have an image $\left(0\right|\alpha_\nu \alpha_\mu$.}
Therefore we conclude that
\begin{equation}
\left(\widetilde{n}\right|H_{BF}=\sum_{n'}\left(\widetilde{n'}\right|h_{n,n'}^*.\label{so49}
\end{equation}
For the mapped system, the matrix $\tilde{h}$ was uniquely defined such that 
$\tilde{h}_{n,n'}=\left(\widetilde{n'}\right|H_{BF}\Big|\widetilde{n}\Big)$ from 
which then follows that $\tilde{h}_{n,n'}=h_{n',n}^*$. Thus the matrix $h$ that we work with 
on the fermion side is the hermitian conjugate of the matrix $\tilde{h}$ that we work with on
the boson side. 

Bearing this in mind, consider the following claim, quoted from the introduction
of \cite{NGDD1}:
\begin{quotation}
It has often been recognized and stated that boson mapping is not only relevant to discussions
about the relationship between phenomenological boson models and shell model type fermion models,
but also constitute an attractive many-body formalism in its own right $[\ldots]$ The latter 
pronouncement is against the background that one might profit from the use of boson variables
in the description of many-fermion problems through the potentially simpler algebraic structures
associated with bosons. This philosophy has its roots in the very simple observation that bosons
commute to $c$-numbers, while fermion pairs commute to a non-trivial operator. Correspondingly
the construction of an orthonormal basis is in general then simpler for bosons than for fermion 
pairs.
\end{quotation}
In the paper in which this claim is made, boson mappings of models
based on the algebras of $SO(12)$ and $Sp(10)$ are considered. In \cite{Dbs1},
where the model considered is a special case of the $SO(5)$ model of this chapter,
it is again claimed that `the boson
mapping may provide a simple technique to solve the fermion problem' (footnote 3, p. 241).
 There are two aspects that need to be considered
when one claims that it is easier to work with the mapped system than the original. 
(This discussion refers to situations when we work within the collective subspace on the
fermion side.) The first concerns the algebraic manipulations needed to compute $\tilde{h}$ on
the boson side, as compared to those needed to compute $h$ on the fermion side.
The mapped Hamiltonian is a function of a few types of boson operators. The result is
that the commutations that must be performed to calculate the matrix $\tilde{h}$ on the boson side
are easier to keep track of than the commutations needed to calculate $h$ on the fermion side, 
because bosons commute to $c$-numbers. 
None-the-less, the structure of the commutation rules of bi-fermion-type 
algebras are, in my opinion, not vastly more complicated than those on the boson side. If the
remarks made in \cite{Dbs1} and \cite{NGDD1} purely refer to a preference for 
computing the matrix $\tilde{h}$ on the boson side, rather than the matrix $h=\tilde{h}^\dagger$ 
on the fermion side,
which is almost a matter of taste rather than a real simplification, one cannot object.
However, especially in \cite{NGDD1}, a second aspect seems to inform the opinion of the authors.
It seems to be taken for granted that diagonalization on
the fermion side can only be achieved after an orthonormal basis is constructed for the collective
subspace. In holding this view they may have been influenced by the early papers in which the
$SO(5)$ model is introduced and analysed without the aid of a boson mapping. Consider for instance
the following remark from a 1972 paper \cite{CKK1} by Chattopadhyay, Krejs and Klein:
\begin{quotation}
Although not mutually orthogonal, these are complete and form a convenient basis for the 
construction of matrix elements of the generators, and by extension of the Hamiltonian itself,
which can be programmed in the form of recurrence relations. Just as for the problem of small
vibrations, the diagonalization of the Hamiltonian is then carried out in two steps, the
first involving the diagonalization of the metric tensor.
\end{quotation}
The states refered to are simmilar to the ones in eq. \ref{so46} that we worked with to compute
$h$ on the fermion side. The diagonalization of the metric tensor that is mentioned is nothing
but the construction of an orthonormal basis, which precedes diagonalization in \cite{CKK1}.
Because of the less trivial commutators of the building blocks of the states that span
the collective subspace, constructing an orthonormal basis by means of a procedure like the
diagonalization of the metric, is a complicated task. It involves the calculation of the inner
product between fermion states of the form eq. \ref{so46}. 
If this were unavoidable, then indeed, working
with the mapped system would have been preferable by far. However, we have seen that we may
diagonalize the original fermion Hamiltonian from the knowledge of its action on a non-orthogonal, 
possibly linearly dependent set of states. The calculations involved in this procedure mirror
those in the mapped system to the extent that we end up with essentially the same matrix in both
systems. The complications arising from the non-orthogonality of states in the original system
is captured in the mapped system by the non-unitarity of the Dyson mapping. 
The complications associated with linear dependencies between states on the fermion side,
are equivalent to those associated with the identification of ghost states in the mapped system.

Our task is now to diagonalize the matrix $\tilde{h}$ and find the ghost states if any, in its 
spectrum. For general values of the dimensionless parameter $\theta$, this is done numerically.
Later we will return to the specific values $\theta=\pm1$ where $g_1=0$ and 
$\theta=0$ where $g_2=0$ and, in trying to understand the behaviour of the ghost states, derive 
some results without resorting to numerics. 

In Section \ref{SS1.9} we looked at two ways of identifying ghost states. The first procedure
identifies the physical subspace by looking for a subspace of boson-fermion space that is left
invariant by all physical operators. This procedure is implemented as follows for the 
system currently under consideration: First we diagonalize the matrix $\tilde{h}$ for
given values of $g_1$ and $g_2$. This
matrix, and therefore the Hamiltonian $H_{BF}$ as well, happen to have a non-degenerate spectrum, 
so that left and right eigenstates can uniquely be labeled by their eigenvalues. We adopt 
notation that denotes the right-eigenvector of $H_{BF}$ with eigenvalue $\lambda$ by 
$\left|\lambda_R\right)$. The left-eigenvector of $H_{BF}$ with eigenvalue $\lambda$ is denoted
$\left|\lambda_L\right)$. We normalize these states such that 
$\big(\lambda_L\big|\lambda_R'\big)=\delta_{\lambda,\lambda'}$. In terms of the usual basis
elements $\left|\widetilde{n_0}\right)$, the states $\left|\lambda_R\right)$ and the functionals
$\left(\lambda_L\right|$ can be expanded as 
\begin{eqnarray}
\left|\lambda_R\right)&=&\sum_{n_0=0}^{\min(n_p,n_n)}c_{\lambda,n_0}\left|n_0\right),\label{so50}\\
\left(\lambda_L\right|&=&\sum_{n_0=0}^{\min(n_p,n_n)}\left(n_0\right|d_{n_0,\lambda}.\label{so51}
\end{eqnarray}
If we adopt the usual convention that the first index labels rows and the second index columns,
then the matrix $c$ with entries $c_{\lambda,n_0}$ contains on the row labeled by $\lambda$ the
components of the
left-eigenvector associated with eigenvalue $\lambda$ of the matrix $\tilde{h}$. The matrix
$d$ with entries $d_{n_0,\lambda}$ is the inverse of $c$. The column of $d$ labeled by $\lambda$
contains the components of the right-eigenvector with eigenvalue $\lambda$ of $\tilde{h}$.

For the algorithm we are 
implementing, we need another physical operator $Q$ such that the only subspace left invariant
by both $Q$ and $H_{BF}$ is the physical subspace.
An operator that should fit the bill is 
$H_{BF}$ with
a different choice of the parameters $g_1$ and $g_2$.\footnote{In general, a set of operators  
$Q_\alpha$ may be needed. However, numerical results show that we are fortunate here to get
away with only a single operator $Q$. If more operators were required, we could have chosen
the mapped Hamiltonian, with several different choices of the parameters.}
We take $Q$ to be this operator. 
We define $\tilde{q}$ as the matrix with entries $\tilde{q}_{m_0,n_0}=\left(\widetilde{n_0}\right|
Q\left|\widetilde{m_0}\right)$. Now
we let $q$ be the matrix $c\tilde{q}d$, so that the entry $q_{\lambda,\lambda'}$ is given by
\begin{equation}
q_{\lambda,\lambda'}=\sum_{m_0,n_0}c_{\lambda,m_0}\tilde{q}_{m_0,n_0}d_{n_0,\lambda'}.\label{so52}
\end{equation}
It is not difficult to see that the following holds: if we expand $Q\left|\lambda_R\right)$ in
terms of the right-eigenvectors of $H_{BF}$, we find
\begin{equation}
Q\left|\lambda_R\right)=\sum_{\lambda'}q_{\lambda,\lambda'}\left|\lambda_R'\right),\label{so53}
\end{equation}
so that $Q\left|\lambda_R\right)$ contains components along all the eigenbasis vectors 
$\left|\lambda_R'\right)$ with $\lambda'$ such that $q_{\lambda,\lambda'}$ is non-zero. 
It is then by examining the entries in the matrix $q$ in the way prescribed in Section \ref{SS1.9}
that the eigenstates of $H_{BF}$ that span the physical subspace are identified.

Continuing to use the same notation that we used above, the second procedure for identifying
ghost states can also be discussed. This procedure involves the operator $\mathcal R$ whose 
null-space is the orthogonal complement of the physical subspace. The $\mathcal R$ operator
as defined in Section \ref{SS1.9} maps boson-fermion space basis states onto the states in the
collective subspace of the fermion system according to the simple rule\footnote{Recall that
the operator $\mathcal R$ simply replaces each $B^j$ with $A^j$ and $\alpha_\nu^\dagger$ with
$a_\nu^+$ in a state like $\prod_j \left(B^j\right)^{n_j}\alpha_\nu^\dagger\left|0\right)$.}
\begin{equation}
\mathcal R \left|\widetilde{n_0}\right)=\left|n_0\right>.\label{so54}
\end{equation}
Ghost states in the spectrum of $H_{BF}$ are now identified by applying the operator $\mathcal R$
to the left-eigenstates $\left|\lambda_L\right)$ to find
\begin{equation}
\mathcal R\left|\lambda_L\right)=\sum_{n_0=0}^{\min(n_p,n_n)}d_{n_0,\lambda}^*\left|n_0\right>.
\label{so55}
\end{equation}
If and only if this last expression is equal to zero, is the eigenvector associated with eigenvalue
$\lambda$ a ghost state. Note here that this last result also follows from the analysis we did on
the fermion side.
Compare eq. \ref{so55} with eq. \ref{so33} and the subsequent 
discussion, remembering that the matrix $h$ is the hermitian conjugate of $\tilde{h}$.

A problem with this method of identifying ghost states is that we need an economical method for 
determining whether the vector on the right of eq. \ref{so55} is zero. The most economical way 
that I can see, is to take the the inner product of the vector with itself. For this to work,
we must be able to calculate inner products of the form $\left<m_0\right|\left.n_0\right>$,
which we can do, if we can calculate the value of a general expression
\begin{equation}
\left<0\right|a_{m,\sigma}\left(E_{-1-1}\right)^{m_1}\left(E_{-10}\right)^{m_0}\left(E_{-11}
\right)^{m_{-1}}\left(E_{11}\right)^{n_1}\left(E_{10}\right)^{n_0}\left(E_{1-1}\right)^{n_{-1}}
a_{n,\sigma'}^+\left|0\right>\equiv C_{\sigma,m_1,m_0,m_{-1}}^{\sigma',n_1,n_0,n_{-1}}.
\label{so56}
\end{equation}
This expression can be evaluated after commuting all pair annihilation operators past 
all pair creation operators, using the commutator
identities of the $SO(5)$ algebra. However, the number of commutations that need to be made, and
hence the number of terms generated, is staggering. A better strategy is to commute just one
of the pair annihilation operators past the pair creation operators, thereby expressing
the inner-product between two (say) $N$ particle states in terms of inner-products between
$N-2$ particle states. This strategy leads to recursion relations for the inner-products.
These recursion relations are fairly complicated to derive for the $SO(5)$ algebra already, and
will no doubt be even harder to derive for more elaborate algebras. 
Since I do not intend to develop further the method of $\mathcal R$-projection for finding 
ghost states in a general setting, I do not write down these recursion relations here. It
is far more practical to use the preceding method for finding ghost states in general. The
$\mathcal R$-projection method is of limited value in numerical work. It is however 
sometimes useful in an analytical as opposed to a numerical setting, as I hope to 
show in a little while. 

\begin{figure}[h]
\resizebox{16cm}{!}{\includegraphics{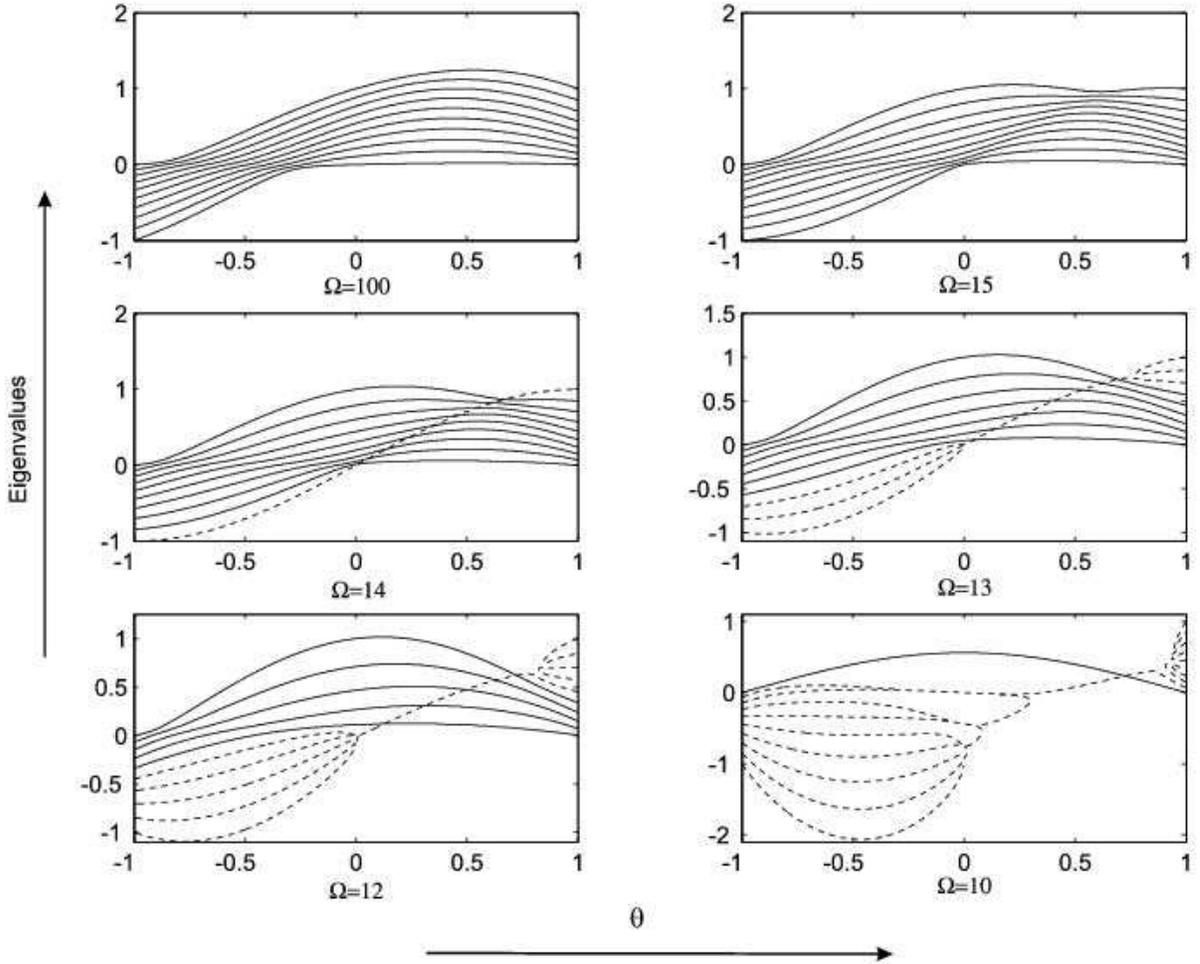}}
\caption{\label{fig1}\em Spectra for 9 protons and 20 neutrons, for various values of $\Omega$.
Real eigenvalues associated with ghost states are included as dashed lines.}
\end{figure}

Everything has now been done except for actually diagonalizing our mapped Hamiltonian, finding
its ghost states and interpreting the results. Therefore we do this now. In Figure \ref{fig1}
we display the results of diagonalization and finding the ghost states. We chose a system
with $9$ protons and $20$ neutrons. When ghost states have real eigenvalues, the eigenvalues are
plotted as dashed lines. When ghost states have complex eigenvalues, these eigenvalues are
no longer displayed. When we analyse the results we display here, the focus is on understanding
the `ins and outs' of the mapping, not on interpreting possible physical phenomena encoded in the
results. 

Let us examine the spectra displayed in Figure \ref{fig1}. At $\theta=-1$ the coupling constant
$g_1$ is switched off and the operator $H_{BF}$ is semi-negative definite. As $\theta$ increases,
the $\left(\frac{1}{\Omega}E_{10}E_{-10}\right)_{BF}$ term is switched on, with a positive coupling
constant, and the eigenvalues generally become less negative. At $\theta=1$ the operator $H_{BF}$
is just the negative of what it was at $\theta=-1$. At the points $\theta=\pm1$ and $\theta=0$
the operator $H_{BF}$ can be diagonalized exactly, using algebraic 
techniques and the fact that these points in parameter space are $SU(2)$ symmetry limits. 
Alternatively, for the $\theta=0$ limit, one can diagonalize $H_{BF}$ by inspecting the matrix 
$\tilde{h}$. This matrix happens to be lower-triangular at the $\theta=0$ point in parameter space,
so that its eigenvalues are the diagonal entries, and eigenvectors are particularly easy to find.
Without going into further details, at $\theta=\pm1$ the spectrum is
\begin{equation}
E_k(\theta=\pm1)=\pm\frac{1}{n_p\left(n_n+1\right)}\left\{k\left(n_n-n_p\right)+k\left(k+1\right)
\right\},\label{so57}
\end{equation}
where $k=0,1,\ldots,n_p$. In this equation, and from here on further, we assume, without any real 
loss of generality, that
there are fewer protons than neutrons in the system. Note that in the $\theta=\pm1$ limits the
mapped Hamiltonian $H_{BF}$ is hermitian and the eigenvalues do not depend on $\Omega$.

In the $\theta=0$ limit the spectrum is
\begin{equation}
E_k(\theta=0)=\frac{k}{n_p}\left(\frac{2\Omega-n_p-n_n+k+1}{2\Omega-n_n+1}\right),\label{so58}
\end{equation}
where $k=0,1,\ldots,n_p$, provided there are no ghost states present. 
This spectrum is almost equidistant for $\Omega\gg n_p+n_n$. These
results, which can be derived independently of all the arguments that lead to the numerically
obtained spectra of Figure \ref{fig1}, fit the numerically obtained spectra perfectly, thus
confirming the correctness of our treatment.

The next thing to consider is the identification of ghost states. We would like to have some
confirmation that our procedure for finding ghost states works. Also, we would like to understand
how ghost states behave in dependence of both $\Omega$ and $\theta$. Looking at decreasing $\Omega$
in Figure \ref{fig1}
we see the first ghost state appearing at $\Omega=14$. The eigenvalue associated with this single
ghost state is real for all $\theta$. Each time we further decrease $\Omega$ by one, two more
ghost states appear. Close to $\theta=\pm1$ these ghost states all have real eigenvalues. As
we get closer to $\theta=0$, the eigenvalues associated with all but one of the ghost states 
become complex. The eigenvalues become complex in pairs. 

An elaborate analysis is possible, in which we derive all these results about the ghost states
from scratch. However, in the interest of not boring readers more than is necessary, and of saving
myself some writing, I take a less strenuous approach, that still paints the above picture emerging
from the numerics, albeit in broader strokes.

Firstly, let us consider the behaviour of the ghost states as we vary $\theta$. The parameter
$\theta$ determines the form of the operator $H_{BF}$ that is diagonalized, but has no
bearing on the mapping itself. The same mapping is used for all $\theta$, and hence the number
of ghost states does not change if we change $\theta$. 
Furthermore, the fact that the operator $H_{BF}$ is hermitian
for $\theta=\pm1$ suggests that there are regions close to $\theta=\pm1$ where the spectrum of 
$H_{BF}$, ghost states included, is real. When there is only one ghost state present, its 
eigenvalue must be real for all $\theta$. This is because the remaining eigenvalues are physical
and hence real, and furthermore the operator $H_{BF}$ has a real trace.

Now we consider the behaviour of the ghost states as we decrease $\Omega$. It is not hard to 
understand why decreasing $\Omega$ might increase the number of ghost states. It has everything
to do with the exclusion principle and the fact that $\Omega$ is proportional to the number of
one-particle states in the fermion space from which we map. If we decrease $\Omega$ while keeping
the number of particles fixed, we decrease also the number of ways to arrange the original fermions
among the one particle states they may occupy. No similar reduction happens in the boson space into
which we map, and thus more of the linearly independent boson states must lie outside the physical
subspace, whose dimension decreases as we decrease $\Omega$. A little more concretely we may recall
that ghost states start entering on the boson side when the set of states 
$v=\left\{\left|n_0\right>\right\}_{n_0=0}^{n_p}$ that spans the collective fermion subspace becomes
linearly dependent. Now, the state $\left|n_0\right>$ is defined as
\begin{equation}
\left|n_0\right>=\frac{1}{\sqrt{\left(n_{-1}\right)!\left(n_0\right)!\left(n_1\right)!}}
\left(A^{-1}\right)^{n_{-1}}\left(A^0\right)^{n_0}\left(A^1\right)^{n_1}a^+_{m\sigma}
\left|0\right>.\label{so59}
\end{equation}
with $n_1$, $n_{-1}$ and $\sigma$ given by eq. \ref{so27} in terms of $n_0$, $n_n$ and $n_p$.
For our present purposes it is necessary to know that of the three numbers $n_1$, $n_0$ and
$n_{-1}$, it is $n_{-1}$ that may take on the largest value (assuming still that $n_n>n_p$).
This value is $n_{-1}=\frac{n_n}{2}$ (for even numbers of neutrons). This happens for $n_0=0$. 
Furthermore, thanks to the exclusion principle we have that
\begin{eqnarray}
\left(A^{\pm1}\right)^n=0&\mbox{if}&n\geq\Omega,\nonumber\\
\left(A^{0}\right)^n=0&\mbox{if}&n\geq2\Omega.\label{so60}
\end{eqnarray}
Thus, we conclude that, when $\Omega$ is smaller than or equal to half the number of neutrons, the
state $\left|n_0=0\right>$ is the zero vector. This means that when $\Omega$ hits the value
$\frac{n_n}{2}$ the set $v$ is certainly linearly dependent and there must be ghost states. For
smaller $\Omega$ even more of the vectors $\left|n_0\right>$ are zero and more linear dependencies
should occur. However, the value $\Omega=\frac{n_n}{2}$ is an under-estimation for the threshold
value of $\Omega$ where ghost states enter the spectrum. This means that the set $v$ becomes
linearly dependent before such time as the state $\left|n_0=0\right>$ is the zero vector. To
get the correct value of $\Omega$ below which ghost states enter, we dust off the $\mathcal R$
operator. From our numerical results we see that in the $\theta=1$ limit, ghost states start
entering the spectrum from above. If $n$ ghost states are present, they are associated with
the $n$ largest eigenvalues. (This is one of the results we would have derived from scratch if
we did the more complete analysis.) So when there is just one ghost state present, it is the
eigenvector of $H_{BF}(\theta=1)$ with eigenvalue $E_{k=n_0}=1$. It is not too hard to check
that this eigenvector is given by 
\begin{equation}
\left|\theta=1,E=1\right)=\left(E_{01}\right)^{n_p}_{BF}\left(B^{-1}\right)^{\frac{n_p+n_n-1}{2}}
\left|\frac{1}{2}\right),\label{so61}
\end{equation}
where the state $\left|\frac{1}{2}\right)$ contains one ideal neutron and 
$\left(E_{01}\right)_{BF}$ is the image under the mapping of the operator $E_{01}$. To verify
that $\left|\theta=1,E=1\right)$ is indeed the eigenstate we are after, one 
uses the fact that $\left(E_{01}\right)_{BF}$ and $\left(E_{0-1}\right)_{BF}$ are $SU(2)$ 
ladder operators, and that $\left(E_{0-1}\right)_{BF}$ annihilates the state  
$\left(B^{-1}\right)^{\frac{n_p+n_n-1}{2}}\left|\frac{1}{2}\right)$. We do not distinguish between
left- and right-eigenvectors because $H_{BF}$ is hermitian when $\theta=1$. If and only if 
the state 
$\left|\theta=1,E=1\right)$ is not a physical state, then, when we act on it with the operator
$\mathcal R$, we must get the zero vector. Acting with $\mathcal R$ we find
\begin{equation}
\mathcal R\left|\theta=1,E=1\right)=\left(E_{01}\right)^{n_p}\left(A^{-1}\right)^{\frac{n_p+n_n-1}
{2}}\left|\frac{1}{2}\right>,\label{so62}
\end{equation}
where $\left|\frac{1}{2}\right>$ contains one real neutron. Looking at the exponent of the $A^{-1}$
operator, we see that ghost states start to appear for values of $\Omega$ smaller than or
equal to $\frac{n_n+n_p-1}{2}$. For a system of $9$ protons and $20$ neutrons, this means that
we expect ghost states for $\Omega\leq14$, a prediction that is confirmed by the numerical results.

\section{Conclusion}
\markright{\bf Section \thesection: Conclusion}
On this note, we conclude our analysis of the $SO(5)$ toy model, that, in outline, included
the following. A mapping of a collective fermion representation of the $SO(5)$ algebra together
with single fermion creation and annihilation operators was derived. The mapping is valid in
the whole collective subspace of the fermion system, as opposed to the even part of the collective
subspace only. This means that a single unpaired fermion may be present in the
mapped system. A specific Hamiltonian, typical of algebraic nuclear models, was introduced.
The diagonalization of the original and mapped Hamiltonians was considered. Ghost states were
identified by finding the subspace of boson-fermion space that is left invariant by all physical
operators. Numerical results were compared to some analytical results for corroboration and also
to understand what goes on `under the hood' of the mapping. 
A result that has emerged from our analysis is that, when we work in the collective subspace, there
is a closer similarity between the original and mapped systems than was perhaps previously 
realised. 

What is still lacking is an example of how a generalized Dyson mapping may be a useful
technique to understand the behaviour of certain many-body systems. We try to address this in
the next chapter.

\chapter{An Application: The Richardson Model for Superconductivity}
\label{Ch3}
\section{Introduction and Motivation}
\markright{\bf Section \thesection: Introduction and Motivation}
In this chapter we apply the techniques of Chapter One to a system with dynamics given
by the Richardson Hamiltonian. This Hamiltonian describes fermions interacting in such a
way that Cooper pairs are formed. In the sixties there appeared a set of papers in which R.W.
Richardson used the model to understand pairing correlations in nuclear wave-functions. See
for instance \cite{RS1}. (In the bibliography of \cite{vDR1} a complete list can be found of the 
papers in which Richardson analyses the model that bears his name. I do not reproduce the list in
the bibliography of this thesis, for fear of creating the impression that I have personally 
read all Richardson's papers.) What is notable about Richardson's analysis, is that it involves
the exact solution of the model. 
In the context of superconductivity in 
metals, the Richardson Hamiltonian is also thought relevant.
Although the famous BCS paper \cite{BCS1} deduces the 
phenomenon of superconductivity from a more general Hamiltonian, for some calculations the
problem is simplified by rather using the Richardson Hamiltonian. (In the notation of \cite{BCS1},
the interaction matrix elements $V_{\rm \bf k,k'}$ are set equal to their average $V$, for 
${\rm \bf |k|,|k'|}\leq \hbar\omega$, and zero elsewhere. This simplifying assumption is made
in several important calculations in the BCS paper.) 
Lately, authors have argued that there are fundamental reasons for considering the Richardson
Hamiltonian to describe very accurately the dynamics of electrons in mesoscopic metallic grains
\cite{KAA1}. A thorough review of experimental and theoretical developments in this field can be
found in \cite{vDR1}. 
Furthermore, the Richardson Hamiltonian has a simple but non-trivial algebraic structure:
The superalgebra, the generators of which are the building blocks of the Hamiltonian, has an even 
sector that is nothing but the $SU(2)$ algebra. 

On the one hand, applying a generalized Dyson mapping to a system described by the Richardson 
Hamiltonian, 
shows that the mapping can survive in the `real world', particularly when the mapped Hamiltonian
cannot be expressed as a function of only the operators in the even sector of the superalgebra.
On the other hand, aspects
of the generalized Dyson mapping, such as the validity of the similarity transform of Section 
\ref{SS1.7}, for which our previous $SO(5)$ example was too intricate, can be dealt with. 

There is another reason to do a mapping of the Richardson model: 
The Richardson model is exactly solvable, in the sense that expressions for the eigenstates
and eigenvalues of the system can be derived in terms of a set of parameters that are 
solutions to a system of equations known as Richardson's equations. To find these solutions one 
has to resort to highly nontrivial numerics. This can only be done if the system does not contain
too many particles.
Also, in the limit of essentially infinitely many particles, the model simplifies to a form that
is tractable \cite{Rich1}.
If neither of these circumstances apply, or if one wants analytical rather than numerical results,
there are two ways to 
proceed. The first more purely mathematical option is to focus on Richardson's equations and
develop analytical techniques to solve these equations to a given degree of accuracy, with the
minimum numerical work required. This was done with success by Yuzbashyan and co-workers
in \cite{YBA1}. The other option is to forget about the Richardson equations and simply
apply familiar quantum many-body approximation techniques to the Richardson Hamiltonian. The
BCS variational treatment of superconducting systems in the thermodynamic limit (where the
number of conduction electrons tend to infinity) is an example of such an approach \cite{BCS1}. 
The approximation method we will consider is conventional perturbation theory in both its 
time-independent and time-dependent forms. This approximation method can potentially yield 
information about systems with more than a few but less than infinitely many particles.
The quantity with respect to which we expand is the inverse of a coupling constant, from there
the term `Strong Coupling Expansion' in the title of \cite{YBA1}. 
This quantity, which we will denote $\frac{1}{G}$ is not very small in a superconducting metal, but 
the eigenstates and energies of the system are analytical functions of it, so that in principal
we can get accurate answers by going to high enough orders in the expansion. Obtaining high orders
however involves huge effort. In practice we therefore settle for truncating expansions at low 
orders. At best (if the properties of the system at small $\frac{1}{G}$ are not too vastly different
from the properties at moderate $\frac{1}{G}$), this will give us a qualitative understanding of
realistic systems. This does not make the second option inferior to the option
chosen by Yuzbashyan and co-workers per se. In fact their method just gives an order for order 
solution 
in $\frac{1}{G}$ to the Richardson equations. Hence their method gives results equivalent to the
usual time independent perturbation theory for the eigenstates and eigenenergies of the system.
It is just the way in which they obtain these results that is specifically designed around the
Richardson equations.
Their method also runs into the difficulty that high order corrections
are required for realistic systems. If we settle on the conventional way of doing a perturbation 
expansion, rather
than the method of Yuzbashyan, we encounter another problem though. Because of electron-electron
correlations in the eigenstates of the zero'th order problem, the calculation of matrix-elements
of the perturbation with respect to these states, which is the stock in trade of any perturbation
expansion, is a formidable task. Here the boson-fermion mapping that was developed in Chapter
One might come in handy. It turns out that this mapping transforms away the bothersome
correlations, so that for the zero'th order mapped system, the eigenstates look like
free particle states. Matrix elements of the perturbation are then more simply calculated in the 
mapped system.

We will do both a time-independent expansion to obtain approximate eigenstates and eigenvalues and
a time-dependent expansion to obtain approximate transition matrix elements. We compare the
time-independent expansion for the mapped system with the results published in \cite{YBA1}. 
We find exact agreement. The method in \cite{YBA1} might however be more straight
forward to implement when higher order corrections than we calculated, are required. The value
of this component of our work then, is that it shows that it is feasible to do calculations using
a mapped system obtained through the boson-fermion mapping, rather than that it allows for
calculating properties of the system with unprecedented ease. 
More speculatively, the intuitive
physical picture behind our method is enticing. Firstly, the Dyson mapping high-lights a few 
previously hidden features of the model in the strong pairing limit. Secondly, our results are 
obtained by means of a 
transformation that, crudely speaking, replaces Cooper pairs with bosons. Thinking about a 
superconducting system in this way might shed some light on the current hot topic of 
the Bose-Einstein condensation to BCS cross-over in cold atomic gases \cite{FS1}. Or it might not. 
This last speculation does not receive a systematic analysis in the present text.

Since it seems that a time-independent perturbation expansion will not yield anything that
is not within the scope of previous analyses, we will also do a time-dependent expansion
for the ground-state to ground-state transition amplitude, in the strong coupling limit.
In doing this expansion I am not so much concerned with what useful information or measurable
effect is exposed by the result. Rather, I illustrate that this quantity is easily calculated
with the aid of the mapped system, whereas I suspect that it involves quite a complicated procedure
to calculate it by any other means.

Some of the matter of this chapter grew from work I presented in the form of a talk at the
second international workshop on pseudo-hermitian Hamiltonians in Quantum Physics held
in Prague (14-16 June 2004). A paper \cite{SG1} based on the talk is to appear in the proceedings
of the workshop. In that paper some of the simpler results from time-independent perturbation 
theory are derived. The paper constitutes a brief (and almost representative) sampling of the work 
contained in this chapter.
\section{The Model}
\markright{\bf Section \thesection: The Model}
\label{SS2.1}
We start by introducing the Richardson model, including the fermion Fock space on which it is 
defined. Therefore, let us consider a one-particle Hilbert space for which the orthonormal states 
\begin{equation}
\fvec{\nu},\hspace{5mm}\nu=-j,-j+1,\ldots,j-1,j,\label{BCS1}
\end{equation}
form a basis, where $j$ is an odd multiple of a half. We also define a convenient integer
\begin{equation}
\Omega=j+\frac{1}{2}.\label{BCS2}
\end{equation}
The single discrete quantum number $\nu$ 
combines an electron spin index and a Bloch wave number in such a way that the state
$(-1)^{j-\nu}\fvec{-\nu}$ is the time-reversed state of the state $\fvec{\nu}$.
Note that $j-\nu$ is an integer in the range $0,1,2,\ldots,2\Omega-1$. This convention,
which can be seen as a particular choice for the phase in the definition of 
the normalization constant for time-reversed states, is different
from the one that is usually employed, but leaves measurable quantities unchanged.
The reason for the unconventional convention is that it makes 
formulas for the structure constants of the emerging superalgebra simpler. 

With each state $\fvec{\nu}$ is then associated anti-commuting fermion creation and 
annihilation operators $a_\nu^+$ and $a_\nu$ that act in the fermion Fock space $\Hil_F$, 
and which are each other's hermitian conjugates.
In terms of these operators, and without the summation convention that we often employed in 
previous chapters, the Richardson Hamiltonian is
\begin{equation}
H=\sum_{\nu=-j}^{j}\epsilon_\nu a_\nu^+ a_\nu 
-\frac{G}{4\Omega} \sum_{\mu,\nu=-j}^{j}(-1)^{2j-\mu-\nu}a_\mu^+ a_{-\mu}^+ 
a_{-\nu} a_\nu.\label{BCS4}
\end{equation}
The $\epsilon_\nu$ are the one-particle energies. We will assume time-reversal symmetry so
that $\epsilon_{-\nu}=\epsilon_\nu$.
The one-body term accounts as well as possible for everything that
electrons do inside a normal metal while the interaction term contains only the mechanism for
Cooper pairing. This interaction involves the scattering
of time reversed pairs of conduction electrons among each other and is governed by a single 
coupling constant $G$ with the dimension of energy. 
In order to apply a boson-fermion mapping to the system, we must identify a superalgebra such that
the Hamiltonian can conveniently be expressed in terms of the superalgebra's generators.
With this in mind, we define the collective fermion pair operator 
\begin{equation}
S^+=\frac{1}{2\sqrt{\Omega}}\sum_{\nu=-j}^{j}(-1)^{j-\nu}a_\nu^+ 
a_{-\nu}^+,\label{BCS5}
\end{equation}
that contains all time-reversed pairs with amplitudes differing only by a phase. Conjugated to
it is a pair annihilation operator
\begin{equation}
S=\frac{1}{2\sqrt{\Omega}}\sum_{\nu=-j}^{j}(-1)^{j-\nu}a_{-\nu} a_\nu.\label{BCS6}
\end{equation}
Using summation convention, we may also write
\begin{eqnarray}
S^+&=&\chi_{\mu\nu}a^\mu a^\nu,\nonumber\\
S&=&\chi^{\mu\nu}a_\nu a_\mu,\label{BCS7}
\end{eqnarray}
where $\chi_{\mu\nu}=\frac{1}{2\sqrt{\Omega}}(-1)^{j-\nu}\delta_{\mu,-\nu}$ and 
$\chi^{\mu\nu}=\left(\chi_{\mu\nu}\right)^*=\chi_{\mu\nu}$ and $a^\mu=\left(a^+_\mu\right)$.
The operators $S$, $S^+$ and $K=\com{S}{S^+}$ are
the generators of a representation of the $SU(2)$ algebra. These, together with the single-
fermion operators $a_\nu$ and $a_\nu^+$ form the generators of a single plus bi-fermion 
superalgebra of the type discussed in Chapter One. The important commutators are:
\begin{eqnarray}
\com{S}{S^+}&=&1-\sum_{\nu=-j}^j\frac{a_\nu^+ a_\nu}{\Omega},\label{BCS8}\\
\com{S^+}{\com{S}{S^+}}&=&\frac{2}{\Omega}S^+,\label{BCS9}\\
\com{a_\nu}{S^+}&=&\frac{1}{\sqrt{\Omega}}(-1)^{j-\nu}a_{-\nu}^+,\label{BCS10}\\
\com{a_\nu^+}{S}&=&\frac{-1}{\sqrt{\Omega}}(-1)^{j-\nu}a_{-\nu}.\label{BCS11}
\end{eqnarray}
The Richardson Hamiltonian can be expressed in terms of these generators as
\begin{equation}
H=\sum_{\nu=-j}^{j}\epsilon_\nu a_\nu^+ a_\nu-G S^+ S.\label{BCS12}
\end{equation}
Since the operator $S^+ S$ is semi-positive definite, we take $G$ to be positive so that the
interaction $-G S^+ S$ lowers the energy. If we neglect the one-body term and solve the
pairing-only problem $H=-GS^+ S$ for $2N$ fermions, we find an unnormalized ground 
state\footnote{Throughout this section of the text, unless otherwise indicated, eigenstates will
not be normalized.}
$\left(S^+\right)^N\left|0\right>$. This means that the operator $S^+$ creates 
Cooper pairs, while $S$ annihilates them.
\section{Properties of the Eigenstates of the Richardson Hamiltonian}
\markright{\bf Section \thesection: Properties of the Eigenstates}
\label{SS2.2}
Before we map the Richardson model onto a boson-fermion model, I briefly state a few facts 
about its eigenstates. The discussion is not intended to be exhaustive, nor is it essential for
performing a boson-fermion mapping. (The boson-fermion mapping and subsequent analysis of the 
system do not require anything more than the fermion Hamiltonian to be known.) The results
I mention here should rather be seen as a reference point which we will refer back to when
we interpret the behavior and properties of the mapped equivalent of the Richardson model.

The first thing to consider is the so-called blocking effect, which effectively reduces the problem
of finding the eigenstates of the Hamiltonian in the whole fermion Fock space $\Hil_F$ to the
problem of finding the eigenstates in a certain subspace of $\Hil_F$. 

To explain the blocking effect, the following terminology is introduced. 
We refer to the one-particle states $\left|-\nu\right>$
and $\left|\nu\right>$ together as the one-particle level $\nu$. The fermion Fock space 
${\mathcal H}_F$ is
a many-particle space in which each one-particle level $\nu$ can be unoccupied, singly occupied or 
doubly 
occupied. Electrons in singly occupied levels do not take part in the dynamics in any other
way than that, due to the exclusion principal, they prohibit scattering into the levels that they 
occupy.  
For instance, consider the two-electron state $a^+_\mu a^+_\nu \left|0\right>$ with
$\left|\mu\right|\not=\left|\nu\right|$. Since the operator $S$ can only remove particles from 
doubly occupied levels, the pairing interaction does not notice the two unpaired electrons, and
we have $S^+ S a_\mu^+ a_\nu^+ \left|0\right>=0$. Hence the Richardson
Hamiltonian endows the state $a_\mu^+ a_\nu^+\left|0\right>$ with the same dynamics
as if the electrons were not interacting:
$H a_\mu^+ a_\nu^+\left|0\right>
=\left(\epsilon_\mu+\epsilon_\nu\right)a_\mu^+ a_\nu^+\left|0\right>$.

To make these ideas more precise and extend them to an arbitrary number of particles we introduce
a specific kind of basis for $M$-particle fermion Fock space. For each element of the basis we
are talking about, we specify two sets of indices. The first set $B$ is called the set of blocked
or singly occupied levels. Each set $B$ contains no more than $M$ of the indices 
$\nu=-j,\ldots,j$, with the condition imposed that if the index $\nu$ is an element of $B$, 
then the index $-\nu$ is not an element of $B$. Let $b$ be the number of elements of $B$, then
we insist that $b$ has the same parity as $M$. The reason is that we want the indices in
$B$ to denote one-particle levels in ${\mathcal H}_F$ that contain only one electron. The 
second set $D$ specifies the doubly occupied levels. It contains $\frac{M-b}{2}$ indices
from the set 
$U=\left\{\frac{1}{2},\ldots,j\right\}\setminus\left\{\left|\nu\right|:\nu\in B\right\}$
of levels that are not singly occupied. These are called the unblocked levels.
Each basis element is then uniquely specified as
\begin{equation}
\left|B,D\right>=\left(\prod_{\mu\in B} a_\mu^+\right)\left(\prod_{\nu\in D}
a_\nu^+ a_{-\nu}^+\right)\left|0\right>.\label{blo1}
\end{equation}
The collection of all possible $B$ partitions the basis into disjoint subsets $P_B$ 
that are defined as follows.
For a given set $B$ of singly occupied levels, let $P_B$ be the set of all states
$\left|B,D\right>$ such that the set $D$ of doubly occupied levels is consistent with the 
choice of singly occupied levels in $B$. The Richardson Hamiltonian leaves each subspace 
$span\left(P_B\right)$ invariant. In fact, a little algebra reveals that
\begin{equation}
H\left|B,D\right>=\left(\sum_{\nu\in B}\epsilon_\nu\right)\left|B,D\right>
+\left(\prod_{\mu\in B} a_\mu^+\right)H'\left(\prod_{\nu\in D}a_\nu^+ 
a_{-\nu}^+\right)\left|0\right>,\label{blo2}
\end{equation}
where the operator $H'$ is the Richardson Hamiltonian redefined on the levels that are not 
singly occupied, i.e. $H'=\sum_{\mu:\left|\mu\right|\in U}\epsilon_\mu a_\mu^+ a_\mu
+G'S'^+ S'$, with $S'^+=\frac{1}{\sqrt{\Omega-b}}\sum_{\nu\in U} (-)^{j-\nu}
a_\nu^+ a_{-\nu}^+$ and $G'=G\frac{\Omega-b}{\Omega}$. This means that to
diagonalize $H$ on the whole fermion Fock space ${\mathcal H}_F$, we only have to diagonalize
the various $H'$ (each of which looks just like $H$ but with fewer levels and a different coupling 
constant) for the electrons distributed among doubly occupied levels, and afterwards add the 
electrons in singly occupied levels and their energies.

The problem considered in this text then, is diagonalizing $H$ for an arbitrary set of 
$\Omega$ one-particle levels and 
with an
arbitrary coupling constant $G$, for an arbitrary even number of particles ($2N$), 
but with no singly 
occupied levels. The space in which $H$ is diagonalized has dimension $\Omega \choose N$.
If we solve this problem, the problem of diagonalizing $H$ with
singly occupied levels taken into account, is solved without further complication.
 
The second thing I now discuss is a solution of sorts to the problem of diagonalizing the
Richardson Hamiltonian when there are no singly-occupied levels. We consider a system of $2N$
electrons. Let $E_k$, $k=1,2,\ldots,N$ be any set of complex numbers that satisfy the so-called
Richardson equations
\begin{equation}
\frac{-G}{\Omega}=\sum_{\nu=\frac{1}{2}}^j\frac{1}{E_k-2\epsilon_\nu}
+\sum_{l=1\not=k}^N\frac{1}{E_l-E_k},\hspace{2mm}k=1,2,\ldots,N.\label{ric1}
\end{equation}
There is a correspondence between solutions of eq. \ref{ric1} and the eigenstates of $H$ with
no singly occupied levels. With every eigenstate $\left|E\right>$ of
$H$ with no singly occupied levels, such that $H\left|E\right>=E\left|E\right>$, is associated
a solution $E_1,\ldots,E_N$ of eq. \ref{ric1} (and vice versa, with every solution of eq. \ref{ric1}
is associated an eigenstate of $H$ with no singly occupied levels), such that the energy of the
state is given by
\begin{equation}
E=\sum_{k=1}^N E_k,\label{ric2}
\end{equation}
and the eigenstate has the form
\begin{equation}
\left|E\right>=\prod_{k=1}^N\left(\sum_{\nu=\frac{1}{2}}^j\frac{(-)^{j-\nu}a_\nu^+ a^+_{-\nu}}
{2\epsilon_\nu-E_k}\right)\left|0\right>.\label{ric3}
\end{equation}
To derive this is not hard but will not be done here, for two reasons. Firstly, a good derivation
is available in the literature \cite[Appendix B]{vDR1}. Secondly, we will shortly encounter an 
operator that is very
similar to the Richardson Hamiltonian. Its eigenstates have properties very much like those
of the Richardson Hamiltonian and the derivation of these properties is accomplished using the
same strategy as with the Richardson Hamiltonian. Since this new operator has not yet featured
in the literature, I rather do the derivation for its eigenstates.

To claim that the above result completely solves the problem of finding the eigenstates 
of the Richardson Hamiltonian with no 
singly occupied levels, is overly optimistic. The reason is that
we still need to find all solutions of eq. \ref{ric1}. This is equivalent to finding all 
simultaneous roots (of which there are $\Omega \choose N$, one for each dimension of the space 
in which $H$ is diagonalized), of $N$ polynomials of $N$ variables and of degree $N\Omega$. 
Since $2N$ is the number of electrons and $\Omega$ the number of one-particle levels, these two
numbers are both large in the truly many-body regime. This makes the solving of eq. \ref{ric1} a
very daunting task and one, it hardly needs to be said, that can only be treated numerically.

The last matter to be discussed is the behavior of the solutions to the Richardson equations as
the coupling constant $G$ in front of the pairing term becomes large when compared to the
one-particle energies $\epsilon_\nu$. Specifically, we mean
\begin{equation}
\frac{\sum_{\nu=\frac{1}{2}}^j\epsilon_\nu}{G}\longrightarrow 0,\label{ric4}
\end{equation}
so that, if we take the one-particle energies as our yard stick, the left-hand side of 
eq. \ref{ric1} tends to zero. As this happens, some of its solutions have
components that decrease without bound while there are others all of whose components stay
finite \cite{YBA1}. To be precise, of the $\Omega \choose N$ solutions $(E_1,\ldots,E_N)$ to 
eq. \ref{ric1}
there are only ${\Omega \choose N}\times\frac{\Omega-2N+1}{\Omega-N+1}$ whose components $E_k$, 
$k=1,\ldots,N$ all remain finite as $G\longrightarrow \infty$. If we set $\frac{\Omega}{G}=0$
on the left-hand side of eq. \ref{ric1}, only these solutions survive, the others having vanished 
off into negative
infinity. Since the energy of an eigenstate is given by the sum of the components of its
corresponding solution to the Richardson equations, this means that even if the
coupling constant $G$ is infinite, there are still 
${\Omega\choose N}\times\frac{\Omega-2N+1}{\Omega-N+1}$ eigenstates with finite energy. 
This concludes the discussion on the properties of the eigenstates of the original Richardson
Hamiltonian. 
 
\section{The Mapping}
\markright{\bf Section \thesection: The Mapping}
\label{SS2.3}
We now construct the boson-fermion mapping from fermion Fock space $\Hil_F$, or a subspace thereof, 
into the 
boson-fermion Fock space $\Hil_{BF}$, that results from the superalgebra introduced above. First
we must decide which subspace of $\Hil_F$ (which may turn out to be the whole $\Hil_F$) we want
to map into $\Hil_{BF}$. In Section \ref{SS1.8} we saw that we could avoid growing a few gray
hairs when we restricted ourselves to a so-called collective subspace of $\Hil_F$. In the present
case the (even sector of the) collective subspace is spanned by the single state
\begin{equation}
\left(S^+\right)^n\fvec{0}.\label{BCS13}
\end{equation}
The Hamiltonian $H$ only leaves the collective subspace invariant when the one-particle energies
$\epsilon_\nu$ are all set equal to each other.

It is evident that we cannot map the collective subspace alone. Therefore
we set follical concerns aside and map a larger subspace of $\Hil_F$ into $\Hil_{BF}$. 
The program of 
Section \ref{SS1.7} must be followed, where states in $\Hil_F$ that contain more than one fermion
that is not in a collective pair are mapped onto states in $\Hil_{BF}$ using the compounded 
operator $X^{-1}\circ T$. Here $T$ is the mapping of $\Hil_F$ into $\Hil_{BF}$ that results
from the Usui operator constructed in Section \ref{SS1.6} and $X$ is the similarity
transformation defined in Section \ref{SS1.7}. (Without the similarity transformation, the
association between Cooper pairs and bosons is not as close as we would like.)

However, the term `defined' is used loosely, since it was mentioned in Section \ref{SS1.7} that
the defining equation for $X$, eq. \ref{BF79}, is problematic. Our first task then is to determine
if and when $X$ is well-defined for the superalgebra we are currently working with.

We set up our notation as follows. With the boson-fermion Fock space $\Hil_{BF}$ are associated
boson creation and annihilation operators $B^\dagger$ and $B$, as well as fermion creation and
annihilation operators $\alpha_\nu^\dagger$ and $\alpha_\nu$. We refer to these fermion operators as
ideal fermion operators. They commute with the boson operators. We also define ideal collective
fermion pair operators 
\begin{equation}
\Sc=\frac{1}{2\sqrt{\Omega}}\sum_{\nu=-j}^{j}(-1)^{j-\nu}\alpha_\nu^\dagger 
\alpha_{-\nu}^\dagger,\label{BCS15}
\end{equation}
and $\Sa=\left(\Sc\right)^\dagger$
which are the counterparts in $\Hil_{BF}$ of the collective fermion operators $S^+$ and 
$S$ that are defined in $\Hil_{F}$. 
$\Nb=B^\dagger B$ is the boson number operator and $\nf=\sum_\nu\alpha_\nu^\dagger\alpha_\nu$ is 
the ideal fermion number operator. 

The problematic defining equation (eq. \ref{BF79}) for the similarity transformation $X$ now reads
\begin{equation}
X\bvec{\psi,n_F,n_B,\lambda}=\sum_{k=0}^\infty\left\{\left(C_F-\lambda\right)^{-1}\Sc B\right\}^k
\bvec{\psi,n_F,n_B,\lambda}.\label{BCS17}
\end{equation}
In this equation $C_F$ is hermitian and given by $C_F=\Sc\Sa$. The state 
$\bvec{\psi,n_F,n_B,\lambda}\in\Hil_{BF}$ is an eigenstate of $C_F$ with eigenvalue $\lambda$. Since
$C_F$ commutes with the ideal fermion number operator and the boson number operator, 
we choose $\bvec{\psi,n_F,n_b,\lambda}$ to have
a well defined number of ideal fermions, namely $n_F$, and a well-defined number of bosons namely
$n_B$. $\psi$ refers to any other simultaneous
eigenvalue(s) that are required to label the state uniquely. Since $C_F$ has an eigenvalue 
$\lambda$, $C_F-\lambda$ has an eigenvalue zero, and hence it is not invertible on the whole
$\Hil_{BF}$. It is however invertible on the orthogonal complement in $\Hil_{BF}$ of the 
$\lambda$-eigenspace of $C_F$. This space is denoted by $\Lambda_\perp$. The operator 
$\left(C_F-\lambda\right)^{-1}$ refers to this inverse of $C_F-\lambda$ in $\Lambda_\perp$. 
Hence it is an operator that is only defined on the space $\Lambda_\perp$. 

The implication for the attempted definition in eq. \ref{BCS17} can now be tackled. We have to
establish whether every term 
\begin{equation}
\bvec{T_k}=\left\{\left(C_F-\lambda\right)^{-1}\Sc B\right\}^k\bvec{\psi,n_F,n_B,\lambda},
\label{BCS18}
\end{equation}
makes sense. 
It turns out that particle number considerations are important in this argument. To avoid 
muddling further down the line, we recall a few facts and define some notation. Since the Richardson
Hamiltonian preserves the number of fermions, it makes sense to choose a subspace 
$V_F\in\Hil_F$ that is left invariant by the fermion number operator 
$\sum_{\nu=-j}^j a_\nu^+ a_\nu$, as
the domain of the mapping $X^{-1}\circ T$. One can then find a basis for $V_F$ in which
each basis state has a well-defined number of fermions. Let $\fvec{\phi,M}$ be such a state,
containing $M$ fermions.\footnote{In deriving the mapping, we work as generally as possible and
don't make the assumption that we work with an even number of particles. 
This will allow
us to derive images for the single-fermion operators.} A property of the operator $T$, as defined 
in Section \ref{SS1.6} is
that $T\fvec{\phi,M}$ is an eigenstate of $2\Nb+\nf$ with eigenvalue $M$. We therefore
choose a basis for $\Hil_{BF}$, the elements of which have both a well-defined number of
ideal fermions $n_F$ {\em and} a well-defined number of bosons $n_B$. If we then demand that 
states in 
$V_F$ have no more than $M_{max}$ fermions, we are guaranteed that the range of $T$, namely
$T(V_F)$, is contained in the subspace of $\Hil_{BF}$ spanned by states for which
twice the number of bosons plus the number of ideal fermions is no more than $M_{max}$:
\begin{equation}
n_F+2n_B\leq M_{max}.\label{BCS22}
\end{equation}

Now we investigate when eq. \ref{BCS18} is acceptable. Because of the commutator 
$\com{\Sa}{\Sc}=1-\frac{\nf}{\Omega}$, the operator identity
\begin{equation}
C_F\Sc B=\Sc B\left(C_F+1-\frac{\nf}{\Omega}\right),\label{BCS19}
\end{equation}
holds. From this it follows that
\begin{equation}
C_F\Sc B\bvec{\psi,n_F,n_B,\lambda}=\left(\lambda+1-\frac{n_F}{\Omega}\right)\Sc 
B\bvec{\psi,n_F,n_B,\lambda}.
\label{BCS20}
\end{equation}
$\Sc B\bvec{\psi,n_F,n_B,\lambda}$ is an eigenvector of $C_F$ with eigenvalue 
$\lambda+1-\frac{n_F}{\Omega}$. In parenthesis we note that for $\bvec{T_1}$ to be well-defined,
$\Sc B\bvec{\psi,n_F,n_B,\lambda}$ must lie in $\Lambda_\perp$. 
From eq. \ref{BCS20} follows that, because
$C_F$ is hermitian, $\Sc B\bvec{\psi,n_F,n_B,\lambda}$ lies in $\Lambda_\perp$ if and only if 
$n_F\not=\Omega$. This condition is enforced by demanding that the states in the subspace 
$V_F$ of $\Hil_F$ contain less than $\Omega$ fermions. That is to
say, we consider a conduction band at less than half filling.
(States in $\Hil_F$ can in principal contain up to $2\Omega$ fermions.) 
When the conduction band is less than half-filled, it follows that $n_F\leq\Omega$ and
$n_B<\frac{\Omega-n_F}{2}$. Since we get the state $\bvec{T_k}$, if it is well-defined, 
from the state $\bvec{\psi,n_F,n_B,\lambda}$ by removing $k$ bosons (and adding $2k$ 
ideal fermions), $\bvec{T_k}$ is the zero vector in $\Hil_{BF}$ for $k>\frac{\Omega-n_F}{2}$.

Continuing in the same vane as in eq. \ref{BCS20} we can use the identity (\ref{BCS19}) to
compute
\begin{equation}
C_F\left(\Sc B\right)^k\bvec{\psi,n_F,n_B,\lambda}=\left[\lambda+\frac{k}{\Omega}(\Omega-n_F-k+1)
\right]\left(\Sc B\right)^k\bvec{\psi,n_F,n_B,\lambda},\label{BCS21}
\end{equation}
or in other words
$\left(\Sc B\right)^k\bvec{\psi,n_F,n_B,\lambda}$ is an eigenvector of $C_F$ 
(when it is not the zero vector), with eigenvalue 
$\lambda+\frac{k}{\Omega}(\Omega-n_F-k+1)$  which, because of the restrictions on $n_F$ and 
$k$ that we have derived above, is different from $\lambda$. This means that for 
$k\leq\frac{\Omega-n_F}{2}$, the state $\left(\Sc B\right)^k\bvec{\psi,n_F,n_B,\lambda}$ lies in
$\Lambda_\perp$. For 
$k>\frac{\Omega-n_F}{2}$, the state $\left(\Sc B\right)^k\bvec{\psi,n_F,n_B,\lambda}$ is 
the zero vector, which also lies in $\Lambda_\perp$. It follows that we may always act
on $\left(\Sc B\right)^k\bvec{\psi,n_F,n_B,\lambda}$ with $\left(C_F-\lambda\right)^{-1}$, the 
result being
\begin{equation}
\left(C_F-\lambda\right)^{-1}\left(\Sc B\right)^k\bvec{\psi,n_F,n_B,\lambda}=
\frac{\Omega}{k(\Omega-n_F-k+1)}\left(\Sc B\right)^k\bvec{\psi,n_F,n_B,\lambda},\label{BCS23}
\end{equation}
where 
$\left(\Sc B\right)^k\bvec{\psi,n_F,n_B,\lambda}$ becomes the zero vector before the 
eigenvalue becomes singular.
The result in eq. \ref{BCS23} is used to compute
\begin{eqnarray}
\bvec{T_k}&=&\underbrace{\left(C_F-\lambda\right)^{-1}\Sc B\ldots
\left(C_F-\lambda\right)^{-1}\Sc B}_{k\hspace{1.5mm}times}\bvec{\psi,n_F,n_B,\lambda}\nonumber\\
&=&\left\{\prod_{l=1}^k\frac{\Omega}{l(\Omega-n_F-l+1)}\right\}\left(\Sc B\right)^k
\bvec{\psi,n_F,n_B,\lambda}\nonumber\\
&=&\frac{(\Omega-n_F-k)!}{(\Omega-n_F)!k!}\Omega^k\left(\Sc B\right)^k\bvec{\psi,n_F,n_B,\lambda}.
\label{BCS24}
\end{eqnarray}
In the second line we simply allowed the $\left(C_F-\lambda\right)^{-1}$ operators to act 
from right to left on whatever is in front of them. Since the `whatever' is either an eigenvector 
of $\left(C_F-\lambda\right)^{-1}$ or the zero vector,
the $\left(C_F-\lambda\right)^{-1}$ operators were replaced by numbers that were
then collected in the curly brackets.
This last calculation confirms that the similarity transform $X$ is well-defined, provided
the conduction band is less than half-filled. Notice that eq. \ref{BCS24} is independent of 
the value of $\lambda$ and $n_B$ so that linear combinations of states for which these differ
may be taken, with eq. \ref{BCS24} remaining valid. This is done implicitly by taking up these 
labels in the generic label $\psi$. Thus, for a general state $\bvec{\psi,n_F}\in\Hil_{BF}$ 
containing $n_F$ ideal fermions, the action of $X$ can be stated as
\begin{equation}
X\bvec{\psi,n_F}
=\sum_{k=0}^\infty\frac{(\Omega-n_F-k)!}{(\Omega-n_F)!k!}\Omega^k\left(\Sc B\right)^k
\bvec{\psi,n_F}.\label{BCS25}
\end{equation}
Note how fortunate we are to have this last result: The source of our good luck is essentially 
the fact that the commutator $\com{C_F}{\Sc B}$ is a function of the ideal fermion number
operator only (eq. \ref{BCS19}). For more complicated algebras, similar result do not hold.

Having confirmed that the similarity transformation $X$ that was introduced in Section \ref{SS1.7}
does indeed exist for the superalgebra that we deal with here, we can write down the images of the
generators of the superalgebra under the mapping. Using the general formulas of Section \ref{SS1.7}
adapted to the current instance, we get for the images of $S$ and $S^+$:
\begin{eqnarray}
S&\longleftarrow&B,\label{BCS26}\\
S^+&\longleftarrow&B^\dagger\left(1-\frac{\nf+\Nb}{\Omega}\right).\label{BCS27}
\end{eqnarray}
Because of the explicit expression for $X$, that is eq. \ref{BCS24}, we can derive explicit 
expressions for the images of the single fermion operators $a_\nu$ and $a_\nu^+$ 
(which is more
than we were able to do in Section \ref{SS1.7}). First to be considered is the single fermion
annihilation operator $a_\nu$. Its image under $X^{-1}\circ T$ is an operator 
$\left(a_\nu\right)_{X,BF}:\Hil_{BF}\rightarrow\Hil_{BF}$ defined by\footnote{Strictly speaking,
because we work with a conduction band at less than half filling, the domain of this operator
is not the whole $\Hil_{BF}$, but this technicality is innocuous and may be forgotten.}
\begin{equation}
X\left(a_\nu\right)_{X,BF}=\alpha_\nu X.\label{BCS28}
\end{equation}
In order for us to use the explicit expression (eq. \ref{BCS24}) for $X$, we have to let the 
operators $X\left(a_\nu\right)_{X,BF}$ and $\alpha_\nu X$ act on a state 
$\bvec{\psi,n_F}\in\Hil_{BF}$ that contains $n_F$ ideal fermions. I therefore start to manipulate
the state $\alpha_\nu X\bvec{\psi,n_F}$:
\begin{eqnarray}
\alpha_\nu X\bvec{\psi,n_F}&=&\sum_{l=0}^\infty\frac{(\Omega-n_F-l)!}{(\Omega-n_F)!l!}
\Omega^l \alpha_\nu\left(\Sc B\right)^l\bvec{\psi,n_F}\nonumber\\
&=&\underbrace{\sum_{l=0}^\infty\frac{(\Omega-n_F-l)!}{(\Omega-n_F)!l!}
\Omega^l\com{\alpha_\nu}{\left(\Sc B\right)^l}\bvec{\psi,n_F}}_1\nonumber\\
&&\hspace{1cm}+\underbrace{\sum_{l=0}^\infty\frac{(\Omega-n_F-l)!}{(\Omega-n_F)!l!}
\Omega^l\left(\Sc B\right)^l\alpha_\nu\bvec{\psi,n_F}}_2,\label{BCS29}
\end{eqnarray}
We now treat these two terms separately. Term (1) is simplified if we use the commutator
$\com{\alpha_\nu}{\Sc}=\frac{(-1)^{j-\nu}}{\sqrt{\Omega}}\alpha_{-\nu}^\dagger$ to compute
$\com{\alpha_\nu}{(\Sc)^l}=\frac{(-1)^{j-\nu}}{\sqrt{\Omega}}l(\Sc)^{l-1}\alpha_{-\nu}^\dagger$. 
Then
\begin{eqnarray}
&&\sum_{l=0}^\infty\frac{(\Omega-n_F-l)!}{(\Omega-n_F)!l!}
\Omega^l\com{\alpha_\nu}{\left(\Sc B\right)^l}\bvec{\psi,n_F}\nonumber\\
&=&\sum_{k=0}^\infty\frac{(\Omega-n_F-k-1)!}{(\Omega-n_F)!k!}
\Omega^{k+1}\left(\Sc B\right)^k\frac{(-1)^{j-\nu}}{\sqrt{\Omega}}B\alpha_{-\nu}^\dagger
\bvec{\psi,n_F}
\nonumber\\
&=&\sum_{k=0}^\infty\frac{(\Omega-(n_F+1)-k)!}{(\Omega-(n_F-1))!k!}
\Omega^k\left(\Sc B\right)^k\left\{\frac{(-1)^{j-\nu}B\alpha_{-\nu}^\dagger}{\sqrt{\Omega}
\left(1-\frac{n_F}{\Omega}\right)}\bvec{\psi,n_F}\right\}.\label{BCS30}
\end{eqnarray}
As the vector in curly brackets contains $n_F+1$ ideal fermions, the operator that it is
preceded by represents the action of the similarity transform $X$. The final form of term (1) is 
then
\begin{equation}
\sum_{l=0}^\infty\frac{(\Omega-n_F-l)!}{(\Omega-n_F)!l!}
\Omega^l\com{\alpha_\nu}{\left(\Sc B\right)^l}\bvec{\psi,n_F}=
X\left\{\frac{(-1)^{j-\nu}B\alpha_{-\nu}^\dagger}{\sqrt{\Omega}
\left(1-\frac{n_F}{\Omega}\right)}\bvec{\psi,n_F}\right\}.\label{BCS31}
\end{equation}
\newpage
Term (2) is split up as follows
\begin{eqnarray}
&&\sum_{l=0}^\infty\frac{(\Omega-n_F-l)!}{(\Omega-n_F)!l!}
\Omega^l\left(\Sc B\right)^l\alpha_\nu\bvec{\psi,n_F}\nonumber\\
&=&\alpha_\nu\bvec{\psi_,n_F}+\sum_{l=1}^\infty\frac{(\Omega-n_F-l)!}{(\Omega-n_F)!l!}
\Omega^l\left(\Sc B\right)^l\alpha_\nu\bvec{\psi,n_F},\nonumber\\
&=&\alpha_\nu\bvec{\psi_,n_F}+\sum_{k=0}^\infty\frac{(\Omega-(n_F+1)-k)!}{(\Omega-(n_F+1))!k!}
\Omega^{k+1}\left(\Sc B\right)^k\left\{\frac{\Sc B\alpha_\nu}{(\Omega-n_F)(k+1)}\bvec{\psi,n_F}
\right\}.\nonumber\\
\label{BCS32}
\end{eqnarray}
We can write $\alpha_\nu\bvec{\psi,n_F}$ as the first term of $X\alpha_\nu\bvec{\psi,n_F}$, i.e.
\begin{eqnarray}
\alpha_\nu\bvec{\psi,n_F}&=&X\alpha_\nu\bvec{\psi,n_F}-
\sum_{l=1}^\infty\frac{(\Omega-(n_F-1)-l)!}{(\Omega-(n_F-1))!l!}
\Omega^l\left(\Sc B\right)^l\alpha_\nu\bvec{\psi,n_F}\nonumber\\
&=&X\alpha_\nu\bvec{\psi,n_F}\nonumber\\
&&-\sum_{k=0}^\infty\frac{(\Omega-(n_F+1)-k)!}{(\Omega-(n_F+1))!k!}
\Omega^{k+1}\left(\Sc B\right)^k\nonumber\\
&&\hspace{2cm}\times\left\{\frac{(\Omega-n_F-k)\Sc B\alpha_\nu}{(\Omega-n_F)(\Omega-n_F+1)(k+1)}
\right\}
\bvec{\psi,n_F}.\label{BCS33}
\end{eqnarray}
By substituting this expression into eq. \ref{BCS32} one finds for term (2)
\begin{eqnarray}
&&\sum_{l=0}^\infty\frac{(\Omega-n_F-l)!}{(\Omega-n_F)!l!}
\Omega^l\left(\Sc B\right)^l\alpha_\nu\bvec{\psi,n_F}\nonumber\\
&=&X\alpha_\nu\bvec{\psi,n_F}\nonumber\\
&&+\sum_{k=0}^\infty\frac{(\Omega-(n_F+1)-k)!}{(\Omega-(n_F+1))!k!}
\Omega^k\left(\Sc B\right)^k
\left\{\frac{1}{\Omega}\frac{\Sc B\alpha_\nu}{\left(1-\frac{n_F}{\Omega}\right)
\left(1-\frac{n_F-1}{\Omega}\right)}\right\}
\bvec{\psi,n_F}\nonumber\\
&=&X\left\{\alpha_\nu+\frac{1}{\Omega}\Sc B\alpha_\nu
\frac{1}{\left(1-\frac{n_F}{\Omega}\right)
\left(1-\frac{n_F-1}{\Omega}\right)}\right\}
\bvec{\psi,n_F}.\label{BCS34}
\end{eqnarray}
We now combine term (1) and term (2), replacing the number $n_F$ with the operator $\nf$
to conclude that the image of $a_\nu$ is given by
\begin{equation}
a_\nu\longleftarrow\alpha_\nu+\frac{1}{\sqrt{\Omega}}B(-)^{j-\nu}{\alpha_{-\nu}^\dagger}
\frac{1}{1-\frac{\nf}{\Omega}}+\frac{1}{\Omega}\Sc B \alpha_\nu
\frac{1}{\left(1-\frac{\nf}{\Omega}\right)\left(1-\frac{\nf-1}{\Omega}\right)}.\label{BCS35}
\end{equation}
The easiest way to find the image $\left(a_\nu^+\right)_{X,BF}$ of $a_\nu^+$ at this 
point is by exploiting the commutator
\begin{equation}
\left(a_\nu^+\right)_{X,BF}=\sqrt{\Omega}(-)^{j+\nu}\com{\left(a_{-\nu}\right)_{X,BF}}
{\left(S^+\right)_{X,BF}},\label{BCS36}
\end{equation}
which is one of the basic commutators between generators of any representation of the 
superalgebra. Since the right-hand side of the equation involves the images of $a_{-\nu}$ and
$S^+$, both of which we have already computed, we quite simply calculate
\begin{equation}
a_\nu^+\longleftarrow\alpha_\nu^\dagger\frac{1-\frac{\nf+\Nb}{\Omega}}{1-\frac{\nf}{\Omega}}
+\frac{1}{\sqrt{\Omega}}B^\dagger(-)^{j-\nu}{\alpha_{-\nu}}
-\frac{1}{\sqrt{\Omega}}\Sc (-)^{j-\nu} \alpha_{-\nu}
\frac{1-\frac{\nf+\Nb}{\Omega}}{\left(1-\frac{\nf}{\Omega}\right)
\left(1-\frac{\nf-1}{\Omega}\right)}.\label{BCS37}
\end{equation}
The images of the singel-fermion operators of eq. \ref{BCS35} and eq. \ref{BCS37} can be 
compared to those derived in \cite{CG1}, that were derived by essentially the same method. We
see that those expressions differ from eq. \ref{BCS35} and eq. \ref{BCS37} by some 
$\Omega$-dependent factors in front of the various terms in the images. The reason for
the difference is that in \cite{CG1} a different normalization is used for the $S^+$ opperator
that features in the definition of the Usui operator. This means that the boson-fermion
representation we work with here is different from that considered in \cite{CG1}. Both 
representations are equally valid though.

All the building blocks required to write down the image $\left(H\right)_{X,BF}$ of the 
Richardson Hamiltonian 
\begin{equation}
H=\sum_{\nu=-j}^{j}\epsilon_\nu a_\nu^+ a_\nu-G S^+ S.\label{BCS38}
\end{equation}
are now available. The image of the Richardson Hamiltonian is then written as
\begin{equation}
\left(H\right)_{X,BF}=K+H_0,\label{BCS39}
\end{equation}
where the operator $H_0$, the image of the pairing term, is given by 
\begin{equation}
H_0=-\frac{G}{\Omega}\nb\left(\Omega+1-\nf-\nb\right).\label{tipert18}
\end{equation}
The operator $K$, the image of the one-body term, can be written as the sum $K=K_+ +K_0 +K_-$ of 
an operator $K_+$ that annihilates a boson and creates two ideal fermions,
an operator $K_0$ that leaves both the number of bosons and the number of fermions unchanged,
and an operator $K_-$ that creates a boson and annihilates two ideal fermions. After a little
algebra we arrive at the following expressions for these operators:
\begin{eqnarray}
K_{-}&=&B^\dagger\xi,\label{BCS40a}\\
K_{0}&=&\frac{\Omega-2\Nb-\nf}{\Omega-\nf}\left\{\sum_{\nu=-j}^j\epsilon_\nu\alpha_\nu^\dagger
\alpha_\nu-\Sc \xi \frac{\Omega}{\Omega-\nf+2}\right\}+\frac{2\Omega\Nb\bar{\epsilon}}{\Omega-\nf},
\label{BCS40b}\\
K_{+}&=&\xi^\dagger B\frac{\Omega\left(\Omega-\nf-\Nb\right)}{\left(\Omega-\nf-1\right)
\left(\Omega-\nf\right)}\nonumber\\
&&+\Sc B\left(2\sum_{\nu=-j}^j\epsilon_\nu\alpha_\nu^\dagger\alpha_\nu
-2\Omega\bar{\epsilon}\right)\frac{\Omega\left(\Omega-\nf-\Nb\right)}{\left(\Omega-\nf-1\right)
\left(\Omega-\nf\right)^2}\nonumber\\
& &-\left(\Sc\right)^2 B \xi\frac{\Omega^2\left(\Omega-\nf-\Nb\right)}
{\left(\Omega-\nf-1\right)\left(\Omega-\nf\right)^2\left(\Omega-\nf+1\right)}.\label{BCS40c}
\end{eqnarray}
In these expressions we exploited the fact that $\epsilon_\nu=\epsilon_{-\nu}$ to affect some
simplification. The operator $\xi^\dagger$ is defined as
\begin{equation}
\xi^\dagger=\frac{1}{\sqrt{\Omega}}\sum_{\nu=\frac{1}{2}}^j 2\epsilon_\nu \beta_\nu^\dagger.
\label{BCS40d}
\end{equation}
Here the operator
$\beta_\nu^\dagger=(-)^{j-\nu}\alpha_\nu^\dagger\alpha_{-\nu}^\dagger$
creates time-reversed pairs of ideal fermions. 
In deriving the expressions in eq. \ref{BCS40c}, we used the symmetry property 
$\beta^\dagger_{-\nu}=\beta^\dagger_\nu$.
The average one-particle energy is denoted $\bar{\epsilon}=\sum_{\nu=\frac{1}{2}}^j
\frac{\epsilon_\nu}{\Omega}$.

The image of the pairing term is simple. It is only a function of the boson
and ideal fermion number operators. The diagonalization of this term alone is accomplished by 
simply using
a basis for the boson-fermion space ${\mathcal H}_{BF}$ the elements of which have well-defined
numbers of both bosons and ideal fermions. 
If the original system contained $2N$ fermions, the physical subspace is 
contained in the subspace of $\Hil_{BF}$ spanned by all states with $N-s$ bosons and $2s$ ideal 
fermions where
$s$ may take on values $0,1,2,\ldots,N$. Such states may be denoted
$\left|\psi,s\right)$, where $\psi$ refers to all other quantum numbers needed to specify the
state uniquely. In this space $H_0$ has a spectrum
\begin{equation}
E_s^{(0)}=-\frac{G}{\Omega}\left(N-s\right)\left(\Omega+1-N-s\right),\label{BCS40e}
\end{equation}
with $s=0,1,2,\ldots,N$. As mentioned previously, for the mapping to be valid, more than half
of the $2\Omega$ states of the original system must be unoccupied. This brings about the
restriction that $2N<\Omega$ and thus also $2s<\Omega$. With this restriction the energy
$E_s^{(0)}$ is a monotonically increasing function of $s$. The number $s$, which 
in nuclear physics is called the seniority, labels
the energy levels of the unperturbed system, with larger $s$ referring to higher excited energy
levels. All but the lowest ($s=0$, i.e. bosons only) level is degenerate. 
The spacing between level $s$ and 
level $s+1$ is $G\left(1-\frac{2s}{\Omega}\right)$, which is almost equidistant for low-lying
states. What is harder, is identifying the ghost states in this spectrum. It turns out that
the original system does have eigenstates associated with each $E_s^{(0)}$ of eq. \ref{BCS40e}.
The ghost states only affect the degeneracy of the levels. Ghosts are present in every energy
level, except for the ground state level, which is non-degenerate.

On the other hand, what was the trivial one-particle 
contribution in the original system, has now become the non-trivial contribution in the mapped
system. Where, in the original system, we had only fermion number operators $a_\nu^+ a_\nu$,
we now also have contributions that scatter bosons into  
time-reversed fermion pairs and vice versa. Any scattering event can be decomposed into
a series of first order processes, where there are only three possible varieties of these first
order processes. Firstly there is the $K_-$ process, where the system loses two ideal fermions
and gains a boson. Then there is the $K_0$ process where two ideal fermions pair scatter off
each other. Lastly there is the $K_+$ process where a boson decays into two ideal fermions. 
This result is so important that I restate it in a slightly different way:
There are only three first order scattering processes. The first
is represented by the operator $K_-$. It takes states from the energy level $E_s^{(0)}$
to states in the level directly below i.e. the $E_{s-1}^{(0)}$ level.
(The ground state is annihilated by $K_-$.)
The second process is represented by $K_0$. This is an `elastic' process in the sense that it 
leaves states in any level $E_s^{(0)}$ in that same level. Lastly there is the process represented
by $K_+$ which takes states in the level $E_s^{(0)}$ to states in the level directly above, i.e.
the $E_{s+1}^{(0)}$ level.

This structure makes the mapped Hamiltonian well-suited to perturbation expansions where we treat 
the image of the pairing term as the unperturbed Hamiltonian and the image of the one-body term as
a perturbation. In order to be able to truncate such expansions after a finite number of terms, 
we assume that the coupling constant $G$
in front of the pairing term is large compared to the one-particle energies $\epsilon_\nu$. 
Expansions of this type are therefore called strong coupling expansions. 
We will do both a time-independent perturbation expansion for
the spectrum of the system, and a time-dependent expansion for transition amplitudes. 

\section{Time-Independent Perturbation Theory}
\markright{\bf Section \thesection: Time-Independent Perturbation Theory}
\addtocontents{toc}{\protect\nopagebreak}
\label{SS2.4}
First on the agenda is time-independent perturbation theory. We are dealing with a non-hermitian
Hamiltonian. To complicate matters, the unperturbed Hamiltonian has a highly degenerate spectrum.
Therefore I start the discussion by presenting the general formalism for finding perturbatively
the spectrum of a non-Hermitian operator if the spectrum is degenerate when the perturbation
is switched off. The argument is developed along the lines of the non-degenerate hermitian case 
as treated by Sakurai \cite{Sak1}.

We are given an operator $H(\lambda)=H_0+\lambda H_1$. Neither $H_0$ nor $H_1$ has to be hermitian. 
We assume that $H$ is fully diagonalizable in a finite region around $\lambda=0$. For 
$\lambda\not=0$, we assume that $H(\lambda)$ has a non-degenerate spectrum, so that a single 
label $\alpha=1,2,\ldots,\Omega$ may be used to specify eigenvalues and eigenvectors uniquely:
\begin{equation}
H(\lambda)\rvec{\alpha}_\lambda=E_\alpha(\lambda)\rvec{\alpha}_\lambda,\label{tipert1}
\end{equation}
with $E_\alpha(\lambda)\not=E_\beta(\lambda)$ if $\alpha\not=\beta$ and $\lambda\not=0$.
The subscript $R$ is used to indicate that we are dealing with right-eigenvectors, later to be 
contrasted
with left-eigenvectors, that do not coincide with the right eigenvectors when $H(\lambda)$ is not
hermitian.
We take it that all the $E_\alpha(\lambda)$ and $\rvec{\alpha}_\lambda$ are analytical functions
of $\lambda$ in a finite region around $\lambda=0$. Furthermore, if two eigenvalues converge at 
$\lambda=0$, i.e. $E_\alpha(0)=E_\beta(0)$, we assume that these eigenvalues have different
first derivatives at $\lambda=0$. This condition is referred to by saying that all degeneracy is
lifted in the first order.
At $\lambda=0$ the spectrum of $H$ may be degenerate. 
Our notation takes care of this as follows:
At $\lambda=0$ we let $\Omega'$ be the number
of distinct eigenvalues of $H_0$. We then indicate these distinct eigenvalues of $H_0$ as 
$\En_l$, $l=1,2,\ldots,\Omega'$. 
We partition the labels $\alpha=1,2,\ldots,\Omega$ into disjoint sets $P_l$, labeled by integers
$l=1,2,\ldots \Omega'$, by defining $P_l$ to be the set of all indices $\alpha$ such 
that $E_\alpha$ flows
to $\En_l$ as $\lambda$ goes to zero. We can then also define a function $p$ from the index set
$1,2,\ldots,\Omega$ to the index set $1,2,\ldots,\Omega'$ as follows: $p(\alpha)=l$ where $l$ is the
unique label such that $\alpha\in P_l$. We refer to the space spanned by all eigenvectors of $H_0$
with eigenvalue $\En_l$ as the $\En_l$-eigenspace of $H_0$. It is assumed we know a
basis for each $\En_l$ eigenspace of $H_0$. The basis elements that span the $\En_l$ eigenspace
will be denoted $\left|\psi,l_R\right)$ where $\psi$ is a discrete index that runs from $1$ to
some integer $\omega_l$ which is the dimension of the $\En_l$ eigenspace. 
The set $\left\{\left|\psi,l_R\right):l=1,\ldots,\Omega';\psi=1,\ldots,\omega_l\right\}$
forms a basis for the domain of $H(\lambda)$. 
It is in terms of
these basis states that we want to express the eigenstates of $H(\lambda)$ approximately.
Corresponding to this basis we can always find a left basis 
$\left\{\left|\psi,l_L\right):l=1,\ldots,\Omega';\psi=1,\ldots,\omega_l\right\}$
such that the equations
\begin{equation}
\left(\psi,l_L\right|\left.\phi,m_R\right)=\delta_{l,m}\delta_{\psi,\phi}, \label{tipert3c}
\end{equation}
hold for all $l,m,\psi$ and $\phi$. 

We uniquely define a projection operator $\Pi_l$ as the linear operator that maps any vector in the 
$\En_l$-eigenspace of $H_0$ onto itself, while mapping any vector that can be written as 
a linear combination of vectors from the other eigenspaces of $H_0$, onto zero. (This might not
be an {\em orthogonal} projection operator though. The different $\En_l$-eigenspaces are usually not
orthogonal to each other as $H_0$ is generally non-hermitian.)
An operator $\tilde{\Pi}_l$ is defined as $\tilde{\Pi}_l={\rm I}-\Pi_l$. In terms of the states
$\left|\psi,l_L\right)$ and $\left|\phi,m_R\right)$, the identity operator ${\rm I}$, the 
unperturbed operator $H_0$ and the projection operators $\Pi_l$ and $\tilde{\Pi}_l$ can be 
expressed as follows:
\begin{eqnarray}
{\rm I}&=&\sum_{l=1}^{\Omega'}\sum_{\psi=1}^{\omega_l}\left|\psi,l_R\right)\left(\psi,l_L\right|,
\label{tipert4a}\\
{H_0}&=&\sum_{l=1}^{\Omega'}\En_l\sum_{\psi=1}^{\omega_l}\left|\psi,l_R\right)\left(\psi,l_L\right|,
\label{tipert4b}\\
\Pi_l&=&\sum_{\psi=1}^{\omega_l}\left|\psi,l_R\right)\left(\psi,l_L\right|,
\label{tipert4}\\
\tilde{\Pi}_l&=&\sum_{k=1\not=l}^{\Omega'}\sum_{\psi=1}^{\omega_k}\left|\psi,k_R\right)
\left(\psi,k_L\right|,
\label{tipert5}
\end{eqnarray}
Note that $\Pi_l\Pi_m=\delta_{lm}\rm \Pi_m$.

With the above preliminaries out of the way we now expand the eigenvalues 
$E_\alpha(\lambda)$ and
eigenvectors $\rvec{\alpha}_\lambda$ of $H(\lambda)$ in terms of $\lambda$:
\begin{eqnarray}
E_\alpha(\lambda)&=&\En_{p(\alpha)}+\lambda\Delta_\alpha^{(1)}+\lambda^2\Delta_\alpha^{(2)}
+\ldots,\nonumber\\
\rvec{\alpha}_\lambda&=&\rvec{\alpha^{(0)}}+\lambda\rvec{\alpha^{(1)}}+\lambda^2\rvec{\alpha^{(2)}}
+\ldots.\label{tipert2}
\end{eqnarray}
This definition fixes the directions of the $\lambda$-independent zero'th order eigenvectors
$\rvec{\alpha^{(0)}}_0$ uniquely as the direction to which the $\alpha$-eigenvector of $H(\lambda)$ 
converges as $\lambda$ goes to zero. This direction is well-defined since we assumed that 
$H(\lambda)$ has a non-degenerate spectrum for $\lambda\not=0$. The magnitudes of the eigenvectors
are left arbitrary.

The set of eigenvectors $\left\{\rvec{\alpha^{(0)}}\right\}_{\alpha=1}^{\Omega}$ also forms a 
basis for
the domain of $H(\lambda)$. We use it to define a second basis 
(sometimes called the contra-variant basis) $\left\{\lvec{\alpha^{(0)}}\right\}_{\alpha=1}^\Omega$
through the set of equations
\begin{equation}
\lfunc{\alpha}\left.\beta^{(0)}_R\right)=\delta_{\alpha,\beta}.\label{tipert3}
\end{equation}
Eq.\ref{tipert3} implies that $\lfunc{\alpha}H_0=\En_{p(\alpha)}\lfunc{\alpha}$.
Using the states $\left\{\rvec{\alpha^{(0)}}\right\}_{\alpha=1}^{\Omega}$ and functionals
$\left\{\left(\alpha_L^{(0)}\right|\right\}_{\alpha=1}^{\Omega}$ we define another set of projection
operators that will come in handy during our calculations. Define a projection operator
\begin{equation}
\pi_\alpha=\left|\alpha_R^{(0)}\right)\left(\alpha_L^{(0)}\right|,\label{tipert4x}
\end{equation}
that singles out the $\left|\alpha_R^{(0)}\right)$ componet of the decomposition of any vector in 
the $\left\{\rvec{\alpha^{(0)}}\right\}_{\alpha=1}^{\Omega}$ basis. Let $\tilde{\pi}_\alpha$ be the
complementary projection operator of $\pi_\alpha$ in the $\mathcal E_{p(\alpha)}$ subspace by 
setting 
\begin{equation}
\tilde{\pi}_\alpha=\sum_{\beta\in P_{p(\alpha)}\setminus\{\alpha\}}\left|\beta_R^{(0)}\right)
\left(\beta_L^{(0)}\right|=\Pi_{p(\alpha)}-\pi_\alpha.\label{tipert5xx}
\end{equation}
Thus, the identity operator may be decomposed as
\begin{equation}
{\rm I}=\tilde{\Pi}_{p(\alpha)}+\tilde{\pi}_\alpha+\pi_\alpha.\label{tipert5x}
\end{equation}
The next step is to choose normalization for the eigenstate 
$\left|\alpha_R\right)_\lambda$. The most convenient choice turns out to be 
$\Big(\alpha_L^{(0)}\Big|\alpha_R\Big)_\lambda=1$. With this choice it follows that
\begin{equation}
\pi_\alpha\left|\alpha_R\right)_\lambda=\left|\alpha_R^{(0)}\right).\label{tipert6x}
\end{equation}
Now we rewrite the eigenvalue equation in the form
\begin{equation}
\left(\mathcal E_{p(\alpha)}-H_0\right)\left|\alpha_R\right)_\lambda=\left(\lambda H_1-
\Delta_\alpha(\lambda)\right)\left|\alpha_R\right)_\lambda,\label{tipert7x}
\end{equation}
where $\Delta_{\alpha}(\lambda)=\sum_{k=1}^\infty\Delta_\alpha^{(k)}\lambda^k$. 

The first thing
we do with this equation is multiply it by $\Pi_{p(\alpha)}$, noting that \newline 
$\Pi_{p(\alpha)}\left(\mathcal E_{p(\alpha)}-H_0\right)=0$, to find
\begin{equation}
\Pi_{p(\alpha)}\left(\lambda H_1-\Delta_\alpha(\lambda)\right)\left|\alpha_R\right)_\lambda=0,
\label{tipert8x}
\end{equation}
which has to hold order for order in $\lambda$.
Note that, by definition, the zero'th order term in the $\lambda$-expansion of 
$\left|\alpha_R\right)_\lambda$, namely $\left|\alpha_R^{(0)}\right)$, is an element of the 
$\mathcal E_{p(\alpha)}$ eigenspace of $H_0$ and hence 
$\Pi_{p(\alpha)}\left|\alpha_R^{(0)}\right)=\left|\alpha_R^{(0)}\right)$. Keeping this in mind, we
look at the terms in eq. \ref{tipert8x} of lowest order in $\lambda$ to find
\begin{equation}
\Pi_{p(\alpha)}H_1\left|\alpha_R^{(0)}\right)=\Delta_\alpha^{(1)}\left|\alpha_R^{(0)}\right).
\label{tipert9x}
\end{equation}
This eigenvalue equation enables us to find the zero'th order vectors $\left|\alpha_R^{(0)}\right)$
as the eigenvectors of $\Pi_{p(\alpha)}H_1$ in the $\mathcal E_{p(\alpha)}$ eigenspace. Since we
assume all degeneracy to be lifted in the first order, no two eigenvalues $\Delta_\alpha^{(1)}$
and $\Delta_\beta^{(1)}$ with $\alpha$, $\beta\in P_{p(\alpha)}$ are the same, and eq. 
\ref{tipert9x} is necessary and sufficient to determine the various $\left|\alpha_R^{(0)}\right)$
in the $\mathcal E_{p(\alpha)}$ eigenspace.

In order to proceed, we define an operator
\begin{equation}
Q_k=\sum_{l\not=k}\frac{1}{\mathcal E_k-\mathcal E_l}\Pi_l.\label{tipert10x}
\end{equation}
We return to eq. \ref{tipert7x} and multiply it with $Q_{p(\alpha)}$, noting that 
$Q_{p(\alpha)}\left(\mathcal E_{p(\alpha)}-H_0\right)=\tilde{\Pi}_{p(\alpha)}.$ Thus we arrive at
one of the most important formulas in the present discussion, namely
\begin{equation}
\tilde{\Pi}_{p(\alpha)}\left|\alpha_R\right)_\lambda=Q_{p(\alpha)}\left(\lambda H_1
-\Delta_\alpha(\lambda)\right)\left|\alpha_R\right)_\lambda.\label{tipert11x}
\end{equation}
This equation has, on the left-hand side, the part of the eigenvector 
$\left|\alpha_R\right)_\lambda$ that can be writen as a linear combination of vectors 
in $\mathcal E_l$ eigenspaces with $l\not=p(\alpha)$. We want a similar equation for that part
of $\left|\alpha_R\right)_\lambda$ that lies inside the $\mathcal E_{p(\alpha)}$-eigenspace.

For this purpose, we firstly rewrite eq. \ref{tipert8x} to read
\begin{equation}
\left(\Delta_\alpha(\lambda)-\lambda\Pi_{p(\alpha)}H_1\Pi_{p(\alpha)}\right)\Pi_{p(\alpha)}
\left|\alpha_R\right)_\lambda-\lambda\Pi_{p(\alpha)}H_1\tilde{\Pi}_{p(\alpha)}\left|\alpha_R\right)
_\lambda=0,\label{tipert12x}
\end{equation}
by recalling that $\Pi_{p(\alpha)}+\tilde{\Pi}_{p(\alpha)}={\rm I}$ and 
$\left(\Pi_{p(\alpha)}\right)^2=\Pi_{p(\alpha)}$. Now we substitute eq. \ref{tipert11x} into the
second term of eq. \ref{tipert12x} to find
\begin{equation}
\left(\Delta_\alpha(\lambda)-\lambda\Pi_{p(\alpha)}H_1\Pi_{p(\alpha)}\right)\Pi_{p(\alpha)}
\left|\alpha_R\right)_\lambda-\lambda\Pi_{p(\alpha)}H_1Q_{p(\alpha)}\left(\lambda H_1
-\Delta_\alpha(\lambda)\right)\left|\alpha_R\right)_\lambda=0.\label{tipert13x}
\end{equation}
Define an operator 
\begin{equation}
q_\alpha=\sum_{\beta\in P_{p(\alpha)}\setminus\{\alpha\}}\frac{1}{\Delta_\alpha^{(1)}-
\Delta_\beta^{(1)}}\left|\beta_R^{(0)}\right)\left(\beta_R^{(0)}\right|.\label{tipert14x}
\end{equation}
Note that $q_\alpha\left(\Delta_\alpha^{(1)}-\Pi_{p(\alpha)}H_1\Pi_{p(\alpha)}\right)=
\tilde{\pi}_\alpha$ and that $q_\alpha\Pi_{p(\alpha)}=q_\alpha$. Hence, by multiplying eq. 
\ref{tipert13x} with $q_\alpha$ and deviding by $\lambda$, to find
\begin{equation}
\tilde{\pi}_\alpha\left|\alpha_R\right)_\lambda=q_\alpha H_1Q_{p(\alpha)}\left(\lambda H_1-
\Delta_\alpha(\lambda)\right)\left|\alpha_R\right)_\lambda-\sum_{k=1}^\infty \Delta_\alpha^{(k+1)}
\lambda^k q_\alpha\left|\alpha_R\right)_\lambda.\label{tipert15x}
\end{equation}
This is the counterpart of eq. \ref{tipert11x} that we needed for the part of 
$\left|\alpha_R\right)_\lambda$ that lies inside the $\mathcal E_{p(\alpha)}$-subspace. When
we combine the results of eq. \ref{tipert6x}, eq. \ref{tipert11x} and eq. \ref{tipert15x}, 
remembering that the identity operator may be decomposed as ${\rm I}=\tilde{\Pi}_{p(\alpha)}
+\tilde{\pi}_\alpha+\pi_\alpha$ we find
\begin{equation}
\left|\alpha_R\right)_\lambda=\left|\alpha_R^{(0)}\right)+\left(1+q_\alpha H_1\right) Q_{p(\alpha)}
\left(\lambda H_1-\Delta_\alpha(\lambda)\right)\left|\alpha_R\right)_\lambda-\sum_{k=1}^\infty
\Delta_{\alpha}^{(k+1)}\lambda^k q_\alpha\left|\alpha_R\right)_\lambda.\label{tipert16x}
\end{equation}
Collecting terms of order $N+1$ we find, for $N\geq0$,
\begin{eqnarray}
\left|\alpha_R^{(N+1)}\right)&=&\left(1+q_\alpha H_1\right)Q_{p(\alpha)}\left\{H_1\left|
\alpha_R^{(N)}\right)-\sum_{M=1}^{N+1}\Delta_{\alpha}^{(M)}\left|\alpha_R^{(N+1-M)}\right)\right\}
\nonumber\\
&&\hspace{10mm}-\sum_{M=2}^{N+2}\Delta_\alpha^{(M)}q_\alpha\left|\alpha_R^{(N+2-M)}\right).
\label{tipert17x}
\end{eqnarray}
Note that $Q_{p(\alpha)}\left|\alpha_R^{(0)}\right)=0$ and $q_\alpha\left|\alpha_R^{(0)}\right)=0$.
Thus, for $N\geq1$, the upper bounds on the two summations in the above expression may respectively 
be decreased from $N+1$ to $N$ and from $N+2$ to $N+1$, while for the case where $N=0$, the two 
summations may be left out entirely. To complete the expansions, we need an expression for 
$\Delta_\alpha^{(M)}$. This we find by multiplying eq. \ref{tipert7x} with the functional 
$\left(\alpha_R^{(0)}\right|$, recalling our normalization convention 
$\Big(\alpha_L^{(0)}\Big|\alpha_R\Big)_\lambda=1$, and collecting terms of order $M$ in $\lambda$
to find
\begin{equation}
\Delta_\alpha^{(M)}=\left(\alpha_L^{(0)}\right|H_1\left|\alpha_R^{(M-1)}\right).\label{tipert18x}
\end{equation}
Our final results are then:
\begin{eqnarray}
\left|\alpha_R^{(N+1)}\right)&=&\left(1+q_\alpha H_1\right)Q_{p(\alpha)}\left\{H_1\left|
\alpha_R^{(N)}\right)-\sum_{M=1}^{N}\Delta_{\alpha}^{(M)}\left|\alpha_R^{(N+1-M)}\right)\right\}
\nonumber\\
&&\hspace{10mm}-\sum_{M=2}^{N+1}\Delta_\alpha^{(M)}q_\alpha\left|\alpha_R^{(N+2-M)}\right),
\label{tipert19x}
\end{eqnarray}
and
\begin{equation}
\Delta_\alpha^{(N+1)}=\left(\alpha_L^{(0)}\right|H_1\left|\alpha_R^{(N)}\right).\label{tipert20x}
\end{equation}
It is understood that, in the case where $N=0$ and the lower bounds of the summations in
eq. \ref{tipert19x} exceed the upper bounds, the summation contains no terms, i.e. 
$\left|\alpha_R^{(1)}\right)=\left(1+q_\alpha H_1\right)Q_{p(\alpha)}H_1\left|\alpha_R^{(0)}
\right)$. The formulas of eq. \ref{tipert19x} and eq. \ref{tipert20x} are all we need to calculate
the corrections of order $N+1$ to both the eigenvectors and eigenvalues of $H$, if the corrections
of orders $0,1,2,\ldots,N$ are known. Together with the initial conditions provided by the 
eigenvalue equation
\begin{equation}
\Pi_{p(\alpha)}H_1\left|\alpha_R^{(0)}\right)=\Delta_\alpha^{(1)}\left|\alpha_R^{(0)}\right).
\label{tipert21x}
\end{equation}
this allows us to write down in principal the expansion of any eigenvector and its eigenvalue to
arbitrary order in $\lambda$. Concerning eq. \ref{tipert20x} for the eigenvalue-correction of order 
$N+1$, note that $\left(\alpha_L^{(0)}\right|H_1q_\alpha=0$. This is true because the states
$\left|\alpha_R^{(0)}\right)$ and functionals $\left(\beta_L^{(0)}\right|$ with $\alpha$, 
$\beta\in P_l$ were chosen to diagonalize $H_1$ in the $\mathcal E_{p(\alpha)}$ subspace, 
in the sense that
$\left(\beta_L^{(0)}\right|H_1\left|\alpha_R^{(0)}\right)=\Delta_\alpha^{(1)}
\delta_{\alpha,\beta}$. This means that, when we substitute from eq. \ref{tipert19x} for
$\left|\alpha_R^{(N)}\right)$ in eq. \ref{tipert20x}, the terms that are multiplied from
the left with an operator $q_\alpha$ disappear and we are left with
\begin{equation}
\Delta_\alpha^{(N+2)}=\left(\alpha_L^{(0)}\right|H_1Q_{p(\alpha)}\left\{H_1\left|
\alpha_R^{(N)}\right)-\sum_{M=1}^{N}\Delta_{\alpha}^{(M)}\left|\alpha_R^{(N+1-M)}\right)\right\},
\label{tipert22x}
\end{equation}
which is an expression for the eigenvalue-correction, with all the dead-wood cut away.

If the an unperturbed eigenstate $\left|\alpha_R^{(0)}\right)$ has a non-degenerate eigenvalue,
i.e. $H_0$ does not contain other eigenvectors with the same eigenvalue as 
$\left|\alpha_R^{(0)}\right)$, the situation and hence the formulas are simpler than above. In
this case, the proper operator to uses for $q_\alpha$ is simply the zero-operator. With this
adjustment the general formulas reduce to formulas valid for the specific case of no degeneracy.

Below, we write down explicitly the first order correction to an eigenstate and the second
order correction to its eigenvalue, for the general case where degeneracy is present in the 
unperturbed system:
\begin{eqnarray}
\left|\alpha_R^{(1)}\right)&=&\sum_{l\not=p(\alpha)}\frac{\Pi_lH_1\left|\alpha_R^{(0)}\right)}
{\mathcal E_{p(\alpha)}-\mathcal E_l}+\sum_{{l\not=p(\alpha)}\atop{\beta\in P_{p(\alpha)}
\setminus\{\alpha\}}}\frac{\left(\beta_L^{(0)}\right|H_1\Pi_lH_1\left|\alpha_R^{(0)}\right)
\left|\beta^{(0)}_R\right)}{\left(\mathcal E_{p(\alpha)}-\mathcal E_l\right)
\left(\Delta_\alpha^{(1)}-\Delta_\beta^{(1)}\right)},\label{tipert23x}\\
\Delta^{(2)}_\alpha&=&\sum_{l\not=p(\alpha)}\frac{\lfunc{\alpha}H_1\Pi_l
H_1\rvec{\alpha^{(0)}}}{\En_{p(\alpha)}-\En_l}\label{tipert24x}
\end{eqnarray}
Note that the formula for the second order correction to the eigenvalue, eq. \ref{tipert24x}
looks the same for the degenerate case as for the non-degenerate case. The only difference is
that, in the degenerate case the zero'th order eigenvector $\left|\alpha_R^{(0)}\right)$ is
a simultaneous eigenvector of $H_0$ and $\Pi_{p(\alpha)}H_1$ lying in the 
$\mathcal E_{p(\alpha)}$-eigenspace, whereas in the non-degenerate case it is uniquely fixed
by diagonalizing $H_0$.

We now make a few remarks concerning the results we obtained. 
Firstly, when we apply this perturbation expansion method in conjunction with the boson-fermion
mapping that we developed earlier, there is still the matter of the physical subspace to address.
The question is: If we approximate the eigenvectors of $H(\lambda)$ by only taking a few terms
in the perturbation expansion, do we lose the notion of physical eigenstates that lie exactly
in the physical subspace? Recall that if we diagonalize $H(\lambda)$ exactly, we can express the
physical subspace as the space spanned by a certain subset of the eigenvectors of $H(\lambda)$.
If we only approximate the eigenvectors by the first few terms of their $\lambda$ expansions, 
can we still find a subset of physical approximate eigenvectors that lie exactly in the
physical subspace and in fact span the physical subspace? Or do the $\lambda$-expansions for the
physical eigenvectors, if truncated after a few terms, only lie close to, but not necessarily in
the physical subspace? Perhaps not unexpectedly, the answer is that the
approximate physical eigenvectors still lie exactly in the physical subspace. This can be seen 
as follows. Recall that any operator that leaves the physical subspace invariant is said to be
a physical operator. Then because we assume that $H(\lambda)$ is a physical operator for all 
$\lambda$, it follows that $H_0$ and $H_1$ are both physical operators. Furthermore, it
is a simple matter to show that the operators $\Pi_l$ and $Q_l$
are physical operators. This means that $\Pi_l H_1$ 
is also a physical operator. We find the zero'th order eigenvectors $\rvec{\alpha^{(0)}}$ by
diagonalizing the operator $\Pi_{p(\alpha)} H_1$ in the $\En_{p(\alpha)}$ eigenspace of $H_0$.
Since $\Pi_{p(\alpha)} H_1$ is a physical operator with a non-degenerate spectrum, the eigenvectors
thus obtained can be separated into a set that spans the overlap of physical subspace
and the $\En_{p(\alpha)}$ eigenspace, and a set of ghost states. Since the operator $q_\alpha$
is defined in such a way that it has the same eigenvectors in the 
$\mathcal E_{p(\alpha)}$ eigenspace as the physical operator $\Pi_{p(\alpha)}H_1$, it follows
that $q_\alpha$ is a physical operator. Furthermore, if we start the recursion
of eq. \ref{tipert10x} with a state from the physical sector, the corrections that the recursion
generates remain inside the physical subspace. This follows inductively from the fact that
higher order corrections in eq. \ref{tipert19x} are generated by acting with physical operators
on lower corrections. If the zero'th order eigenvector is physical, this implies that the first
order correction is physical. If the zero'th order eigenvector and the first order correction are
both physical, then so too is the second order correction, etc.
This implies that the exact physical eigenstates, when expanded in terms of $\lambda$, consist
of terms that all lie in the physical subspace themselves. Truncating the expansions after a
finite number of terms therefore still leaves one in the physical subspace.

Secondly, as we saw, we know how to express the operators $\Pi_l$, 
the states $\left|\alpha_R^{(0)}\right)$ and as a result also the states 
$\left|\alpha_L^{(0)}\right)$ in terms of the known states $\left|\psi,l_L\right)$ and 
$\left|\psi,l_R\right)$. This means that all quantities in the results we just derived are
expressible in terms of known objects. For instance, for the second order corrections to
the eigenvalues we have
\begin{equation}
\Delta^{(2)}_\alpha=\sum_{l=1\not=p(\alpha)}^{\Omega'}
\sum_{\psi=1}^{\omega_l}\frac{\lfunc{\alpha}H_1\Big|\psi,l_R\Big)\Big(\psi,l_L\Big|
H_1\rvec{\alpha^{(0)}}}{\En_{p(\alpha)}-\En_{p(\psi)}}
\label{tipert17y}.
\end{equation}
Remarkably, it follows from the derivation of the perturbation expansion
that it is not necessary for the $\left|\psi,l_R\right)$
basis to be separable into physical and unphysical states. (The basis states enter only as a
decomposition of the $\Pi_l$ projection operators, whose properties are basis-independent.)
Thus we are free to choose 
the basis in which we express the $\Pi_l$ operators such that the calculation
of matrix elements of the perturbation is as simple as possible, without worrying whether we mix 
up physical and unphysical states in a degenerate subspace.

We are now ready to start our time-independent perturbation expansion for the mapped Hamiltonian.
Since all energies but that of the ground state have degenerate eigenspaces associated with them,
our first task is to find the correct zero'th order eigenstates along with the first order 
corrections to the energies. We take it the original system has $2N$ fermions. Then each degenerate
eigenspace of the unperturbed system is labeled uniquely by the integer $s$, and characterized as
the space of all states with $2s$ ideal fermions and $N-s$ bosons. In order to find the correct 
zero'th order eigenstates and the first order corrections to the energies of the states in the
$s$-eigenspace, we must diagonalize the operator $\Pi_s K$ in this
eigenspace. The operator $\Pi_s$ is in this case quite simply the projection operator that 
selects states with $2s$ ideal fermions and $N-s$ bosons. This makes the decomposition 
$K=K_{+}+K_{0}+K_{-}$ very useful, as already mentioned. More precisely, if $\left|s\right)$ is
any state in the $s$-eigenspace of $H_0$, it holds that
\begin{equation}
\Pi_{s'}K\left|s\right)=\left\{\begin{array}{lcl}K_-\left|s\right)&{\rm if}&s'+1=s\\
K_0\left|s\right)&{\rm if}&s'=s\\
K_+\left|s\right)&{\rm if}&s'=s+1\\
0&{\rm otherwise}&\end{array}\right. .\label{tipert17a}
\end{equation}
For the particular case we are interested in, it follows that
\begin{eqnarray}
\Pi_s K\left|s\right)&=&K_{0}\left|s\right)\nonumber\\
&=&\left\{\frac{\Omega-2N}{\Omega-2s}\left(\sum_{\nu=-j}^j\epsilon_\nu\alpha_\nu^\dagger\alpha_\nu
-\frac{2}{\Omega-2s+2}\sum_{\mu,\nu=\frac{1}{2}}^j\epsilon_\nu\beta_\mu^\dagger\beta_\nu\right)
\right.\nonumber\\
&&\left.+\frac{\Omega\left(N-s\right)}{\Omega-2s}2\bar{\epsilon}\right\}\left|s\right).
\label{tipert19y}
\end{eqnarray}
The operator acting on $\left|s\right)$ introduces scattering between the ideal fermions
of the states in the $s$-eigenspace. 
The diagonalization of this operator is simplified by the blocking effect
in much the same way as was the case for the original Richardson Hamiltonian. A simple way
to consider the blocking effect here is to take it into account before doing the mapping, i.e.
by restricting the fermion space from which we map to the subspace of full fermion Fock space
in which no levels are singly occupied. 
This implies that in the mapped system we only work with states of $\Hil_{BF}$ that have no 
ideal fermions singly occupying levels. To verify the above statement, note that the
image of $a_\nu^+ a_{-\nu}^+$ creates no ideal fermion in the state $\mu$ without
either also annihilating it or creating an accompanying ideal fermion in the state $-\mu$ as well.
Therefore, because of the blocking effect, the operator in eq. \ref{tipert19y} only has
to be diagonalised in the space spanned by states of the form
\begin{equation}
\frac{\left(B^\dagger\right)^{N-s}}{\sqrt{(N-s)!}}\left(\prod_{k=1}^s\beta^\dagger_{\nu_k}\right)
\left|0\right),\label{tipert20a}
\end{equation}
which is what we now do. 

The operator in eq. \ref{tipert19y} treats the bosons as spectators. 
Furthermore, the term proportional to identity and the scaling factor 
$\frac{\Omega-2N}{\Omega-2s}$ in the first term are unimportant for the diagonalization procedure.
This allows us to construct the eigenstates of the operator in eq. \ref{tipert19y} as follows:
First we find every eigenstate $\left|\psi\right)$ (with eigenvalue $\eta_\psi$) of the operator
\begin{equation}
K_{0,\rm eff}=\sum_{\nu=-j}^j\epsilon_\nu\alpha_\nu^\dagger\alpha_\nu
-\frac{2}{\Omega-2s+2}\sum_{\mu,\nu=\frac{1}{2}}^j\epsilon_\nu\beta_\mu^\dagger\beta_\nu
\label{tipert20y}
\end{equation}
in the space of states with $2s$ ideal fermions with 
no singly occupied levels and no bosons.\pagebreak
The eigenstates of the operator in eq. \ref{tipert19y} are then 
$\frac{\left(B^\dagger\right)^{N-s}}{\sqrt{(N-s)!}}\left|\psi\right)$ with eigenvalues 
$\frac{\Omega-2N}{\Omega-2s}\eta_\psi+\frac{N-s}{\Omega-2s}2\Omega\bar{\epsilon}$. 
The real problem to solve is diagonalizing the operator $K_{0,\rm eff}$ in eq. \ref{tipert20y} 
in the
space of $2s$ ideal fermions with no singly occupied levels and no bosons. 
This last operator is sufficiently similar to the Richardson Hamiltonian
that we started with that its diagonalization proceeds with the same ease or difficulty as the
diagonalization of the Richardson Hamiltonian for $2s$ electrons. What we gain, for
low lying states  at least, is that the number of ideal fermions ($2s$) for which the operator 
$K_{0,\rm eff}$ in eq. \ref{tipert20y} must be diagonalized, is much smaller than the number of 
electrons ($2N$) for which the Richardson Hamiltonian must be diagonalized. 

After playing around with the operator $K_{0,\rm eff}$, and for instance diagonalizing it in
the space of a two-particle system, one starts to suspect that it has a lot in common with
the original Richardson Hamiltonian. We therefore try to diagonalize it following the same 
procedure that was followed in \cite[Appendix B]{vDR1}.
This involves making an ansatz for the eigenstates of the operator $K_{0,\rm eff}$. 
We assume that every eigenstate $\left|\psi\right)$ has the form
\begin{equation}
\left|\psi\right)=\left(\prod_{k=1}^s A_{\psi,k}^\dagger\right)\left|0\right),\label{tipert21y}
\end{equation}
where $A_{\psi,k}^\dagger=\sum_{\nu=\frac{1}{2}}^j\frac{1}{2\epsilon_\nu-\eta_{\psi,k}}
\beta_\nu^\dagger$ and the complex numbers $\eta_{\psi,k}$ still have to be fixed.
Because the ansatz for the form of the eigenstates uses the $\beta_\nu^\dagger$ operators, it
is useful 
to collect
a few facts about these operators, namely 
\begin{eqnarray}
\left(\beta_\nu\right)^2&=&0,\label{tipert22y}\\
\com{\beta_\mu^\dagger}{\beta_\nu}&=&\delta_{\mu,\nu}\left(\alpha_\nu^\dagger\alpha_\nu+
\alpha_{-\nu}^\dagger\alpha_{-\nu}-1\right),\label{tipert23y}\\
\com{\com{\beta_\mu^\dagger}{\beta_\nu}}{\beta_\rho^\dagger}&=&2\delta_{\mu,\nu}\delta_{\nu,\rho}
\beta_\rho^\dagger,\label{tipert24y}
\end{eqnarray}
and to rewrite the operator in eq. \ref{tipert20y} in terms of the $\beta$'s as
\begin{equation}
\sum_{\mu,\nu=\frac{1}{2}}^j\epsilon_\nu\left(\com{\beta_\mu^\dagger}{\beta_\nu}+\delta_{\mu,\nu}
-\frac{2}{\Omega-2s+2}\beta_\mu^\dagger\beta_\nu\right),\label{tipert25}
\end{equation}
where we used the fact that $\epsilon_\nu=\epsilon_{-\nu}$.
With this form of the operator we check the ansatz. First we compute the commutator
\begin{eqnarray}
\com{K_{0,\rm eff}}{A^\dagger_{\psi,k}}&=&
\sum_{\mu,\nu,\rho=\frac{1}{2}}^j\frac{\epsilon_\nu}{2\epsilon_\rho-\eta_{\psi,k}}
\left(\com{\com{\beta_\mu^\dagger}{\beta_\nu}}{\beta_\rho^\dagger}
-\frac{2}{\Omega-2s+2}
\underbrace{\com{\beta_\mu^\dagger\beta_\nu}{\beta_\rho^\dagger}}_{=\delta_{\nu,\rho}
\beta_\mu^\dagger\com{\beta_\rho}{\beta_\rho^\dagger}}\right)\nonumber\\
&=&\sum_{\rho=\frac{1}{2}}^j\underbrace{\frac{2\epsilon_\rho}
{2\epsilon_\rho-\eta_{\psi,k}}}_{=1+\frac{\eta_{\psi,k}}{2\epsilon_\rho-\eta_{\psi,k}}}
\beta_\rho^\dagger-\frac{1}{\Omega-2s+2}\sum_{\mu,\rho=\frac{1}{2}}^j
\frac{2\epsilon_\rho}{2\epsilon_\rho-\eta_{\psi,k}}\beta_\mu^\dagger\com{\beta_\rho^\dagger}
{\beta_\rho}\nonumber\\
&=&\eta_{\psi,k}A^\dagger_{\psi,k}+\sqrt{\Omega}\Sc\left\{1-\frac{1}{\Omega-2s+2}
\sum_{\rho=\frac{1}{2}^\Omega}\frac{2\epsilon_\rho}{2\epsilon_\rho-\eta_{\psi,k}}
\com{\beta_\rho^\dagger}{\beta_\rho}\right\},\nonumber\\
\label{tipert26}
\end{eqnarray}
where, in the last line, we used the relation $\sqrt{\Omega}\Sc=\sum_{\rho=\frac{1}{2}}^j
\beta_\rho^\dagger$. Now we compute
\begin{eqnarray}
K_{0,\rm eff}\left|\psi\right)&=&\com{Q}{\prod_{k=1}^s A_{\psi,k}^\dagger}\left|0\right)\nonumber\\
&=&\sum_{k=1}^s\left(\prod_{l=1}^{k-1}A_{\psi,l}^\dagger\right)\com{Q}{A_{\psi,k}^\dagger}
\left(\prod_{l=k+1}^s A_{\psi,l}^\dagger\right)\left|0\right)\nonumber\\
&=&\eta_\psi\left|\psi\right)+\sum_{k=1}^s\sqrt{\Omega}\Sc\prod_{l=1\not=k}^s A_{\psi,l}^\dagger
\left|0\right)\nonumber\\
&&+\frac{\sqrt{\Omega}}{\Omega-2s+2}\Sc
\sum_{k=1}^s\left(\prod_{l=1}^{k-1}A_{\psi,l}^\dagger\right)
\left\{\sum_{\rho=\frac{1}{2}}^j\frac{2\epsilon_\rho}
{2\epsilon_\rho-\eta_{\psi,k}}
\com{\beta_\rho^\dagger}{\beta_\rho}\right\}
\left(\prod_{l=k+1}^s A_{\psi,l}^\dagger\right)\left|0\right),\nonumber\\
\label{tipert27}
\end{eqnarray}
where $\eta_\psi=\sum_{k=1}^s\eta_{\psi,k}$. In order to proceed, we note that
\begin{eqnarray}
\com{\beta_\rho^\dagger}{\beta_\rho}\prod_{l=k+1}^s A_{\psi,l}^\dagger\left|0\right)
&=&\com{\com{\beta_\rho^\dagger}{\beta_\rho}}{\prod_{l=k+1}^s A_{\psi,l}^\dagger}\left|0\right)
+\prod_{l=k+1}^s A_{\psi,l}^\dagger
\underbrace{\com{\beta_\rho^\dagger}{\beta_\rho}\left|0\right)}_{=-\left|0\right)}\nonumber\\
&=&\sum_{l=k+1}^s\left(\prod_{m=k+1}^{l-1}A_{\psi,m}^\dagger\right)
\com{\com{\beta_\rho^\dagger}{\beta_\rho}}{A_{\psi,l}^\dagger}
\left(\prod_{m=l+1}^sA_{\psi,m}^\dagger\right)\left|0\right)\nonumber\\
&&-\prod_{l=k+1}^sA_{\psi,l}^\dagger\left|0\right).\label{tipert28}
\end{eqnarray}
This last result we substitute into eq. \ref{tipert27} to find
\begin{eqnarray}
K_{0,\rm eff}\left|\psi\right)
&=&\eta_\psi\left|\psi\right)+\sqrt{\Omega}\Sc\sum_{k=1}^s\left(1-\frac{1}{\Omega-2s+2}
\sum_{\rho=\frac{1}{2}}^j\frac{2\epsilon_\rho}{2\epsilon_\rho-\eta_{\psi,k}}\right)
\prod_{l=1\not=k}^s A_{\psi,l}^\dagger
\left|0\right)\nonumber\\
&&+\frac{\sqrt{\Omega}\Sc}{\Omega-2s+2}
\sum_{k=1}^s\sum_{l=k+1}^s\left(\prod_{m=1\not=k}^{l-1}A_{\psi,m}^\dagger\right)\nonumber\\
&&\hspace{3cm}\times\sum_{\rho=\frac{1}{2}}^j
\underbrace{\frac{2\epsilon_\rho
\com{\com{\beta_\rho^\dagger}{\beta_\rho}}{A_{\psi,l}^\dagger}}
{2\epsilon_\rho-\eta_{\psi,k}}}
\left(\prod_{m=l+1}^s A_{\psi,m}^\dagger\right)\left|0\right).\label{tipert29}
\end{eqnarray}
The trick now is to deal correctly with the indicated commutator.
\begin{eqnarray}
\com{\sum_{\rho=\frac{1}{2}}^j\frac{2\epsilon_\rho\com{\beta_\rho^\dagger}{\beta^\rho}}
{2\epsilon_\rho-\eta_{\psi,k}}}{A_{\psi,l}^\dagger}&=&\sum_{\rho=\frac{1}{2}}^j
\frac{2\epsilon_\rho}{\left(2\epsilon_\rho-\eta_{\psi,k}\right)
\left(2\epsilon_\rho-\eta_{\psi,l}\right)}
\beta_\rho^\dagger\nonumber\\
&=&\frac{\eta_{\psi,k}A_{\psi,k}^\dagger-\eta_{\psi,l}A_{\psi,l}^\dagger}
{\eta_{\psi,k}-\eta_{\psi,l}},\label{tipert30}
\end{eqnarray}
where the second line follows from the first if we notice that
\begin{equation}
\frac{2\epsilon_\rho}{\left(2\epsilon_\rho-\eta_{\psi,k}\right)\left(2\epsilon_\rho-\eta_{\psi,l}
\right)}
=\frac{1}{\eta_{\psi,k}-\eta_{\psi,l}}\left(\frac{\eta_{\psi,k}}{2\epsilon_\rho-\eta_{\psi,k}}-
\frac{\eta_{\psi,l}}{2\epsilon_\rho-\eta_{\psi,l}}\right).\label{tipert31}
\end{equation}
The last manipulation is not valid when both $\eta_{\psi,k}$ and $\eta_{\psi,l}$ are zero.
We side-step this problem by assuming that none of the $\eta_{\psi,k}$, $k=1,2,\ldots,s$ are zero, 
and later treat the case 
where some are. (The case where only one of the $\eta$'s is zero could have been included in the
present analysis, but fits more naturally with the treatment for the case where any number of
the $\eta$'s are zero.) By substituting eq. \ref{tipert30} into eq. \ref{tipert29} we find
\begin{eqnarray}
K_{0,\rm eff}\left|\psi\right)
&=&\eta_\psi\left|\psi\right)+\sqrt{\Omega}\Sc\sum_{k=1}^s\left(1-\frac{1}{\Omega-2s+2}
\sum_{\rho=\frac{1}{2}}^j\frac{2\epsilon_\rho}{2\epsilon_\rho-\eta_{\psi,k}}\right)
\prod_{l=1\not=k}^s A_{\psi,l}^\dagger
\left|0\right)\nonumber\\
&&+\frac{\sqrt{\Omega}\Sc}{\Omega-2s+2}\left\{
\sum_{k=1}^s\sum_{l=k+1}^s\frac{\eta_{\psi,k}}{\eta_{\psi,k}-\eta_{\psi,l}}
\left(\prod_{m=1\not=l}^s A_{\psi,m}^\dagger\right)\left|0\right)\right\}\nonumber\\
&&-\frac{\sqrt{\Omega}\Sc}{\Omega-2s+2}\left\{
\sum_{k=1}^s\sum_{l=k+1}^s\frac{\eta_{\psi,l}}{\eta_{\psi,k}-\eta_{\psi,l}}
\left(\prod_{m=1\not=k}^s A_{\psi,m}^\dagger\right)\left|0\right)\right\}.\label{tipert32}
\end{eqnarray}
We now enumerate differently the first $k,l$ summation by noting that for the set $\mathcal I$ of
indices over which the summation runs, it holds that
\begin{eqnarray}
\mathcal I&=&\left\{(k,l):k=1,2,\ldots,s;\hspace{2mm}l=k+1,k+2,\ldots,s\right\}\nonumber\\
&=&\left\{(k,l):l=1,2,\ldots,s;\hspace{2mm} k=1,2,\ldots,l-1\right\}.\label{tipert33}
\end{eqnarray}
and rename some dummy-indices to find
\begin{eqnarray}
K_{0,\rm eff}\left|\psi\right)
&=&\eta_\psi\left|\psi\right)\nonumber\\
&&+\sqrt{\Omega}\Sc\sum_{k=1}^s\left\{1-\frac{1}{\Omega-2s+2}\left(
\sum_{\rho=\frac{1}{2}}^j\frac{2\epsilon_\rho}{2\epsilon_\rho-\eta_{\psi,k}}
+2\sum_{l=\frac{1}{2}}^j\frac{\eta_{\psi,l}}{\eta_{\psi,l}-\eta_{\psi,k}}
\right)\right\}\nonumber\\
&&\hspace{9.6cm}\times\prod_{m=1\not=k}^s A_{\psi,m}^\dagger
\left|0\right)\nonumber.\\\label{tipert34}
\end{eqnarray}
If we choose the $\eta_{\psi,k}$ to satisfy 
\begin{equation}
1-\frac{1}{\Omega-2s+2}\left(
\sum_{\rho=\frac{1}{2}}^j\frac{2\epsilon_\rho}{2\epsilon_\rho-\eta_{\psi,k}}
+2\sum_{l=1\not=k}^s\frac{\eta_{\psi,l}}{\eta_{\psi,l}-\eta_{\psi,k}}
\right)=0,\label{tipert35}
\end{equation}
for $k=1,2,\ldots,s$ we have $K_{0,\rm eff}\left|\psi\right)=\eta_\psi\left|\psi\right)$.
Provided none of the $\eta_{\psi,k}$ are zero\footnote{The derivation involves dividing the 
equation enumerated by $k$ by $\eta_{\psi,k}$.}, this can be rewritten as
\begin{equation}
0=\sum_{\rho=\frac{1}{2}}^j\frac{1}{\eta_{\psi,k}-2\epsilon_\rho}+2\sum_{l=1\not=k}^s
\frac{1}{\eta_{\psi,l}-\eta_{\psi,k}},\hspace{2mm} k=1,2,\ldots,s,\label{tipert36}
\end{equation}
which are just Richardson's equations in the limit where the coupling constant dominates the 
one-particle energies ($\frac{1}{G}\longrightarrow0$).
Now we return to the case where one or more of the $\eta$ are assumed zero. Define an operator
$A_0^\dagger=\sum_{\rho=\frac{1}{2}}^j\frac{1}{2\epsilon_\rho}\beta_\rho^\dagger$, which is just
$A_{\psi,k}^\dagger$, with $\eta_{\psi,k}$ set to zero. The commutator between $K_{0,\rm eff}$ 
and $A_0^\dagger$ reduces from that given in eq. \ref{tipert26} to 
\begin{equation}
\com{K_{0,\rm eff}}{A_0}=\sqrt{\Omega}\Sc\left\{1-\frac{\nf-\Omega}{\Omega-2s+2}\right\}.
\label{tipert37}
\end{equation}
If $\left|2s-2\right)$ is any state with $2s-2$ ideal fermions, then 
$\left\{1-\frac{\nf-\Omega}{\Omega-2s+2}\right\}\left|2s-2\right)=0$ so that 
\begin{equation}
K_{0,\rm eff}A_0^\dagger\left|2s-2\right)=A_0^\dagger K_{0,\rm eff}\left|2s-2\right).
\label{tipert38}
\end{equation}
Therefore, the eigenstates of $K_{0,\rm eff}$ in the space with $2s-2$ ideal fermions, multiplied by
$A_0^\dagger$ are the eigenstates of $K_{0,\rm eff}$ with $2s$ ideal fermions and at least one of 
the $\eta_{\psi,k}$ zero. The implication is that $\eta_{\psi,k}=0$ solutions do exist and should
be treated as follows: The solutions with $t\leq s$ of the $\eta$ zero are obtained by finding
the eigenstates of $K_{0,\rm eff}$ with $2(s-t)$ fermions such that none of the $\eta_{\psi,k}$, 
$k=1,2,\ldots,s-t$ are zero. This is done by solving eq. \ref{tipert36} with $s$ replaced 
by $s-t$. The corresponding eigenstates of $K_{0,\rm eff}$ with $2s$ fermions and $t$ of the $\eta$ 
zero, are
\begin{equation}
\left|\psi\right)=\left(A_0^\dagger\right)^t\prod_{k=1}^{s-t}A_{\psi,k}^\dagger\left|0\right),
\label{tipert39}
\end{equation}
with eigenvalues $\eta_\psi=\sum_{k=1}^{s-t}\eta_{\psi,k}$.

We are now ready to write down the first order corrections to the energies, and the correct
zero'th order eigenstates that the perturbation selects, for each degenerate $s$-eigenspace of
$H_0$. For given $s$ and $t$ we enumerate the different solutions to 
\begin{equation}
0=\sum_{\rho=\frac{1}{2}}^j\frac{1}{\eta_{\psi,k}-2\epsilon_\rho}+2\sum_{l=1\not=k}^{s-t}
\frac{1}{\eta_{\psi,l}-\eta_{\psi,k}},\hspace{2mm} k=1,2,\ldots,s-t\label{tipert40}
\end{equation}
with different $\psi$. Corresponding to each $\psi$ is an eigenstate $\left|\psi,s,t\right)$
of $H_0$ given by
\begin{equation}
\left|\psi,s,t\right)=\frac{\left(B^\dagger\right)^{N-s}}{\sqrt{\left(N-s\right)!}}
\left(A_0^\dagger\right)^t\prod_{k=1}^{s-t}
\left(\sum_{\rho=\frac{1}{2}}^j\frac{\beta_\rho^\dagger}{2\epsilon_\rho-\eta_{\psi,k}}\right)
\left|0\right).\label{tipert41}
\end{equation}
These are the correct zero'th order eigenstates selected by the perturbation. To zero'th
order the state $\left|\psi,s,t\right)$ has an energy
\begin{equation}
E^{(0)}_s=-\frac{G}{\Omega}\left(N-s\right)\left(\Omega+1-N-s\right).\label{tipert42}
\end{equation}
To this is added a first order correction due to the perturbation
\begin{equation}
\Delta_{\psi,s,t}^{(1)}=\frac{\Omega-2N}{\Omega-2s}\sum_{k=1}^{s-t}\eta_{\psi,k}
+\left(N-s\right)\frac{\Omega}{\Omega-2s}2\bar{\epsilon}.\label{tipert43}
\end{equation}
From the eigenstates in eq. \ref{tipert40}, the subset that spans the physical subspace must now 
be selected. The procedure we employ to do this, is improvised specifically for this model.
We shift all the one-particle energies $\epsilon_\nu$ by the 
same amount $\epsilon$. On the level of the original Richardson Hamiltonian, this amounts
to adding a multiple of the total electron number operator, i.e. 
$H\longrightarrow H+\epsilon\sum_{\nu=-j}^ja_\nu^+ a_\nu$. This means that the energies
of the $2N$ electron eigenstates of the Richardson Hamiltonian are shifted by an amount
$2N\epsilon$. In the mapped system, only the physical eigenstates have anything to do with
the original Richardson Hamiltonian, and so the eigenvalues of physical eigenvectors must 
still shift by an amount $2N\epsilon$. States that do not lie in the physical subspace, do not
have to behave this way. (Although, there is nothing that says they cannot shift their eigenvalues
by an amount $2N\epsilon$). Therefore, if we find an eigenstate of the mapped system whose 
eigenvalue does not shift by an amount $2N\epsilon$ when each $\epsilon_\nu$ is shifted by an
amount $\epsilon$, we know it is a ghost state. Let us therefore calculate the shift in the 
first order correction $\Delta_{\psi,s,t}^{(1)}$ to the energy of the state 
$\left|\psi,s,t\right)$ when we replace each $\epsilon_\nu$ with 
$\epsilon_\nu'=\epsilon_\nu+\epsilon$. The average one-particle energy shifts from $\bar{\epsilon}$
to $\bar{\epsilon}+\epsilon$. The argument about the $\eta_{\psi,k}$ that are zero does not 
involve the values of $\epsilon_\nu$. Thus, for given $t$, there are still solutions with
only $s-t$ of the $\eta$ non-zero or in other words the $\eta$ that were zero before the shift
remains zero after the shift.
We we will denote the new non-zero $\eta$ by $\eta_{\psi,k}'$,
$k=1,2,\ldots,s-t$. They must satisfy eq. \ref{tipert40} with each $\epsilon_\rho$ replaced with
$\epsilon_\rho+\epsilon$. It is trivial to see that if $\eta_{\psi,k}$ satisfied eq. \ref{tipert40}
before we made the shift, then $\eta_{\psi,k}'=\eta_{\psi,k}+2\epsilon$ satisfies it after the
shift is made. Hence, the non-zero $\eta$ all get shifted by $2\epsilon$. This means that the
the first-order correction to the energy of state $\left|\psi,s,t\right)$ is
\begin{eqnarray}
\Delta_{\psi,s,t}^{(1)'}&=&\frac{\Omega-2N}{\Omega-2s}\sum_{k=1}^{s-t}\left(\eta_{\psi,k}+2\epsilon
\right)+\left(N-s\right)\frac{\Omega}{\Omega-2s}2\left(\bar{\epsilon}+\epsilon\right)
\nonumber\\
&=&\Delta_{\psi,s,t}^{(1)}+2N\epsilon-2t\frac{\Omega-2N}{\Omega-2s}.\label{tipert44}
\end{eqnarray}
To get a shift of $2N\epsilon$ we have to set $t=0$, which means that the states 
$\left|\psi,s,t\right)$ with $t\not=0$ are ghost states. 

The question now is whether these are all the ghost states. The answer is affirmative. To show this
we count the number of eigenstates we are left with after removing the $t\not=0$ states, and 
compare this number to the dimension of the physical subspace which we know to be 
$\Omega\choose N$. Let $\omega_s$ therefore be the number of $t=0$ states in the 
$s$-eigenspace. According to eq. \ref{tipert36}, these states are associated with finite solutions of
Richardson's equations for $2s$ ideal fermions in the limit of infinite coupling. If we accept
the result that was stated without proof in Section \ref{SS2.2}, that for given $s$ there
are ${\Omega \choose s}\times\frac{\Omega-2s+1}{\Omega-s+1}$ such solutions and therefore as many
$t=0$ linearly independent eigenstates in the $s$-eigenspace, we can compute the combined number of 
eigenstates left over in all the $s$-eigenspaces as follows:
\begin{eqnarray}
\sum_{s=0}^N \omega_s&=&\sum_{s=0}^N {\Omega \choose s}\times\frac{\Omega-2s+1}{\Omega-s+1}
\nonumber\\
&=&1+\sum_{s=1}^N\left\{{\Omega\choose s} - {\Omega \choose {s-1}}\right\}\nonumber\\
&=&\Omega \choose N.\label{tipert45}
\end{eqnarray}
This result says that the number of $t=0$ states just makes up the dimension of the physical 
subspace, which confirms that the $t=0$ states span the physical subspace.

At this point, let us collect the main results together (for the benefit of readers
who skipped over the preceding intricate arguments) and briefly discuss their implications.
We exclude ghost-states from our discussion. The unnormalized zero'th order physical eigenstates 
of the mapped Hamiltonian, selected by the perturbation $K$, are
\begin{equation}
\left|\psi,s\right)=\frac{\left(B^\dagger\right)^{N-s}}{\sqrt{\left(N-s\right)!}}
\prod_{k=1}^s
\left(\sum_{\rho=\frac{1}{2}}^j\frac{\beta_\rho^\dagger}{2\epsilon_\rho-\eta_{\psi,k}}\right)
\left|0\right),\label{tipert45a}
\end{equation}
where the parameters $\eta_{\psi,k}$, $k=1,\ldots,s$ refer to one of the 
$\omega_s=\Omega \choose N\times\frac{\Omega-2s+1}{\Omega-s+1}$ finite solutions of 
the strong coupling limit Richardson equations
\begin{equation}
0=\sum_{\rho=\frac{1}{2}}^j\frac{1}{\eta_{\psi,k}-2\epsilon_\rho}+2\sum_{l=1\not=k}^s
\frac{1}{\eta_{\psi,l}-\eta_{\psi,k}},\hspace{2mm} k=1,2,\ldots,s,\label{tipert45b}
\end{equation}
and different $\psi$ refer to different solutions. The quantum number $s$ takes on values from
$0$ for the ground state to $N$ for the most excited states. The energy of the state 
$\left|\psi,s\right)$ with the first order correction included is
\begin{equation}
E_{\psi,s}=\underbrace{-\frac{G}{\Omega}\left(N-s\right)\left(\Omega+1-N-s\right)}_{(1)}+
\underbrace{\left(N-s\right)\frac{\Omega}{\Omega-2s}2\bar{\epsilon}}_{(2)}
+\underbrace{\frac{\Omega-2N}{\Omega-2s}\sum_{k=1}^{s}\eta_{\psi,k}}_{(3)}.
\label{tipert45c}
\end{equation}
The results for the non-degenerate ground state is obtained by setting $s=0$ in the
the formulas for the eigenvector, eq. \ref{tipert45a}, and energy, eq. \ref{tipert45c}, and
dropping all products and summations where the lower bound exceeds the upper bound.
This is the same result as was found in \cite{YBA1}. The way we arrived at it though,
suggests an amusing new interpretation not discussed in \cite{YBA1}, for the low-lying spectrum to
this order in perturbation theory. We assume the number of one-particle levels $\Omega$ to
be large, so that, for low-lying states, terms of order $\frac{s}{\Omega}$ may be neglected. 
In that case term (1)
represents an equidistant spectrum with level-spacing $G$. It counts the number of 
bosons in the state $\left|\psi,s\right)$, assigning an energy $-\frac{G}{\Omega}N
\left(\Omega-N+1\right)$ if there are $N$ bosons, and approximately increases the energy by
an amount $G$ for each boson less than $N$. Thus, term (1) approximately constitutes an additive
`pairing' energy decrease of $G$ per boson. Term (2) can be seen as the total
one-particle energy contribution of the pairs of electrons that went into making the 
`Cooper pair bosons', loosely speaking.
This interpretation follows from the fact that the average one-particle energy of a single Cooper
pair is $\left<0\right|S\sum_{\nu=-j}^j\epsilon_\nu a_\nu^+ a_\nu S^+\left|0\right>=
2\bar{\epsilon}$ together with the fact that the state $\left|\psi,s\right)$ contains $N-s$
bosons. So term (2) represents an additive `internal' energy of $2\bar{\epsilon}$ per boson. 
(The factor $\frac{\Omega}{\Omega-s}$ is close to one for small $s$.) 
For given $s$, term (3) represents the energy 
contribution of $2s$ ideal fermions, interacting through the Richardson Hamiltonian, where
the energy scale is scaled by
a factor of approximately $1-\frac{2N}{\Omega}$ and   
the coupling constant $G$ is sent to infinity while insisting that only 
finite energy eigenstates are accessible to the ideal fermions. This interpretation of
term (3) is corroborated by the form of the eigenstate $\left|\psi,s\right)$ in eq. \ref{tipert45a}.

The fact that the ideal fermions are arranged in eigenstates of the original (hermitian)
Richardson Hamiltonian with a strong pairing interaction, implies that they are mutually orthogonal.
This is an amazing result:
Because the unperturbed mapped Hamiltonian is hermitian, the various $s$-eigenspaces are
orthogonal to each other. However, since the perturbation $K$ and more specifically the component
$K_0$ that determines the zero'th order eigenstates, is non-hermitian, the 
zero'th order eigenbasis for an $s$-eigenspace that it selects, will not be orthogonal. Yet, for
the mapped Richardson Hamiltonian the non-orthogonality disappears when the ghost states are
removed. In other words, only the ghost states break the orthogonality of the basis selected by
$K$, so that the orthogonality of the zero'th order physical eigenstates is preserved by the 
mapping. This makes the task of finding the left basis that goes along with the states 
$\left|\psi,s\right)$ simple. It should however be noted that these remarks do not imply
that the boson-fermion mapping is unitary. Although the orthogonality of the states 
$\left|\psi,s\right)$ are preserved, the mapping still introduces stretching of the norms
of these states.

To proceed to higher orders in the perturbation expansion now `only' involves the calculation of
matrix elements of the perturbation $K$ and the summation of terms 
containing products of these matrix elements. The calculations involved soon become sufficiently
inelegant to convince one that one is only interested in the first order energy corrections
and that the correct zero'th order eigenstates will do just fine. Yet, should one proceed to
higher orders in perturbation theory by 
calculating the necessary matrix elements, the least unpleasant route
to  follow is, I believe, to work with the mapped system. Then one can express the $\Pi_l$ 
operators in terms of free particle basis states of
the form $\frac{\left(B^\dagger\right)^{N-s}}{\sqrt{(N-s)!}}\prod_{k=1}^s\beta_{\nu_s}^\dagger
\left|0\right)$, for which the action of the perturbation is not too hard to calculate.  
Honesty compels me to mention at this point that
the method of \cite{YBA1} has an advantage over the above procedure, namely that the calculation 
of matrix elements is completely circumvented. However, the method is designed only for the 
Richardson Model. There might thus be models for which the only way to proceed is conventional 
perturbation theory, as we have done, in which case the Dyson mapping might be a useful tool.

In the work we have
done so far we have not yet used the full image of the Richardson Hamiltonian. We have only used
the expressions for $H_0$ and $K_{0}$. The expressions for $K_{+}$ and $K_{-}$
are as yet untested. Therefore I calculate the second order correction to the ground state
energy. This calculation is not too complicated and uses the full image of the Richardson
Hamiltonian.

To zero'th order, the ground state is $\left|{\rm GS}\right)=\frac{1}{\sqrt{N!}}
\left(B^\dagger\right)^N\left|0\right)$. According to eq. \ref{tipert24x} the second order
correction to the ground state energy is
\begin{equation}
\Delta_{\rm GS}^{(2)}=\sum_{s=0}^N \frac{\left({\rm GS}\right|H_1\Pi_sH_1\left|{\rm GS}\right)}
{E_0^{(0)}-E_s^{(0)}},\label{tipert46}
\end{equation}
where $\Pi_s$ is the projection operator that selects states with $2s$ ideal fermions and 
$N-s$ bosons, i.e. states in the $s$-eigenspace of $H_0$, and $H_1$ is the 
perturbation $K$. Again the decomposition $K=K_{+}+K_{0}+K_{-}$
is very useful. Because $K_{0}$ leaves every $s$-eigenspace invariant, while $K_{\pm}$
takes states from the $s$ eigenspace to the $s\pm1$ eigenspace, the expression for the second
order correction to the ground state energy reduces to
\begin{equation}
\Delta_{\rm GS}^{(2)}=\frac{\left({\rm GS}\right|K_{-}
\Pi_{s=1} K_{+}\left|{\rm GS}\right)}
{E_0^{(0)}-E_1^{(0)}},\label{tipert47}
\end{equation}
We conveniently choose the orthonormal basis $\left|\nu\right)=\frac{\left(B^\dagger\right)^{N-1}}
{\sqrt{(N-1)!}}\beta_\mu^\dagger\left|0\right)$ for the $(s=1)$-eigenspace, so that the operator
$\Pi_{s=1}$ can be expressed as $\Pi_{s=1}=\sum_{\mu=\frac{1}{2}}^j\left|\mu\right)
\left(\mu\right|$. If we substitute this into the expression for the second order correction
to the ground state energy, we see that we have managed to express the correction in terms of the
matrix elements $\left({\rm GS}\right|K_{-}\left|\mu\right)$ and
$\left(\mu\right|K_{+}\left|{\rm GS}\right)$. These matrix elements
are not hard to calculate, the result being
\begin{eqnarray}
\left({\rm GS}\right|\sum_\nu\epsilon_\nu K_{\nu-}\left|\mu\right)&=&2\epsilon_\mu\sqrt{\frac{N}
{\Omega}},\nonumber\\
\left(\mu\right|\sum_\nu\epsilon_\nu K_{\nu+}\left|{\rm GS}\right)&=&2\left(\epsilon_\nu-
\bar{\epsilon}\right)\sqrt{\frac{N}{\Omega}}\frac{\Omega-N}{\Omega-1}.\label{tipert48}
\end{eqnarray}
The energy difference in the denominator is $E_0^{(0)}-E_1^{(0)}=-G$. Combining everything
we find a second order correction to the ground state energy
\begin{equation}
\Delta_{\rm GS}^{(2)}=-\frac{4N\left(\Omega-N\right)}{G\left(\Omega-1\right)}{\rm var}(\epsilon),
\label{tipert49}
\end{equation}
where ${\rm var}(\epsilon)=\frac{\sum_{\nu=\frac{1}{2}}^j\epsilon_\nu^2}{\Omega}-\bar{\epsilon}^2$.
Note here how we used a basis 
$\left\{\left|\nu\right)\right\}_{\nu=\frac{1}{2}}^j$ for the 
$(s=1)$-eigenspace that mixes physical and ghost state. Again, the result agrees with the result
obtained by \cite{YBA1}. (The method of the authors of \cite{YBA1} allowed them to calculate
this particular correction to the seventh order.)

\section{Time-Dependent Perturbation Theory}
\markright{\bf Section \thesection: Time-Dependent Perturbation Theory}
\addtocontents{toc}{\protect\nopagebreak}
The good news about the results of eq. \ref{tipert45c} and eq. \ref{tipert49}, obtained through 
time-independent perturbation theory is
that they agree with the results obtained by \linebreak Yuzbashyan and co-workers in \cite{YBA1}. 
In a sense this is, however, also the bad news: results already exist to compare with everything we 
calculated. As mentioned in the introduction, the time-independent
perturbation expansion that uses the boson-fermion mapping does not yield anything that is
not in the scope of previous analyses. In an attempt to calculate something with the aid of
the boson-fermion mapping that would otherwise
be hard to calculate, we therefore consider a time-dependent perturbation expansion for transition
amplitudes between states of the unperturbed system. The reason I believe this calculation
to be simplified by the mapping will be highlighted at the appropriate point in the derivation
of the expansion.

We start by reviewing some basics of time-dependent perturbation theory.
We have a quantum system with a Hamiltonian $\tilde{H}=\tilde{H}_0+\tilde{H}_1$ where both 
$\tilde{H}_0$ 
and $\tilde{H}_1$ are hermitian.
It is assumed that the orthonormal eigenstates of $\tilde{H}_0$ are known. We denote these 
eigenstates
$\left|\alpha\right>$, with $\tilde{H}_0\left|\alpha\right>=E_\alpha\left|\alpha\right>$. 
Here $\alpha$
stands for all the quantum numbers necessary to specify a state uniquely, and it is immaterial
whether the spectrum is degenerate. We assume that $\tilde{H}_1$ is a small perturbation to 
$\tilde{H}_0$. For this system let $\tilde{U}(t)$ be the time-evolution operator
which satisfies $i\hbar\frac{d\tilde{U}(t)}{dt}=\tilde{H}\tilde{U}(t)$, with the boundary condition
$\tilde{U}(0)=\rm I$. We define an interaction picture time-evolution operator
\begin{equation}
\tilde{U}_I(t)=\exp\left(-\frac{i\tilde{H}_0t}{\hbar}\right)\tilde{U}(t),\label{tdpert1}
\end{equation}
which is close to identity when $\tilde{H}_1$ is small. Hence we have the hope that it is possible
to expand $\tilde{U}_I(t)$ in powers of $\tilde{H}_1$ around identity. To find this expansion we 
go about as follows. First we note that $\tilde{U}_I(t)$ obeys a differential equation similar to 
that obeyed by $\tilde{U}(t)$, namely
\begin{equation}
i\hbar\frac{d\tilde{U}_I(t)}{dt}=\tilde{H}_{1I}(t)\tilde{U}_I(t),\label{tdpert2}
\end{equation}
if we define $\tilde{H}_{1I}=\exp\left(i\frac{\tilde{H}_0t}{\hbar}\right)\tilde{H}_1\exp
\left(-i\frac{\tilde{H}_0t}{\hbar}\right)$.
This differential equation can be converted into an integral equation, 
\begin{equation}
\tilde{U}_I(t)={\rm I}-\frac{i}{\hbar}\int_0^tdt'\hspace{1mm}\tilde{H}_{1I}(t')
+\left(-\frac{i}{\hbar}\right)^2
\int_0^tdt'\hspace{1mm}\int_0^{t'}dt''\hspace{1mm}\tilde{H}_{1I}(t')\tilde{H}_{1I}(t'')+\ldots,
\label{tdpert3}
\end{equation}
taking into account the boundary condition $\tilde{U}_I(t=0)={\rm I}$. Our goal is to calculate the
transition amplitude at time $t$ between eigenstates of $\tilde{H}_0$, i.e. we are interested
in a quantity $\left<\beta\right|\tilde{U}(t)\left|\alpha\right>$. Using the definition of 
$\tilde{U}_I(t)$ we can rewrite this in terms of a matrix element of $\tilde{U}_I(t)$ as follows:
\begin{equation}
\left<\beta\right|\tilde{U}(t)\left|\alpha\right>=\exp\left(-i\frac{E_\beta t}{\hbar}\right)
\left<\beta\right|\tilde{U}_I(t)\left|\alpha\right>.\label{tdpert4}
\end{equation}
The task that remains is to calculate the matrix element 
$\left<\beta\right|\tilde{U}_I(t)\left|\alpha\right>$ to some order in 
$\tilde{H}_{1I}$ with the aid of 
the power expansion for $\tilde{U}_I(t)$ in eq. \ref{tdpert3}. This involves calculating matrix
elements of operators like $\tilde{H}_{1I}(t')\tilde{H}_{1I}(t'')$ with respect to eigenstates of 
$\tilde{H}_0$. 

Let us suppose at this point that the Hamiltonian $\tilde{H}$ that we started with is a fermion
Hamiltonian for which a boson-fermion mapping is available, and let us further suppose that
it is a simpler task to compute the above-mentioned matrix elements for the mapped system. 
In this section, we use the following notation: For the similarity transformed operators
we will use the notation $Q=T\tilde{Q}T^{-1}$. We have here kept quiet about the fact that
we have at our disposal extended images of operators and functionals. Rather untidily we will from
now on understand $H_0$ to denote the operator defined on the whole boson-fermion Fock
space which, when restricted to the physical subspace, is equivalent to $\tilde{H}_0$. We can choose
as bi-orthonormal basis for the whole boson-fermion Fock space, the left and right eigenstates of 
this operator, properly normalized:
\begin{eqnarray}
H_0\left|\nu_R\right)&=&E_{\nu}\left|\nu_R\right),\nonumber\\
\left(\mu_L\right|H_0&=&\left(\mu_L\right|E_\mu,\nonumber\\
\left(\mu_L\right.\left|\nu_R\right)&=&\delta_{\mu\nu}.
\label{tdpert5}
\end{eqnarray}
The $\left|\nu_R\right)$ and $\left|\nu_L\right)$ bases contain states outside
the physical subspace, and thanks to the possible degeneracy of physical and ghost states may
even mix physical and ghost states together. None the less, the identity in the whole boson-fermion
Fock space, in terms of these states, is
\begin{equation}
{\rm I}=\sum_\nu\left|\nu_R\right)\left(\nu_L\right|.\label{tdpert6}
\end{equation}
In terms of our mapped operators and states, we find to second order in $H_{1I}$
\begin{eqnarray}
\left<\beta\right|\tilde{U}_I(t)\left|\alpha\right>&=&\delta_{\alpha\beta}
-\frac{i}{\hbar}\int_0^tdt'\hspace{1mm}\exp\left(i\frac{(E_\beta-E_\alpha)t'}{\hbar}\right)
\left<\beta_L\right|T^{-1}H_1T\left|\alpha_R\right>\nonumber\\
&&+\left(-\frac{i}{\hbar}\right)^2\sum_{\tilde{\nu}}\int_0^tdt'\hspace{1mm}\int_0^{t'}dt''
\hspace{1mm}
\exp\left(i\frac{(E_\beta-E_\nu)t'}{\hbar}\right)\exp\left(i\frac{(E_\nu
-E_\alpha)t''}{\hbar}\right)\nonumber\\
&&\hspace{70mm}\times\left<\beta_L\right|T^{-1}H_1\left|\nu_R\right)
\left(\nu_L\right|H_1T\left|\alpha_R\right>\nonumber\\
&&+\ldots,\label{tdpert7}
\end{eqnarray}
We now apply these formulas to the mapped equivalent $\left(H\right)_{X,BF}=K+H_0$ of the 
Richardson Hamiltonian that we derived in Section \ref{SS2.3}. As usual, we take the
unperturbed Hamiltonian to be $H_0$ that only includes the pairing effect, and
treat $K$, the one-particle energy contribution, as the perturbation. We will
calculate the ground state to ground state transition matrix element 
to second order in the perturbation
$K$. If the normalized ground state of the original 
system is $\left|{\rm GS}\right>$ then we know that for the mapped ground state we have
\begin{eqnarray}
\left|{\rm GS}_R\right)&=&T\left|{\rm GS}\right>\nonumber\\
&=&c\frac{\left(B^\dagger\right)^N}{\sqrt{N!}}\left|0\right),\label{tdpert8}
\end{eqnarray}
where $c$ is a (real) number which we will not need to calculate. Similarly we have
\begin{eqnarray}
\left({\rm GS}_L\right|&=&\left<{\rm GS}\right|T^{-1}\nonumber\\
&=&\left(0\right|\frac{B^N}{\sqrt{N!}}c^{-1}.\label{tdpert9}
\end{eqnarray}
The next step is to choose a convenient eigenbasis of the unperturbed Hamiltonian 
in which to express the identity operator. 
Because we only work to second order in the perturbation, and
because of the frequently mentioned property that the perturbation takes states in the 
$s$-eigenspace to states with components only in the $(s-1)$-, $s$- and $(s+1)$-eigenspaces, we 
need to specify this basis only up to the $(s=1)$-eigenspace for the ground state ($s=0$) to
ground state transition amplitude. We choose to work with the right eigenbasis consisting of
$\left|{\rm GS}_R\right)$ and $\left|\nu\right)=\frac{\left(B^\dagger\right)^{N-1}}
{\sqrt{(N-1)!}}\beta^\dagger_\nu\left|0\right)$, $\nu=\frac{1}{2},\ldots,j$ and the corresponding
left eigenbasis
$\left({\rm GS}_L\right|$ and $\left(\nu\right|=\left(0\right|\beta_\nu\frac{B^{N-1}}
{\sqrt{(N-1)!}}$, $\nu=\frac{1}{2},\ldots,j$. With this choice eq. \ref{tdpert7} reduces to
\begin{eqnarray}
\left<{\rm GS}\right|\tilde{U}_I(t)\left|{\rm GS}\right>&=&1
-\frac{it}{\hbar}
\left({\rm GS}_L\right|K_0\left|{\rm GS}_R\right)-\frac{t^2}{2\hbar^2}
\left({\rm GS}_L\right|K_0\left|{\rm GS}_R\right)^2\nonumber\\
&&+\left(-\frac{i}{\hbar}\right)^2\sum_{\nu}\int_0^tdt'\hspace{1mm}\int_0^{t'}dt''
\hspace{1mm}
\exp\left(-i\frac{Gt'}{\hbar}\right)\exp\left(i\frac{Gt''}{\hbar}\right)\nonumber\\
&&\hspace{20mm}\times\left({\rm GS}_L\right|K_-\left|\nu\right)
\left(\nu\right|K_+\left|{\rm GS}_R\right)\nonumber\\
&&+\ldots\label{tdpert10}
\end{eqnarray}
Since, in every term where $\left|GS_R\right)$ appears, there is a corresponding 
$\left(GS_L\right|$, the unknown proportionality factor $c$ is always canceled by its inverse
$c^{-1}$. Forgetting about the $c$'s then, we have already calculated all the matrix elements 
appearing in the above expression. It is here that the advantage of using the boson-fermion
mapping lies. The states with respect to which the matrix elements are calculated for the mapped
system are free particle states. This is to be contrasted with the strongly correlated eigenstates
of $\tilde{H}_0=-GS^+ S$ we would have had to use in the unmapped system. In the
case of time-independent perturbation expansions Yuzbashyan got around the problem of
calculating these matrix elements thanks to the Richardson equations. However, for
time-dependent perturbation expansions there is, as far as I can see, no similar trick
to avoid the calculation of the matrix elements. This makes the mapping very useful.

Evaluation of the double integral gives
\begin{equation}
\int_0^tdt'\hspace{1mm}\int_0^{t'}dt''
\hspace{1mm}
\exp\left(-i\frac{Gt'}{\hbar}\right)\exp\left(i\frac{Gt''}{\hbar}\right)=
\frac{\hbar t}{iG}-\frac{\hbar^2}{G^2}\left(\exp\frac{-itG}{\hbar}-1\right).\label{tdpert11}
\end{equation}
Substituting this, and the previously calculated expressions for the matrix elements of 
the perturbation into eq. \ref{tdpert10} we find
\begin{eqnarray}
\left<{\rm GS}\right|\tilde{U}_I(t)\left|{\rm GS}\right>&=&1-\frac{it}{\hbar}2N\bar{\epsilon}
-\frac{2t^2}{\hbar^2}N^2\bar{\epsilon}^2\nonumber\\
&&+\left\{\frac{1}{G^2}\left(\exp\frac{-itG}{\hbar}-1\right)+\frac{it}{G\hbar}\right\}
4N\frac{\Omega-N}{\Omega-1}{\rm var}(\epsilon).\label{tdpert12}
\end{eqnarray}
For this truncation to be valid two conditions have to hold. Firstly, as with the 
time-independent perturbation expansion, the coupling constant $G$ should dominate the 
one-particle energies. Secondly, the expansion is only valid for small enough times: 
the one-particle energies must be much smaller than $\frac{\hbar}{t}$. We can use
eq. \ref{tdpert12} to calculate the probability of finding the system in the unperturbed
ground state at a time $t\ll\frac{\hbar}{\Omega\bar{\epsilon}}$ if it was in the unperturbed
ground state at time $t=0$. The result is
\begin{equation}
\left|\left<{\rm GS}\right|U(t)\left|{\rm GS}\right>\right|^2=1-16N\frac{\Omega-N}{\Omega-1}
\frac{{\rm var}(\epsilon)}{G^2}\sin^2\frac{tG}{2\hbar}.\label{tdpert13}
\end{equation}
The same result could also have been obtained by calculating the total amplitude for scattering to 
excited states. We do not perform the calculation but merely mention the following. 
The first non-zero term in the expansion for the amplitude for scattering to excited states is of 
the order $\frac{1}{G}$. To get the total probability to order $\frac{1}{G^2}$ for excitation to any
state above the ground state, only the first (order $\frac{1}{G}$) non-zero term in the 
excitation amplitude is therefore required. To this order, only excitation to the $s=1$ level is 
possible.  
I contend that such an excitation represents the decay of one Cooper pair. The argument
is as follows:
We map
back to the fermion system to find that the states in
the first excited level (up to normalization) are given by
\begin{equation}
\left(S^+\right)^{N-1}\left(\sum_{\nu=\frac{1}{2}}^j\frac{(-)^{j-\nu}a_\nu^+ 
a_{-\nu}^+}{2\epsilon_\nu-\eta_\phi}\right)\left|0\right>,\label{tdpert14}
\end{equation}
where $\eta_\phi$ is a finite solution to $0=\sum_{\nu=\frac{1}{2}}^j\frac{1}{2\epsilon_\nu
-\eta_\phi}$. (Actually, we map the state in eq. \ref{tdpert14} in the forward direction and
confirm that it is equal to one of the zero'th order eigenstates in the $s=1$ level 
of the mapped Hamiltonian that is selected by the perturbation.)
In such a state, the collective pair operator $S^+$ creates a Cooper pair with `long-range' 
momentum space correlations, in the sense that all the pair operators 
$(-)^{j-\nu}a_{\nu}^+ a_{-\nu}^+$ contribute with equal amplitude. It is this 
type of correlation which defines a Cooper pair.
(The momentum 
we are talking about here is the momentum difference between the two particles in the pair. The
total momentum of the pair is zero. Also, we assume that momentum correlations closely mimic
Bloch wave-number correlations.) In contrast, if we set 
$\left|\nu\right>=(-)^{j-\nu}a_{\nu}^+ a_{-\nu}^+\left|0\right>$ and examine the 
amplitude $f_\phi(\nu)=\left<\nu\right|\sum_{\mu=\frac{1}{2}}^j\frac{(-)^{j-\mu}a_\mu^+ 
a_{-\mu}^+}{2\epsilon_\mu-\eta_\phi}\left|0\right>$ as a function of $\nu$, we find that it 
is sharply peaked. In this sense, the operator 
$\sum_{\mu=\frac{1}{2}}^j\frac{(-)^{j-\mu}a_\mu^+ 
a_{-\mu}^+}{2\epsilon_\mu-\eta_\phi}$ creates a two-electron state with only short-range
momentum correlations, and which is well-localized in momentum space (or more correctly, Bloch
wave-number space). Thus, the two particle state created by such an operator is `close' to an 
eigenstate of the two-particle system with the pairing
switched off and far from the type of correlation that marks a Cooper pair. 
A state in the $s=1$ level, such as that in eq. \ref{tdpert14}, therefore contains $N-1$ Cooper
pairs. The two remaining electrons are, to a good approximation, in eigenstates of the single
particle energy operator $\sum \epsilon_\nu a^+_\nu a_\nu$.

Note that without doing a detailed calculation we could have
known the probability to remain in the ground state is of the form
$1-c \sin^2\frac{tG}{2\hbar}$.
This form follows from the fact that, to second order, the only processes possible are 
(a) the system remains in its ground state and (b) a single Cooper pair decays and recombines, the
energy involved in this virtual transition being $G$. All the detailed information about 
the non-zero matrix elements of the perturbation resides in the coefficient $c$, which in this
case is equal to
$16N\frac{\Omega-N}{\Omega-1}\frac{{\rm var}(\epsilon)}{G^2}$. The fact that the energy dependence
takes the form of a variance, guarantees that shifting all one-particle energies by the same amount
does not influence the transition probability. Furthermore, for large $\Omega$, the expression
grows linearly with the number of particles. In other words, 
$1-\left|\left<{\rm GS}\right|U(t)\left|{\rm GS}\right>\right|^2$, which is the probability for
one Cooper pair to have decayed by time $t$, doubles if we double the number of particles.
This makes good sense. In our intuitive picture, Cooper pairs are independent entities that, to 
a good approximation, propagate without noticing each other.  
They have finite lifetimes because of the perturbation $K$.
In this picture, the probability for one Cooper pair out of $N$ to have decayed after a (short)
time $t$, is extensive, as it is the sum of the independent probabilities for Cooper pairs 
$1,2,\ldots$ or $N$ to have decayed by time $t$. For smaller $\Omega$ this independenc is
clearly inhibited by the factor $\frac{\Omega-N}{\Omega-1}$. 
\section{Conclusion}
\markright{\bf Section \thesection: Conclusion}
\addtocontents{toc}{\protect\nopagebreak}
Much of the analysis presented in this chapter is new, and
contains contributions relevant to both the fields of generalised Dyson mappings, and the
Richardson Hamiltonian. As far as the generalised Dyson mapping is concerned, we analysed
what is perhaps the simplest system for which the Hamiltonian cannot simply be written in terms
of only linear combinations of products of operators in the even sector of the superalgebra.
We also included states with more than one ideal fermion in the mapped system. To
my knowledge, no boson expansion method type analyses has previously been applied in such a 
general setting. It was pointed out
that, when there are $\Omega$ ideal fermions in the mapped system, the similarity transform of
Section \ref{SS1.7} breaks down. We demonstrated how to do perturbation theory, both time-dependent
and time-independent, for a mapped system. The ease with which a perturbation 
expansion can be made in the mapped system provides a motivation for doing a generalised Dyson
mapping.

As far as new insights into the Richardson model goes, it was demonstrated why doing a strong
coupling expansion is feasible. The key insight here is that the one-body term 
$\sum_\nu \epsilon_\nu a_\nu^+ a_\nu$ only introduces scattering between adjacent energy
levels of the infinite pairing system. This result could in principle be derived without doing
a mapping, but is immediately obvious only in the mapped system. 

I conclude with a speculation on possible
future developments involving the mapped version of the Richardson Hamiltonian.
In presenting the results of time-dependent perturbation theory, I have stressed that the aim
is not to understand measurable effects in a superconductor. The reasons for having to make
this cautionary remark are the following. Firstly, the strong coupling limit is not realised
in a superconductor. Secondly, we have little control over the state the electrons are in at 
the start of the experiment. It is not possible to prepare the system in the ground state of
the infinite pairing regime and then to let it evolve with a finite (large) pairing force.

A setup much more suited to the kind of calculation we performed here, is that of ultra-cold
Fermi gases \cite{GRJ1}. There atoms that are believed to obey Fermi statistics are cooled in a 
trap, 
to a point where they condense into what is loosely called a single macroscopic wave-function. 
By means of the clever use of external magnetic fields, a situation can be created in which the
interaction between the fermionic atoms can be adjusted. Depending on the type of interaction
the magnetic field is tuned for, the macroscopic wave-function is either a condensate of 
Cooper pairs, or a Bose-Einstein condensate of diatomic molecules. On a completely practical level,
it is good news that the 
pairing strength of the Cooper pairs can be adjusted. Therefore a system can be
prepared in the infinite pairing regime. In a short time-interval the pairing force may then be
decreased, and the resulting state of the system probed. An analysis of such a situation,
that goes beyond the mean field theory of \cite{FS1}, is
possible with the techniques developed in the section on time-dependent perturbation theory. 
The two challenges to this program are the computation
of high order corrections and the identification of observable
quantities that indicate the change the system undergoes. For instance, with reference to the
result in eq. \ref{tdpert13}, how might it be possible to see if a single Cooper pair has decayed?

On a more conceptual level, the whole question of when a system is a Bose-Einstein condensate
of diatomic molecules and when it is a condensate of Cooper pairs, and what happens in between,
may supply fertile ground for generalised Dyson boson-fermion mappings. Since a textbook 
system representing a Bose-Einstein condensate has many different one-particle levels that the
bosons can occupy, while the mapping we considered in this chapter piles all the bosons into the
same level, more work is required for this analysis. A mapping needs to be devised that allows more
one-particle levels for the bosons. The big stumbling block in this enterprise is that the 
similarity transformation of Section \ref{SS1.7} will then become much harder to deal with.

\backmatter
\markright

\end{document}